\PassOptionsToPackage{pdfencoding=auto,psdextra}{hyperref}

\documentclass[longauth,
  kazakh,
  russian,
  english]{aa}
\usepackage[LGR,T2A,T1]{fontenc}
\usepackage{CJKutf8}
\newcommand{\tinier}{\fontsize{7.5pt}{7.5pt}\selectfont}
\usepackage{setspace}

\usepackage{graphicx}
\usepackage{natbib}
\usepackage{cuted}
\usepackage{scalerel}
\usepackage{rotating}
\usepackage[table]{xcolor}
\usepackage{pdflscape}
\usepackage{placeins}
\usepackage{float}
\usepackage{xcolor}
\usepackage{orcidlink}

\usepackage{subcaption}
\usepackage{tikz}
\usepackage{xspace}

\newcommand{\red}[1]{\textcolor{black}{#1}}

\newcommand{\phildone}[1]{\textcolor{black}{#1}}
\newcommand{\black}[1]{\textcolor{black}{#1}}
\newcommand{\blue}[1]{\textcolor{blue}{#1}}
\newcommand{\comments}[1]{\textcolor{black}{#1}}

\newcommand{\fixed}[1]{\textcolor{black}{#1}}

\newcommand{\old}{$^{\star}$}

\newcommand{\gradeA}{$\num{6073}$\xspace}
\newcommand{\gradeB}{$\num{8999}$\xspace}
\newcommand{\gradeC}{$\num{9010}$\xspace}
\newcommand{\gradeX}{$\num{2580}$\xspace}
\newcommand{\gradeAB}{$\num{15072}$\xspace}
\newcommand{\gradeABC}{$\num{24082}$\xspace}
\newcommand{\gradeABCX}{$\num{26662}$\xspace}

\newcommand{\dspl}{$84$\xspace}
\newcommand{\discs}{$1000$\xspace}

\newcommand{\angelosmodels}{$\num{9538}$\xspace}

\newcommand{\randoms}{$\num{50000}$\xspace}
\newcommand{\drtwo}{$\num{60000}$\xspace}
\newcommand{\ews}{$\num{110000}$\xspace}
\newcommand{\dronearea}{$\num{2000}$\xspace}

\newcommand{\experts}{$\num{67}$\xspace}

\newcommand{\knownA}{$\num{417}$\xspace}
\newcommand{\knownB}{$\num{161}$\xspace}
\newcommand{\knownC}{$\num{55}$\xspace}

\newcommand{\stellarmass}{$\logten (M_\ast /M_\odot) = 11.39^{+0.29}_{-0.33}$\xspace}

\newcommand{\ssfr}{$\log_{10}(\textrm{sSFR/yr}^{-1}) = -11.25^{+0.57}_{-0.37}$\xspace}

\newcommand{\photoz}{$0.68 ^{+0.33}_{-0.29}$\xspace}

\newcommand{\thetaE}{$0.85 ^{+0.44}_{-0.24}$\xspace}

\newcommand{\orcid}[1]{\,\orcidlink{#1}}

\usepackage{txfonts}
\usepackage{hyperref}
\hypersetup{
    colorlinks=true,
    linkcolor=blue,
    filecolor=magenta,      
    urlcolor=blue,
    citecolor=blue
}
\makeatletter
\renewcommand*\aa@pageof{, page \thepage{} of \pageref*{LastPage}}
\makeatother

\usepackage[utf8]{inputenc}

\usepackage[switch, modulo]{lineno}
              
\renewcommand{\linenumbers}[0]{}

\usepackage{euclid}

\begin{document}

%
% Put the title and authors of your (Standard Project) paper here
%

% Hello and welcome to my paper :>

\title{Euclid Data Release 1 (DR1)}
\subtitle{Euclid Lens Identification Pipeline and Scientific Exploration (ECLIPSE) A: Strong lens catalogue overview\thanks{
Results in this paper were obtained with the help of the volunteers of the Space Warps project whose contributions are individually acknowledged on the Space Warps 
\href{https://www.zooniverse.org/projects/aprajita/space-warps-esa-euclid/about/team}{`The Team'} webpage and in Appendix \ref{app:sw}.}}

%% please do not edit the author list once you copy it from the
%% Publication Portal -- contact ECEB Bureau for changes

%%%% Version Wednesday 28th of October 2026 05:58:56 PM UT
%%%% Assumes the new A&A style file from Oct 2025 or later
%%%% Please do not edit the author list -- contact ECEB Bureau for changes
% \newcommand{\orcid}[1]{} %% if already defined in aa.cls: comment, or use renewcommand			   
\author{Euclid Collaboration: N.~E.~P.~Lines\orcid{0009-0004-7751-1914}\inst{\ref{aff1}}
\and P.~Holloway\orcid{0009-0002-8896-6100}\inst{\ref{aff1}}
\and T.~Li\orcid{0009-0005-5008-0381}\inst{\ref{aff1}}
\and T.~E.~Collett\orcid{0000-0001-5564-3140}\inst{\ref{aff1}}
\and S.~Schuldt\orcid{0000-0003-2497-6334}\thanks{\email{stefan.schuldt@utu.fi}}\inst{\ref{aff2},\ref{aff3},\ref{aff4}}
\and S.~H.~Vincken\orcid{0009-0005-7305-2359}\inst{\ref{aff5}}
\and J.~W.~Nightingale\orcid{0000-0002-8987-7401}\inst{\ref{aff6}}
\and A.~Nersesian\orcid{0000-0001-6843-409X}\inst{\ref{aff7},\ref{aff8}}
\and M.~Fogliardi\orcid{0009-0006-4964-5311}\inst{\ref{aff9},\ref{aff10}}
\and A.~Verma\orcid{0000-0002-0730-0781}\inst{\ref{aff11}}
\and C.~J.~Storfer\orcid{0000-0002-0385-0014}\inst{\ref{aff12}}
\and A.~Manj\'on-Garc\'ia\orcid{0000-0002-7413-8825}\inst{\ref{aff13}}
\and J.~Wilde\orcid{0000-0002-4460-7379}\inst{\ref{aff14}}
\and W.~J.~R.~Enzi\orcid{0009-0004-2992-3148}\inst{\ref{aff1},\ref{aff15}}
\and J.~A.~Acevedo~Barroso\orcid{0000-0002-9654-1711}\inst{\ref{aff16}}
\and A.~Sonnenfeld\orcid{0000-0002-6061-5977}\inst{\ref{aff17},\ref{aff18},\ref{aff19}}
\and K.~Rojas\orcid{0000-0003-1391-6854}\inst{\ref{aff20}}
\and E.~Baeten\inst{\ref{aff21}}
\and P.~Bergamini\orcid{0000-0003-1383-9414}\inst{\ref{aff10}}
\and C.~Cornen\orcid{0000-0002-7786-2798}\inst{\ref{aff21}}
\and F.~Courbin\orcid{0000-0003-0758-6510}\inst{\ref{aff14},\ref{aff22},\ref{aff23}}
\and H.~Degaudenzi\orcid{0000-0002-5887-6799}\inst{\ref{aff24}}
\and G.~Despali\orcid{0000-0001-6150-4112}\inst{\ref{aff25},\ref{aff10},\ref{aff26}}
\and R.~Gavazzi\orcid{0000-0002-5540-6935}\inst{\ref{aff27},\ref{aff28}}
\and C.~Macmillan\inst{\ref{aff21}}
\and R.~B.~Metcalf\orcid{0000-0003-3167-2574}\inst{\ref{aff25},\ref{aff10}}
\and D.~J.~Bowden\orcid{0009-0008-6114-1401}\inst{\ref{aff11}}
\and R.~Ca\~nameras\orcid{0000-0002-2468-5169}\inst{\ref{aff27}}
\and J.~H.-H.~Chan\orcid{0000-0001-8797-725X}\inst{\ref{aff29},\ref{aff30}}
\and L.~R.~Ecker\orcid{0009-0005-3508-2469}\inst{\ref{aff31},\ref{aff32},\ref{aff33}}
\and M.~Grespan\orcid{0000-0002-8213-0217}\inst{\ref{aff11},\ref{aff34}}
\and Q.~He\orcid{0000-0003-3672-9365}\inst{\ref{aff35}}
\and D.~P.~Johnson\orcid{0000-0002-0311-2513}\inst{\ref{aff36}}
\and C.~Krawczyk\orcid{0000-0001-9233-2341}\inst{\ref{aff1}}
\and N.~Massari\orcid{0009-0006-8421-930X}\inst{\ref{aff27}}
\and A.~Melo\orcid{0000-0002-6449-3970}\inst{\ref{aff37},\ref{aff38},\ref{aff39}}
\and L.~A.~Moustakas\orcid{0000-0003-3030-2360}\inst{\ref{aff16}}
\and C.~Olave\inst{\ref{aff40}}
\and A.~Paulino-Afonso\orcid{0000-0002-0943-0694}\inst{\ref{aff41},\ref{aff42}}
\and C.~Saulder\orcid{0000-0002-0408-5633}\inst{\ref{aff32},\ref{aff31}}
\and D.~Sluse\orcid{0000-0001-6116-2095}\inst{\ref{aff7}}
\and G.~Smadja\inst{\ref{aff43}}
\and H.~Srivastava\orcid{0009-0002-3254-0952}\inst{\ref{aff44}}
\and D.~Stern\orcid{0000-0003-2686-9241}\inst{\ref{aff16}}
\and T.~T.~Thai\orcid{0000-0002-8408-4816}\inst{\ref{aff45}}
\and C.~Tortora\orcid{0000-0001-7958-6531}\inst{\ref{aff46}}
\and N.~Wielders\orcid{0009-0001-8110-819X}\inst{\ref{aff6}}
\and I.~T.~Andika\orcid{0000-0001-6102-9526}\inst{\ref{aff31}}
\and A.~Angress\orcid{0009-0005-5250-9487}\inst{\ref{aff47},\ref{aff30}}
\and D.~J.~Ballard\orcid{0009-0003-3198-7151}\inst{\ref{aff48}}
\and M.~Bradac\orcid{0000-0001-5984-0395}\inst{\ref{aff49}}
\and D.~Delley\orcid{0000-0002-4958-7469}\inst{\ref{aff32}}
\and P.~Fleury\orcid{0000-0001-9292-3651}\inst{\ref{aff36}}
\and A.~Galan\orcid{0000-0003-2547-9815}\inst{\ref{aff24}}
\and G.~Granata\orcid{0000-0002-9512-3788}\inst{\ref{aff1}}
\and T.~Harvey\orcid{0000-0002-4130-636X}\inst{\ref{aff50},\ref{aff51}}
\and N.~B.~Hogg\orcid{0000-0001-9346-4477}\inst{\ref{aff52},\ref{aff53}}
\and K.~Jahnke\orcid{0000-0003-3804-2137}\inst{\ref{aff51}}
\and K.~Kuijken\orcid{0000-0002-3827-0175}\inst{\ref{aff54}}
\and T.~Li\orcid{0009-0003-5064-1914}\inst{\ref{aff55}}
\and P.~J.~Marshall\orcid{0000-0002-0113-5770}\inst{\ref{aff56},\ref{aff57}}
\and M.~Meneghetti\orcid{0000-0003-1225-7084}\inst{\ref{aff10},\ref{aff26}}
\and A.~More\orcid{0000-0001-7714-7076}\inst{\ref{aff58},\ref{aff59}}
\and B.~C.~Nagam\orcid{0000-0002-3724-7694}\inst{\ref{aff60}}
\and D.~O'Ryan\orcid{0000-0003-1217-4617}\inst{\ref{aff61}}
\and J.~Pearson\orcid{0000-0001-8555-8561}\inst{\ref{aff62}}
\and H.~Qu\orcid{0009-0006-0299-0265}\inst{\ref{aff63}}
\and D.~Ryczanowski\orcid{0000-0002-4429-3429}\inst{\ref{aff1},\ref{aff64}}
\and T.~Saifollahi\orcid{0000-0002-9554-7660}\inst{\ref{aff65},\ref{aff66}}
\and S.~Serjeant\orcid{0000-0002-0517-7943}\inst{\ref{aff62}}
\and C.~Spiniello\orcid{0000-0002-3909-6359}\inst{\ref{aff39},\ref{aff11}}
\and L.~Trobbiani\orcid{0009-0000-8058-834X}\inst{\ref{aff40},\ref{aff15}}
\and X.~Xu\orcid{0009-0002-0985-2309}\inst{\ref{aff55},\ref{aff67}}
\and M.~Zumalacarregui\orcid{0000-0002-9943-6490}\inst{\ref{aff68},\ref{aff69}}
\and A.~L.~Faisst\orcid{0000-0002-9382-9832}\inst{\ref{aff63}}
\and B.~Altieri\orcid{0000-0003-3936-0284}\inst{\ref{aff61}}
\and S.~Andreon\orcid{0000-0002-2041-8784}\inst{\ref{aff70}}
\and N.~Auricchio\orcid{0000-0003-4444-8651}\inst{\ref{aff10}}
\and C.~Baccigalupi\orcid{0000-0002-8211-1630}\inst{\ref{aff71},\ref{aff72},\ref{aff73},\ref{aff74}}
\and M.~Baldi\orcid{0000-0003-4145-1943}\inst{\ref{aff40},\ref{aff10},\ref{aff26}}
\and A.~Balestra\orcid{0000-0002-6967-261X}\inst{\ref{aff75}}
\and S.~Bardelli\orcid{0000-0002-8900-0298}\inst{\ref{aff10}}
\and P.~Battaglia\orcid{0000-0002-7337-5909}\inst{\ref{aff10}}
\and R.~Bender\orcid{0000-0001-7179-0626}\inst{\ref{aff32},\ref{aff31}}
\and A.~Biviano\orcid{0000-0002-0857-0732}\inst{\ref{aff72},\ref{aff71}}
\and M.~Bolzonella\orcid{0000-0003-3278-4607}\inst{\ref{aff10}}
\and E.~Branchini\orcid{0000-0002-0808-6908}\inst{\ref{aff76},\ref{aff77},\ref{aff70}}
\and M.~Brescia\orcid{0000-0001-9506-5680}\inst{\ref{aff78},\ref{aff46}}
\and S.~Camera\orcid{0000-0003-3399-3574}\inst{\ref{aff79},\ref{aff80},\ref{aff81}}
\and G.~Ca\~nas-Herrera\orcid{0000-0003-2796-2149}\inst{\ref{aff54}}
\and V.~Capobianco\orcid{0000-0002-3309-7692}\inst{\ref{aff81}}
\and C.~Carbone\orcid{0000-0003-0125-3563}\inst{\ref{aff4}}
\and V.~F.~Cardone\inst{\ref{aff82},\ref{aff83}}
\and J.~Carretero\orcid{0000-0002-3130-0204}\inst{\ref{aff84},\ref{aff85}}
\and M.~Castellano\orcid{0000-0001-9875-8263}\inst{\ref{aff82}}
\and G.~Castignani\orcid{0000-0001-6831-0687}\inst{\ref{aff10}}
\and S.~Cavuoti\orcid{0000-0002-3787-4196}\inst{\ref{aff46},\ref{aff86}}
\and K.~C.~Chambers\orcid{0000-0001-6965-7789}\inst{\ref{aff12}}
\and A.~Cimatti\inst{\ref{aff87}}
\and C.~Colodro-Conde\inst{\ref{aff88}}
\and G.~Congedo\orcid{0000-0003-2508-0046}\inst{\ref{aff89}}
\and C.~J.~Conselice\orcid{0000-0003-1949-7638}\inst{\ref{aff50}}
\and L.~Conversi\orcid{0000-0002-6710-8476}\inst{\ref{aff90},\ref{aff61}}
\and Y.~Copin\orcid{0000-0002-5317-7518}\inst{\ref{aff43}}
\and A.~Costille\inst{\ref{aff27}}
\and H.~M.~Courtois\orcid{0000-0003-0509-1776}\inst{\ref{aff91}}
\and M.~Cropper\orcid{0000-0003-4571-9468}\inst{\ref{aff92}}
\and G.~De~Lucia\orcid{0000-0002-6220-9104}\inst{\ref{aff72}}
\and C.~Dolding\orcid{0009-0003-7199-6108}\inst{\ref{aff92}}
\and H.~Dole\orcid{0000-0002-9767-3839}\inst{\ref{aff93}}
\and F.~Dubath\orcid{0000-0002-6533-2810}\inst{\ref{aff24}}
\and X.~Dupac\inst{\ref{aff61}}
\and S.~Dusini\orcid{0000-0002-1128-0664}\inst{\ref{aff94}}
\and S.~Escoffier\orcid{0000-0002-2847-7498}\inst{\ref{aff95}}
\and M.~Farina\orcid{0000-0002-3089-7846}\inst{\ref{aff96}}
\and R.~Farinelli\inst{\ref{aff10}}
\and F.~Faustini\orcid{0000-0001-6274-5145}\inst{\ref{aff82}}
\and S.~Ferriol\inst{\ref{aff43}}
\and S.~Fotopoulou\orcid{0000-0002-9686-254X}\inst{\ref{aff97}}
\and M.~Frailis\orcid{0000-0002-7400-2135}\inst{\ref{aff72}}
\and E.~Franceschi\orcid{0000-0002-0585-6591}\inst{\ref{aff10}}
\and M.~Fumana\orcid{0000-0001-6787-5950}\inst{\ref{aff4}}
\and L.~Gabarra\orcid{0000-0002-8486-8856}\inst{\ref{aff11}}
\and S.~Galeotta\orcid{0000-0002-3748-5115}\inst{\ref{aff72}}
\and K.~George\orcid{0000-0002-1734-8455}\inst{\ref{aff98}}
\and B.~Gillis\orcid{0000-0002-4478-1270}\inst{\ref{aff89}}
\and C.~Giocoli\orcid{0000-0002-9590-7961}\inst{\ref{aff10},\ref{aff26}}
\and J.~Gracia-Carpio\orcid{0000-0003-4689-3134}\inst{\ref{aff32}}
\and A.~Grazian\orcid{0000-0002-5688-0663}\inst{\ref{aff75}}
\and F.~Grupp\inst{\ref{aff32},\ref{aff31}}
\and S.~Gwyn\orcid{0000-0001-8221-8406}\inst{\ref{aff99}}
\and W.~G.~Hartley\orcid{0000-0002-4563-5101}\inst{\ref{aff24}}
\and S.~V.~H.~Haugan\orcid{0000-0001-9648-7260}\inst{\ref{aff100}}
\and S.~Hemmati\orcid{0000-0003-2226-5395}\inst{\ref{aff63}}
\and H.~Hoekstra\orcid{0000-0002-0641-3231}\inst{\ref{aff54}}
\and W.~Holmes\orcid{0009-0007-8554-4646}\inst{\ref{aff16}}
\and I.~M.~Hook\orcid{0000-0002-2960-978X}\inst{\ref{aff101}}
\and F.~Hormuth\inst{\ref{aff102}}
\and A.~Hornstrup\orcid{0000-0002-3363-0936}\inst{\ref{aff103},\ref{aff104}}
\and M.~Jhabvala\inst{\ref{aff105}}
\and B.~Joachimi\orcid{0000-0001-7494-1303}\inst{\ref{aff106}}
\and S.~Kermiche\orcid{0000-0002-0302-5735}\inst{\ref{aff95}}
\and A.~Kiessling\orcid{0000-0002-2590-1273}\inst{\ref{aff16}}
\and B.~Kubik\orcid{0009-0006-5823-4880}\inst{\ref{aff43}}
\and M.~K\"ummel\orcid{0000-0003-2791-2117}\inst{\ref{aff31}}
\and M.~Kunz\orcid{0000-0002-3052-7394}\inst{\ref{aff107}}
\and H.~Kurki-Suonio\orcid{0000-0002-4618-3063}\inst{\ref{aff3},\ref{aff108}}
\and A.~M.~C.~Le~Brun\orcid{0000-0002-0936-4594}\inst{\ref{aff109}}
\and S.~Ligori\orcid{0000-0003-4172-4606}\inst{\ref{aff81}}
\and P.~B.~Lilje\orcid{0000-0003-4324-7794}\inst{\ref{aff100}}
\and V.~Lindholm\orcid{0000-0003-2317-5471}\inst{\ref{aff3},\ref{aff108}}
\and I.~Lloro\orcid{0000-0001-5966-1434}\inst{\ref{aff110}}
\and M.~Magliocchetti\orcid{0000-0001-9158-4838}\inst{\ref{aff96}}
\and G.~Mainetti\orcid{0000-0003-2384-2377}\inst{\ref{aff111}}
\and D.~Maino\inst{\ref{aff112},\ref{aff4},\ref{aff113}}
\and O.~Mansutti\orcid{0000-0001-5758-4658}\inst{\ref{aff72}}
\and O.~Marggraf\orcid{0000-0001-7242-3852}\inst{\ref{aff114}}
\and M.~Martinelli\orcid{0000-0002-6943-7732}\inst{\ref{aff82},\ref{aff83}}
\and N.~Martinet\orcid{0000-0003-2786-7790}\inst{\ref{aff27}}
\and F.~Marulli\orcid{0000-0002-8850-0303}\inst{\ref{aff25},\ref{aff10},\ref{aff26}}
\and R.~J.~Massey\orcid{0000-0002-6085-3780}\inst{\ref{aff115}}
\and N.~Mauri\orcid{0000-0001-8196-1548}\inst{\ref{aff87},\ref{aff26}}
\and E.~Medinaceli\orcid{0000-0002-4040-7783}\inst{\ref{aff10}}
\and M.~Melchior\inst{\ref{aff5}}
\and E.~Merlin\orcid{0000-0001-6870-8900}\inst{\ref{aff75}}
\and G.~Meylan\orcid{0000-0001-6503-0209}\inst{\ref{aff116}}
\and P.~Monaco\orcid{0000-0003-2083-7564}\inst{\ref{aff117},\ref{aff72},\ref{aff73},\ref{aff71}}
\and A.~Mora\orcid{0000-0002-1922-8529}\inst{\ref{aff118}}
\and M.~Moresco\orcid{0000-0002-7616-7136}\inst{\ref{aff25},\ref{aff10}}
\and C.~Moretti\orcid{0000-0003-3314-8936}\inst{\ref{aff72},\ref{aff71},\ref{aff73}}
\and L.~Moscardini\orcid{0000-0002-3473-6716}\inst{\ref{aff25},\ref{aff10},\ref{aff26}}
\and R.~Nakajima\orcid{0009-0009-1213-7040}\inst{\ref{aff114}}
\and C.~Neissner\orcid{0000-0001-8524-4968}\inst{\ref{aff119},\ref{aff85}}
\and R.~C.~Nichol\orcid{0000-0003-0939-6518}\inst{\ref{aff120}}
\and S.-M.~Niemi\orcid{0009-0005-0247-0086}\inst{\ref{aff121}}
\and C.~Padilla\orcid{0000-0001-7951-0166}\inst{\ref{aff119}}
\and S.~Paltani\orcid{0000-0002-8108-9179}\inst{\ref{aff24}}
\and F.~Pasian\orcid{0000-0002-4869-3227}\inst{\ref{aff72}}
\and W.~J.~Percival\orcid{0000-0002-0644-5727}\inst{\ref{aff122},\ref{aff123},\ref{aff124}}
\and V.~Pettorino\orcid{0000-0002-4203-9320}\inst{\ref{aff121}}
\and A.~Pezzotta\orcid{0000-0003-0726-2268}\inst{\ref{aff70}}
\and G.~Polenta\orcid{0000-0003-4067-9196}\inst{\ref{aff125}}
\and M.~Poncet\inst{\ref{aff66}}
\and L.~A.~Popa\inst{\ref{aff126}}
\and L.~Pozzetti\orcid{0000-0001-7085-0412}\inst{\ref{aff10}}
\and F.~Raison\orcid{0000-0002-7819-6918}\inst{\ref{aff32}}
\and R.~Rebolo\orcid{0000-0003-3767-7085}\inst{\ref{aff88},\ref{aff127},\ref{aff128}}
\and A.~Renzi\orcid{0000-0001-9856-1970}\inst{\ref{aff129},\ref{aff94},\ref{aff10}}
\and J.~Rhodes\orcid{0000-0002-4485-8549}\inst{\ref{aff16}}
\and G.~Riccio\inst{\ref{aff46}}
\and I.~Risso\orcid{0000-0003-2525-7761}\inst{\ref{aff76},\ref{aff77},\ref{aff70}}
\and F.~Rizzo\orcid{0000-0002-9407-585X}\inst{\ref{aff72}}
\and E.~Romelli\orcid{0000-0003-3069-9222}\inst{\ref{aff72}}
\and M.~Roncarelli\orcid{0000-0001-9587-7822}\inst{\ref{aff10}}
\and B.~Rusholme\orcid{0000-0001-7648-4142}\inst{\ref{aff63}}
\and R.~Saglia\orcid{0000-0003-0378-7032}\inst{\ref{aff31},\ref{aff32}}
\and Z.~Sakr\orcid{0000-0002-4823-3757}\inst{\ref{aff69},\ref{aff130},\ref{aff131}}
\and D.~Sapone\orcid{0000-0001-7089-4503}\inst{\ref{aff132}}
\and B.~Sartoris\orcid{0000-0003-1337-5269}\inst{\ref{aff32},\ref{aff72}}
\and M.~Schirmer\orcid{0000-0003-2568-9994}\inst{\ref{aff51}}
\and P.~Schneider\orcid{0000-0001-8561-2679}\inst{\ref{aff114}}
\and T.~Schrabback\orcid{0000-0002-6987-7834}\inst{\ref{aff133}}
\and A.~Secroun\orcid{0000-0003-0505-3710}\inst{\ref{aff95}}
\and E.~Sihvola\orcid{0000-0003-1804-7715}\inst{\ref{aff134}}
\and P.~Simon\inst{\ref{aff114}}
\and C.~Sirignano\orcid{0000-0002-0995-7146}\inst{\ref{aff129},\ref{aff94}}
\and G.~Sirri\orcid{0000-0003-2626-2853}\inst{\ref{aff26}}
\and L.~Stanco\orcid{0000-0002-9706-5104}\inst{\ref{aff94}}
\and J.-L.~Starck\orcid{0000-0003-2177-7794}\inst{\ref{aff135}}
\and P.~Tallada-Cresp\'{i}\orcid{0000-0002-1336-8328}\inst{\ref{aff84},\ref{aff85}}
\and A.~N.~Taylor\inst{\ref{aff89}}
\and H.~I.~Teplitz\orcid{0000-0002-7064-5424}\inst{\ref{aff136}}
\and I.~Tereno\orcid{0000-0002-4537-6218}\inst{\ref{aff137},\ref{aff138}}
\and N.~Tessore\orcid{0000-0002-9696-7931}\inst{\ref{aff92}}
\and S.~Toft\orcid{0000-0003-3631-7176}\inst{\ref{aff139},\ref{aff140}}
\and R.~Toledo-Moreo\orcid{0000-0002-2997-4859}\inst{\ref{aff141},\ref{aff142}}
\and F.~Torradeflot\orcid{0000-0003-1160-1517}\inst{\ref{aff85},\ref{aff84}}
\and A.~Tsyganov\inst{\ref{aff143}}
\and I.~Tutusaus\orcid{0000-0002-3199-0399}\inst{\ref{aff144},\ref{aff145},\ref{aff130}}
\and J.~Valiviita\orcid{0000-0001-6225-3693}\inst{\ref{aff3},\ref{aff108}}
\and T.~Vassallo\orcid{0000-0001-6512-6358}\inst{\ref{aff72},\ref{aff98}}
\and Y.~Wang\orcid{0000-0002-4749-2984}\inst{\ref{aff63}}
\and J.~Weller\orcid{0000-0002-8282-2010}\inst{\ref{aff31},\ref{aff32}}
\and A.~Zacchei\orcid{0000-0003-0396-1192}\inst{\ref{aff72},\ref{aff71}}
\and G.~Zamorani\orcid{0000-0002-2318-301X}\inst{\ref{aff10}}
\and F.~M.~Zerbi\orcid{0000-0002-9996-973X}\inst{\ref{aff70}}
\and E.~Zucca\orcid{0000-0002-5845-8132}\inst{\ref{aff10}}
\and M.~Ballardini\orcid{0000-0003-4481-3559}\inst{\ref{aff9},\ref{aff146},\ref{aff10}}
\and S.~Borgani\orcid{0000-0001-6151-6439}\inst{\ref{aff117},\ref{aff71},\ref{aff72},\ref{aff73},\ref{aff147}}
\and E.~Bozzo\orcid{0000-0002-8201-1525}\inst{\ref{aff24}}
\and C.~Burigana\orcid{0000-0002-3005-5796}\inst{\ref{aff15},\ref{aff148}}
\and R.~Cabanac\orcid{0000-0001-6679-2600}\inst{\ref{aff130}}
\and A.~Cappi\inst{\ref{aff149},\ref{aff10}}
\and F.~Caro\orcid{0009-0003-1053-0507}\inst{\ref{aff82}}
\and T.~Castro\orcid{0000-0002-6292-3228}\inst{\ref{aff150},\ref{aff72},\ref{aff71}}
\and J.~A.~Escartin~Vigo\inst{\ref{aff32}}
\and J.~Garc\'ia-Bellido\orcid{0000-0002-9370-8360}\inst{\ref{aff69}}
\and T.~Gasparetto\orcid{0000-0002-7913-4866}\inst{\ref{aff82}}
\and V.~Gautard\inst{\ref{aff151}}
\and C.~M.~Gutierrez\orcid{0000-0001-7854-783X}\inst{\ref{aff88},\ref{aff128}}
\and J.~Macias-Perez\orcid{0000-0002-5385-2763}\inst{\ref{aff152}}
\and R.~Maoli\orcid{0000-0002-6065-3025}\inst{\ref{aff153},\ref{aff82}}
\and M.~P\"ontinen\orcid{0000-0001-5442-2530}\inst{\ref{aff3}}
\and E.~Sarpa\orcid{0000-0002-1256-655X}\inst{\ref{aff72}}
\and V.~Scottez\orcid{0009-0008-3864-940X}\inst{\ref{aff154},\ref{aff155}}
\and M.~Sereno\orcid{0000-0003-0302-0325}\inst{\ref{aff10},\ref{aff26}}
\and M.~Tenti\orcid{0000-0002-4254-5901}\inst{\ref{aff26}}
\and M.~Tucci\inst{\ref{aff24}}
\and M.~Wiesmann\orcid{0009-0000-8199-5860}\inst{\ref{aff100}}
\and Y.~Akrami\orcid{0000-0002-2407-7956}\inst{\ref{aff69},\ref{aff156}}
\and G.~Angora\orcid{0000-0002-0316-6562}\inst{\ref{aff46},\ref{aff9}}
\and S.~Anselmi\orcid{0000-0002-3579-9583}\inst{\ref{aff94},\ref{aff129},\ref{aff157}}
\and M.~Archidiacono\orcid{0000-0003-4952-9012}\inst{\ref{aff112},\ref{aff113}}
\and G.~Aric\`o\orcid{0000-0002-2802-2928}\inst{\ref{aff26}}
\and F.~Atrio-Barandela\orcid{0000-0002-2130-2513}\inst{\ref{aff158}}
\and M.~Baes\orcid{0000-0002-3930-2757}\inst{\ref{aff8}}
\and L.~Bazzanini\orcid{0000-0003-0727-0137}\inst{\ref{aff9},\ref{aff10}}
\and D.~Bertacca\orcid{0000-0002-2490-7139}\inst{\ref{aff129},\ref{aff75},\ref{aff94}}
\and M.~Bethermin\orcid{0000-0002-3915-2015}\inst{\ref{aff65}}
\and F.~Beutler\orcid{0000-0003-0467-5438}\inst{\ref{aff89}}
\and L.~Blot\orcid{0000-0002-9622-7167}\inst{\ref{aff159},\ref{aff109}}
\and M.~L.~Brown\orcid{0000-0002-0370-8077}\inst{\ref{aff50}}
\and S.~Bruton\orcid{0000-0002-6503-5218}\inst{\ref{aff160}}
\and A.~Calabro\orcid{0000-0003-2536-1614}\inst{\ref{aff82}}
\and B.~Camacho~Quevedo\orcid{0000-0002-8789-4232}\inst{\ref{aff71},\ref{aff74},\ref{aff72}}
\and C.~S.~Carvalho\inst{\ref{aff138}}
\and F.~Cogato\orcid{0000-0003-4632-6113}\inst{\ref{aff25},\ref{aff10}}
\and S.~Conseil\orcid{0000-0002-3657-4191}\inst{\ref{aff43}}
\and A.~R.~Cooray\orcid{0000-0002-3892-0190}\inst{\ref{aff55}}
\and P.~Corcho-Caballero\orcid{0000-0001-6327-7080}\inst{\ref{aff35}}
\and B.~Csizi\orcid{0000-0003-3227-6581}\inst{\ref{aff133}}
\and O.~Cucciati\orcid{0000-0002-9336-7551}\inst{\ref{aff10}}
\and H.~Dannerbauer\orcid{0000-0001-7147-3575}\inst{\ref{aff88},\ref{aff128}}
\and T.~de~Boer\orcid{0000-0001-5486-2747}\inst{\ref{aff12}}
\and F.~De~Paolis\orcid{0000-0001-6460-7563}\inst{\ref{aff161},\ref{aff162},\ref{aff163}}
\and G.~Desprez\orcid{0000-0001-8325-1742}\inst{\ref{aff35}}
\and A.~D\'iaz-S\'anchez\orcid{0000-0003-0748-4768}\inst{\ref{aff13}}
\and S.~Di~Domizio\orcid{0000-0003-2863-5895}\inst{\ref{aff76},\ref{aff77}}
\and J.~M.~Diego\orcid{0000-0001-9065-3926}\inst{\ref{aff164}}
\and P.-A.~Duc\orcid{0000-0003-3343-6284}\inst{\ref{aff65}}
\and V.~Duret\orcid{0009-0009-0383-4960}\inst{\ref{aff95}}
\and M.~Y.~Elkhashab\orcid{0000-0001-9306-2603}\inst{\ref{aff72},\ref{aff73},\ref{aff117},\ref{aff71}}
\and A.~Enia\orcid{0000-0002-0200-2857}\inst{\ref{aff151},\ref{aff10}}
\and Y.~Fang\orcid{0000-0002-0334-6950}\inst{\ref{aff31}}
\and A.~Farina\orcid{0009-0000-3420-929X}\inst{\ref{aff70},\ref{aff77}}
\and A.~Finoguenov\orcid{0000-0002-4606-5403}\inst{\ref{aff3}}
\and A.~Franco\orcid{0000-0002-4761-366X}\inst{\ref{aff161},\ref{aff162},\ref{aff163}}
\and Y.~Fu\orcid{0000-0002-0759-0504}\inst{\ref{aff54},\ref{aff35}}
\and K.~Ganga\orcid{0000-0001-8159-8208}\inst{\ref{aff165}}
\and E.~Gaztanaga\orcid{0000-0001-9632-0815}\inst{\ref{aff144},\ref{aff145},\ref{aff1}}
\and Z.~Ghaffari\orcid{0000-0002-6467-8078}\inst{\ref{aff72},\ref{aff71}}
\and F.~Giacomini\orcid{0000-0002-3129-2814}\inst{\ref{aff26}}
\and F.~Gianotti\orcid{0000-0003-4666-119X}\inst{\ref{aff10}}
\and G.~Gozaliasl\orcid{0000-0002-0236-919X}\inst{\ref{aff166},\ref{aff3}}
\and A.~Gruppuso\orcid{0000-0001-9272-5292}\inst{\ref{aff10},\ref{aff26}}
\and M.~Guidi\orcid{0000-0001-9408-1101}\inst{\ref{aff40},\ref{aff10}}
\and A.~Hall\orcid{0000-0002-3139-8651}\inst{\ref{aff89}}
\and C.~Hern\'andez-Monteagudo\orcid{0000-0001-5471-9166}\inst{\ref{aff128},\ref{aff88}}
\and T.~Heuberger\orcid{0009-0009-3474-8467}\inst{\ref{aff20}}
\and H.~Hildebrandt\orcid{0000-0002-9814-3338}\inst{\ref{aff167}}
\and J.~Hjorth\orcid{0000-0002-4571-2306}\inst{\ref{aff168}}
\and M.~Jauzac\orcid{0000-0003-1974-8732}\inst{\ref{aff169},\ref{aff115},\ref{aff170},\ref{aff171},\ref{aff130}}
\and J.~J.~E.~Kajava\orcid{0000-0002-3010-8333}\inst{\ref{aff172},\ref{aff2},\ref{aff173}}
\and Y.~Kang\orcid{0009-0000-8588-7250}\inst{\ref{aff133}}
\and V.~Kansal\orcid{0000-0002-4008-6078}\inst{\ref{aff174},\ref{aff175}}
\and D.~Karagiannis\orcid{0000-0002-4927-0816}\inst{\ref{aff9},\ref{aff176}}
\and J.~Kim\orcid{0000-0003-2776-2761}\inst{\ref{aff11}}
\and C.~C.~Kirkpatrick\inst{\ref{aff134}}
\and A.~Kov\'acs\orcid{0000-0002-5825-579X}\inst{\ref{aff177},\ref{aff178}}
\and I.~Kova{\v{c}}i{\'{c}}\orcid{0000-0001-6751-3263}\inst{\ref{aff8}}
\and K.~Koyama\orcid{0000-0001-6727-6915}\inst{\ref{aff1}}
\and S.~Kruk\orcid{0000-0001-8010-8879}\inst{\ref{aff61}}
\and M.~C.~Lam\orcid{0000-0002-9347-2298}\inst{\ref{aff89}}
\and M.~Lattanzi\orcid{0000-0003-1059-2532}\inst{\ref{aff146}}
\and L.~Legrand\orcid{0000-0003-0610-5252}\inst{\ref{aff179},\ref{aff93}}
\and G.~Leroy\orcid{0009-0004-2523-4425}\inst{\ref{aff169},\ref{aff115}}
\and G.~F.~Lesci\orcid{0000-0002-4607-2830}\inst{\ref{aff25},\ref{aff10}}
\and J.~Lesgourgues\orcid{0000-0001-7627-353X}\inst{\ref{aff180}}
\and L.~Linke\orcid{0000-0002-2622-8113}\inst{\ref{aff133}}
\and S.~J.~Liu\orcid{0000-0001-7680-2139}\inst{\ref{aff96}}
\and G.~Maggio\orcid{0000-0003-4020-4836}\inst{\ref{aff72}}
\and E.~A.~Magnier\orcid{0000-0002-7965-2815}\inst{\ref{aff12}}
\and F.~Mannucci\orcid{0000-0002-4803-2381}\inst{\ref{aff181}}
\and F.~R.~Marleau\orcid{0000-0002-1442-2947}\inst{\ref{aff133}}
\and C.~J.~A.~P.~Martins\orcid{0000-0002-4886-9261}\inst{\ref{aff41},\ref{aff42}}
\and M.~Migliaccio\inst{\ref{aff182},\ref{aff183}}
\and M.~Miluzio\inst{\ref{aff61},\ref{aff184}}
\and G.~Morgante\inst{\ref{aff10}}
\and S.~Nadathur\orcid{0000-0001-9070-3102}\inst{\ref{aff1}}
\and K.~Naidoo\orcid{0000-0002-9182-1802}\inst{\ref{aff1},\ref{aff51}}
\and P.~Natoli\orcid{0000-0003-0126-9100}\inst{\ref{aff9},\ref{aff146}}
\and A.~Navarro-Alsina\orcid{0000-0002-3173-2592}\inst{\ref{aff114}}
\and S.~Nesseris\orcid{0000-0002-0567-0324}\inst{\ref{aff69}}
\and L.~Nicastro\orcid{0000-0001-8534-6788}\inst{\ref{aff10}}
\and M.~Oguri\orcid{0000-0003-3484-399X}\inst{\ref{aff185},\ref{aff186}}
\and F.~Oppizzi\orcid{0000-0003-3904-8370}\inst{\ref{aff77},\ref{aff94},\ref{aff129}}
\and F.~Pace\orcid{0000-0001-8039-0480}\inst{\ref{aff79},\ref{aff80},\ref{aff81},\ref{aff187}}
\and D.~Paoletti\orcid{0000-0003-4761-6147}\inst{\ref{aff10},\ref{aff148}}
\and F.~Passalacqua\orcid{0000-0002-8606-4093}\inst{\ref{aff94}}
\and K.~Paterson\orcid{0000-0001-8340-3486}\inst{\ref{aff51}}
\and L.~Patrizii\inst{\ref{aff26}}
\and C.~Pattison\orcid{0000-0003-3272-2617}\inst{\ref{aff1}}
\and A.~Pisani\orcid{0000-0002-6146-4437}\inst{\ref{aff95}}
\and D.~Potter\orcid{0000-0002-0757-5195}\inst{\ref{aff188}}
\and G.~W.~Pratt\inst{\ref{aff135}}
\and S.~Quai\orcid{0000-0002-0449-8163}\inst{\ref{aff25},\ref{aff10}}
\and M.~Radovich\orcid{0000-0002-3585-866X}\inst{\ref{aff75}}
\and G.~Rodighiero\orcid{0000-0002-9415-2296}\inst{\ref{aff129},\ref{aff75}}
\and W.~Roster\orcid{0000-0002-9149-6528}\inst{\ref{aff32}}
\and S.~Sacquegna\orcid{0000-0002-8433-6630}\inst{\ref{aff189}}
\and M.~Sahl\'en\orcid{0000-0003-0973-4804}\inst{\ref{aff190}}
\and D.~B.~Sanders\orcid{0000-0002-1233-9998}\inst{\ref{aff12}}
\and J.~Schaye\orcid{0000-0002-0668-5560}\inst{\ref{aff54}}
\and A.~Schneider\orcid{0000-0001-7055-8104}\inst{\ref{aff188}}
\and D.~Sciotti\orcid{0009-0008-4519-2620}\inst{\ref{aff82},\ref{aff83}}
\and E.~Sellentin\orcid{0009-0002-2655-3458}\inst{\ref{aff191},\ref{aff54}}
\and F.~Shankar\orcid{0000-0001-8973-5051}\inst{\ref{aff192}}
\and L.~C.~Smith\orcid{0000-0002-3259-2771}\inst{\ref{aff52}}
\and M.~Talia\orcid{0000-0003-4352-2063}\inst{\ref{aff25},\ref{aff10}}
\and K.~Tanidis\orcid{0000-0001-9843-5130}\inst{\ref{aff193}}
\and C.~Tao\orcid{0000-0001-7961-8177}\inst{\ref{aff95}}
\and G.~Testera\orcid{0000-0003-2970-766X}\inst{\ref{aff77}}
\and R.~Teyssier\orcid{0000-0001-7689-0933}\inst{\ref{aff194}}
\and S.~Tosi\orcid{0000-0002-7275-9193}\inst{\ref{aff76},\ref{aff70},\ref{aff77}}
\and A.~Troja\orcid{0000-0003-0239-4595}\inst{\ref{aff72}}
\and C.~Uhlemann\orcid{0000-0001-7831-1579}\inst{\ref{aff195},\ref{aff6}}
\and C.~Valieri\inst{\ref{aff26}}
\and A.~Venhola\orcid{0000-0001-6071-4564}\inst{\ref{aff196}}
\and D.~Vergani\orcid{0000-0003-0898-2216}\inst{\ref{aff10}}
\and G.~Vernardos\orcid{0000-0001-8554-7248}\inst{\ref{aff30},\ref{aff29}}
\and P.~Vielzeuf\orcid{0000-0003-2035-9339}\inst{\ref{aff95}}
\and E.~Vilenius\orcid{0000-0002-6184-7681}\inst{\ref{aff92}}
\and S.~Vinciguerra\orcid{0009-0005-4018-3184}\inst{\ref{aff27}}
\and M.~von~Wietersheim-Kramsta\orcid{0000-0003-4986-5091}\inst{\ref{aff115},\ref{aff169}}
\and N.~A.~Walton\orcid{0000-0003-3983-8778}\inst{\ref{aff52}}
\and L.~Wang\orcid{0000-0002-6736-9158}\inst{\ref{aff197},\ref{aff35}}
\and A.~H.~Wright\orcid{0000-0001-7363-7932}\inst{\ref{aff167}}
\and D.~Scott\orcid{0000-0002-6878-9840}\inst{\ref{aff44}}}
										   
%%%% please do not edit the affiliation list -- contact ECEB Bureau for changes
\institute{Institute of Cosmology and Gravitation, University of Portsmouth, Portsmouth PO1 3FX, UK\label{aff1}
\and
Finnish Centre for Astronomy with ESO (FINCA), Quantum, Vesilinnantie 5, University of Turku, 20014 Turku, Finland\label{aff2}
\and
Department of Physics, P.O. Box 64, University of Helsinki, 00014 Helsinki, Finland\label{aff3}
\and
INAF-IASF Milano, Via Alfonso Corti 12, 20133 Milano, Italy\label{aff4}
\and
University of Applied Sciences and Arts of Northwestern Switzerland, School of Engineering, 5210 Windisch, Switzerland\label{aff5}
\and
School of Mathematics, Statistics and Physics, Newcastle University, Herschel Building, Newcastle-upon-Tyne, NE1 7RU, UK\label{aff6}
\and
STAR Institute, University of Li{\`e}ge, Quartier Agora, All\'ee du six Ao\^ut 19c, 4000 Li\`ege, Belgium\label{aff7}
\and
Universiteit Gent, Department of Physics and Astronomy, Proeftuinstraat 86 N3, 9000 Ghent, Belgium
\label{aff8}
\and
Dipartimento di Fisica e Scienze della Terra, Universit\`a degli Studi di Ferrara, Via Giuseppe Saragat 1, 44122 Ferrara, Italy\label{aff9}
\and
INAF-Osservatorio di Astrofisica e Scienza dello Spazio di Bologna, Via Piero Gobetti 93/3, 40129 Bologna, Italy\label{aff10}
\and
Department of Physics, University of Oxford, Keble Road, Oxford OX1 3RH, UK\label{aff11}
\and
Institute for Astronomy, University of Hawaii, 2680 Woodlawn Drive, Honolulu, HI 96822, USA\label{aff12}
\and
Departamento F\'isica Aplicada, Universidad Polit\'ecnica de Cartagena, Campus Muralla del Mar, 30202 Cartagena, Murcia, Spain\label{aff13}
\and
Institut de Ci\`{e}ncies del Cosmos (ICCUB), Universitat de Barcelona (IEEC-UB), Mart\'{i} i Franqu\`{e}s 1, 08028 Barcelona, Spain\label{aff14}
\and
INAF, Istituto di Radioastronomia, Via Piero Gobetti 101, 40129 Bologna, Italy\label{aff15}
\and
Jet Propulsion Laboratory, California Institute of Technology, 4800 Oak Grove Drive, Pasadena, CA, 91109, USA\label{aff16}
\and
Department of Astronomy, School of Physics and Astronomy, Shanghai Jiao Tong University, Shanghai 200240, China\label{aff17}
\and
State Key Laboratory of Dark Matter Physics, School of Physics and Astronomy, Shanghai Jiao Tong University, Shanghai 200240, China\label{aff18}
\and
Shanghai Key Laboratory for Particle Physics and Cosmology, Shanghai 200240, China\label{aff19}
\and
University of Applied Sciences and Arts of Northwestern Switzerland, School of Computer Science, 5210 Windisch, Switzerland\label{aff20}
\and
Citizen Scientist, Zooniverse c/o University of Oxford,  Keble Road, Oxford OX1 3RH, UK\label{aff21}
\and
Instituci\'o Catalana de Recerca i Estudis Avan\c{c}ats (ICREA), Passeig de Llu\'{\i}s Companys 23, 08010 Barcelona, Spain\label{aff22}
\and
Institut de Ciencies de l'Espai (IEEC-CSIC), Campus UAB, Carrer de Can Magrans, s/n Cerdanyola del Vall\'es, 08193 Barcelona, Spain\label{aff23}
\and
Department of Astronomy, University of Geneva, ch. d'Ecogia 16, 1290 Versoix, Switzerland\label{aff24}
\and
Dipartimento di Fisica e Astronomia "Augusto Righi" - Alma Mater Studiorum Universit\`a di Bologna, via Piero Gobetti 93/2, 40129 Bologna, Italy\label{aff25}
\and
INFN-Sezione di Bologna, Viale Berti Pichat 6/2, 40127 Bologna, Italy\label{aff26}
\and
Aix-Marseille Universit\'e, CNRS, CNES, LAM, Marseille, France\label{aff27}
\and
Institut d'Astrophysique de Paris, UMR 7095, CNRS, and Sorbonne Universit\'e, 98 bis boulevard Arago, 75014 Paris, France\label{aff28}
\and
American Museum of Natural History, Department of Astrophysics, New York, NY 10024, USA\label{aff29}
\and
Department of Physics and Astronomy, Lehman College of the CUNY, Bronx, NY 10468, USA\label{aff30}
\and
Universit\"ats-Sternwarte M\"unchen, Fakult\"at f\"ur Physik, Ludwig-Maximilians-Universit\"at M\"unchen, Scheinerstr.~1, 81679 M\"unchen, Germany\label{aff31}
\and
Max Planck Institute for Extraterrestrial Physics, Giessenbachstr. 1, 85748 Garching, Germany\label{aff32}
\and
Max-Planck-Institut f\"ur Astrophysik, Karl-Schwarzschild-Str.~1, 85748 Garching, Germany\label{aff33}
\and
National Centre for Nuclear Research, ul. Pasteura 7, 02-093, Warsaw, Poland\label{aff34}
\and
Kapteyn Astronomical Institute, University of Groningen, PO Box 800, 9700 AV Groningen, The Netherlands\label{aff35}
\and
Laboratoire univers et particules de Montpellier, Universit\'e de Montpellier, CNRS, 34090 Montpellier, France\label{aff36}
\and
Departamento de F\'isica, Universidad de Santiago de Chile, Av. Victor Jara 3493, Santiago, Chile\label{aff37}
\and
Center for Interdisciplinary Research in Astrophysics Space Science (CIRAS), Universidad de Santiago, Chile\label{aff38}
\and
European Southern Observatory, Karl-Schwarzschild-Str.~2, 85748 Garching, Germany\label{aff39}
\and
Dipartimento di Fisica e Astronomia, Universit\`a di Bologna, Via Gobetti 93/2, 40129 Bologna, Italy\label{aff40}
\and
Centro de Astrof\'{\i}sica da Universidade do Porto, Rua das Estrelas, 4150-762 Porto, Portugal\label{aff41}
\and
Instituto de Astrof\'isica e Ci\^encias do Espa\c{c}o, Universidade do Porto, CAUP, Rua das Estrelas, PT4150-762 Porto, Portugal\label{aff42}
\and
Universit\'e Lyon 1, CNRS/IN2P3, IP2I Lyon, Villeurbanne, France\label{aff43}
\and
Department of Physics and Astronomy, University of British Columbia, Vancouver, BC V6T 1Z1, Canada\label{aff44}
\and
National Astronomical Observatory of Japan, 2-21-1 Osawa, Mitaka, Tokyo 181-8588, Japan\label{aff45}
\and
INAF-Osservatorio Astronomico di Capodimonte, Via Moiariello 16, 80131 Napoli, Italy\label{aff46}
\and
CUNY Graduate Center, City University of New York, 365 5th Ave., New York, NY 10016, USA\label{aff47}
\and
Sydney Institute for Astronomy, School of Physics, University of Sydney, NSW 2006, Australia\label{aff48}
\and
University of Ljubljana, Faculty of mathematics and physics, Jadranska ulica 19, 1000 Ljubljana, Slovenia\label{aff49}
\and
Jodrell Bank Centre for Astrophysics, Department of Physics and Astronomy, University of Manchester, Oxford Road, Manchester M13 9PL, UK\label{aff50}
\and
Max-Planck-Institut f\"ur Astronomie, K\"onigstuhl 17, 69117 Heidelberg, Germany\label{aff51}
\and
Institute of Astronomy, University of Cambridge, Madingley Road, Cambridge CB3 0HA, UK\label{aff52}
\and
Kavli Institute for Cosmology Cambridge, Madingley Road, Cambridge, CB3 0HA, UK\label{aff53}
\and
Leiden Observatory, Leiden University, Einsteinweg 55, 2333 CC Leiden, The Netherlands\label{aff54}
\and
Department of Physics \& Astronomy, University of California Irvine, Irvine CA 92697, USA\label{aff55}
\and
Kavli Institute for Particle Astrophysics \& Cosmology (KIPAC), Stanford University, Stanford, CA 94305, USA\label{aff56}
\and
SLAC National Accelerator Laboratory, 2575 Sand Hill Road, Menlo Park, CA 94025, USA\label{aff57}
\and
The Inter-University Centre for Astronomy and Astrophysics, Post Bag 4, Ganeshkhind, Pune 411007, India\label{aff58}
\and
Kavli Institute for the Physics and Mathematics of the Universe (WPI), University of Tokyo, Kashiwa, Chiba 277-8583, Japan\label{aff59}
\and
Minnesota Institute for Astrophysics, University of Minnesota, 116 Church St SE, Minneapolis, MN 55455, USA\label{aff60}
\and
ESAC/ESA, Camino Bajo del Castillo, s/n., Urb. Villafranca del Castillo, 28692 Villanueva de la Ca\~nada, Madrid, Spain\label{aff61}
\and
School of Physical Sciences, The Open University, Milton Keynes, MK7 6AA, UK\label{aff62}
\and
Caltech/IPAC, 1200 E. California Blvd., Pasadena, CA 91125, USA\label{aff63}
\and
School of Physics and Astronomy, University of Birmingham, Birmingham, B15 2TT, UK\label{aff64}
\and
Universit\'e de Strasbourg, CNRS, Observatoire astronomique de Strasbourg, UMR 7550, 67000 Strasbourg, France\label{aff65}
\and
Centre National d'Etudes Spatiales -- Centre spatial de Toulouse, 18 avenue Edouard Belin, 31401 Toulouse Cedex 9, France\label{aff66}
\and
University of Southern California, 3551 Trousdale Parkway, Los Angeles, CA 90089, USA\label{aff67}
\and
Max Planck Institute for Gravitational Physics (Albert Einstein Institute), Am Muhlenberg 1, D-14476 Potsdam-Golm, Germany\label{aff68}
\and
Instituto de F\'isica Te\'orica UAM-CSIC, Campus de Cantoblanco, 28049 Madrid, Spain\label{aff69}
\and
INAF-Osservatorio Astronomico di Brera, Via Brera 28, 20122 Milano, Italy\label{aff70}
\and
IFPU, Institute for Fundamental Physics of the Universe, via Beirut 2, 34151 Trieste, Italy\label{aff71}
\and
INAF-Osservatorio Astronomico di Trieste, Via G. B. Tiepolo 11, 34143 Trieste, Italy\label{aff72}
\and
INFN, Sezione di Trieste, Via Valerio 2, 34127 Trieste TS, Italy\label{aff73}
\and
SISSA, International School for Advanced Studies, Via Bonomea 265, 34136 Trieste TS, Italy\label{aff74}
\and
INAF-Osservatorio Astronomico di Padova, Via dell'Osservatorio 5, 35122 Padova, Italy\label{aff75}
\and
Dipartimento di Fisica, Universit\`a di Genova, Via Dodecaneso 33, 16146, Genova, Italy\label{aff76}
\and
INFN-Sezione di Genova, Via Dodecaneso 33, 16146, Genova, Italy\label{aff77}
\and
Department of Physics "E. Pancini", University Federico II, Via Cinthia 6, 80126, Napoli, Italy\label{aff78}
\and
Dipartimento di Fisica, Universit\`a degli Studi di Torino, Via P. Giuria 1, 10125 Torino, Italy\label{aff79}
\and
INFN-Sezione di Torino, Via P. Giuria 1, 10125 Torino, Italy\label{aff80}
\and
INAF-Osservatorio Astrofisico di Torino, Via Osservatorio 20, 10025 Pino Torinese (TO), Italy\label{aff81}
\and
INAF-Osservatorio Astronomico di Roma, Via Frascati 33, 00078 Monteporzio Catone, Italy\label{aff82}
\and
INFN-Sezione di Roma, Piazzale Aldo Moro, 2 - c/o Dipartimento di Fisica, Edificio G. Marconi, 00185 Roma, Italy\label{aff83}
\and
Centro de Investigaciones Energ\'eticas, Medioambientales y Tecnol\'ogicas (CIEMAT), Avenida Complutense 40, 28040 Madrid, Spain\label{aff84}
\and
Port d'Informaci\'{o} Cient\'{i}fica, Campus UAB, C. Albareda s/n, 08193 Bellaterra (Barcelona), Spain\label{aff85}
\and
INFN -- Sezione di Napoli, Via Cinthia 6, 80126, Napoli, Italy\label{aff86}
\and
Dipartimento di Fisica e Astronomia "Augusto Righi" - Alma Mater Studiorum Universit\`a di Bologna, Viale Berti Pichat 6/2, 40127 Bologna, Italy\label{aff87}
\and
Instituto de Astrof\'{\i}sica de Canarias, E-38205 La Laguna, Tenerife, Spain\label{aff88}
\and
Institute for Astronomy, University of Edinburgh, Royal Observatory, Blackford Hill, Edinburgh EH9 3HJ, UK\label{aff89}
\and
European Space Agency/ESRIN, Largo Galileo Galilei 1, 00044 Frascati, Roma, Italy\label{aff90}
\and
Universit\'e Lyon 1, CNRS/IN2P3, IUF, IP2I Lyon, Villeurbanne, France \label{aff91}
\and
Mullard Space Science Laboratory, University College London, Holmbury St Mary, Dorking, Surrey RH5 6NT, UK\label{aff92}
\and
Universit\'e Paris-Saclay, CNRS, Institut d'astrophysique spatiale, 91405, Orsay, France\label{aff93}
\and
INFN-Padova, Via Marzolo 8, 35131 Padova, Italy\label{aff94}
\and
Aix-Marseille Universit\'e, CNRS/IN2P3, CPPM, Marseille, France\label{aff95}
\and
INAF-Istituto di Astrofisica e Planetologia Spaziali, via del Fosso del Cavaliere, 100, 00100 Roma, Italy\label{aff96}
\and
School of Physics, HH Wills Physics Laboratory, University of Bristol, Tyndall Avenue, Bristol, BS8 1TL, UK\label{aff97}
\and
University Observatory, LMU Faculty of Physics, Scheinerstr.~1, 81679 Munich, Germany\label{aff98}
\and
National Research Council, Herzberg Astronomy and Astrophysics Research Centre, 5071 W. Saanich Rd. Victoria, BC, V9E 2E7, Canada\label{aff99}
\and
Institute of Theoretical Astrophysics, University of Oslo, P.O. Box 1029 Blindern, 0315 Oslo, Norway\label{aff100}
\and
Department of Physics, Lancaster University, Lancaster, LA1 4YB, UK\label{aff101}
\and
Felix Hormuth Engineering, Goethestr. 17, 69181 Leimen, Germany\label{aff102}
\and
Technical University of Denmark, Elektrovej 327, 2800 Kgs. Lyngby, Denmark\label{aff103}
\and
Cosmic Dawn Center (DAWN), Denmark\label{aff104}
\and
NASA Goddard Space Flight Center, Greenbelt, MD 20771, USA\label{aff105}
\and
Department of Physics and Astronomy, University College London, Gower Street, London WC1E 6BT, UK\label{aff106}
\and
Universit\'e de Gen\`eve, D\'epartement de Physique Th\'eorique and Centre for Astroparticle Physics, 24 quai Ernest-Ansermet, CH-1211 Gen\`eve 4, Switzerland\label{aff107}
\and
Helsinki Institute of Physics, Gustaf H{\"a}llstr{\"o}min katu 2, University of Helsinki, 00014 Helsinki, Finland\label{aff108}
\and
Laboratoire d'etude de l'Univers et des phenomenes eXtremes, Observatoire de Paris, Universit\'e PSL, Sorbonne Universit\'e, CNRS, 92190 Meudon, France\label{aff109}
\and
SKAO, Jodrell Bank, Lower Withington, Macclesfield SK11 9FT, UK\label{aff110}
\and
Centre de Calcul de l'IN2P3/CNRS, 21 avenue Pierre de Coubertin 69627 Villeurbanne Cedex, France\label{aff111}
\and
Dipartimento di Fisica "Aldo Pontremoli", Universit\`a degli Studi di Milano, Via Celoria 16, 20133 Milano, Italy\label{aff112}
\and
INFN-Sezione di Milano, Via Celoria 16, 20133 Milano, Italy\label{aff113}
\and
Universit\"at Bonn, Argelander-Institut f\"ur Astronomie, Auf dem H\"ugel 71, 53121 Bonn, Germany\label{aff114}
\and
Department of Physics, Institute for Computational Cosmology, Durham University, South Road, Durham, DH1 3LE, UK\label{aff115}
\and
Institute of Physics, Laboratory of Astrophysics, Ecole Polytechnique F\'ed\'erale de Lausanne (EPFL), Observatoire de Sauverny, 1290 Versoix, Switzerland\label{aff116}
\and
Dipartimento di Fisica - Sezione di Astronomia, Universit\`a di Trieste, Via Tiepolo 11, 34131 Trieste, Italy\label{aff117}
\and
Telespazio UK S.L. for European Space Agency (ESA), Camino bajo del Castillo, s/n, Urbanizacion Villafranca del Castillo, Villanueva de la Ca\~nada, 28692 Madrid, Spain\label{aff118}
\and
Institut de F\'{i}sica d'Altes Energies (IFAE), The Barcelona Institute of Science and Technology, Campus UAB, 08193 Bellaterra (Barcelona), Spain\label{aff119}
\and
School of Mathematics and Physics, University of Surrey, Guildford, Surrey, GU2 7XH, UK\label{aff120}
\and
European Space Agency/ESTEC, Keplerlaan 1, 2201 AZ Noordwijk, The Netherlands\label{aff121}
\and
Waterloo Centre for Astrophysics, University of Waterloo, Waterloo, Ontario N2L 3G1, Canada\label{aff122}
\and
Department of Physics and Astronomy, University of Waterloo, Waterloo, Ontario N2L 3G1, Canada\label{aff123}
\and
Perimeter Institute for Theoretical Physics, Waterloo, Ontario N2L 2Y5, Canada\label{aff124}
\and
Space Science Data Center, Italian Space Agency, via del Politecnico snc, 00133 Roma, Italy\label{aff125}
\and
Institute of Space Science, Str. Atomistilor, nr. 409 M\u{a}gurele, Ilfov, 077125, Romania\label{aff126}
\and
Consejo Superior de Investigaciones Cientificas, Calle Serrano 117, 28006 Madrid, Spain\label{aff127}
\and
Universidad de La Laguna, Dpto. Astrof\'\i sica, E-38206 La Laguna, Tenerife, Spain\label{aff128}
\and
Dipartimento di Fisica e Astronomia "G. Galilei", Universit\`a di Padova, Via Marzolo 8, 35131 Padova, Italy\label{aff129}
\and
Institut de Recherche en Astrophysique et Plan\'etologie (IRAP), Universit\'e de Toulouse, CNRS, UPS, CNES, 14 Av. Edouard Belin, 31400 Toulouse, France\label{aff130}
\and
Universit\'e St Joseph; Faculty of Sciences, Beirut, Lebanon\label{aff131}
\and
Departamento de F\'isica, FCFM, Universidad de Chile, Blanco Encalada 2008, Santiago, Chile\label{aff132}
\and
Universit\"at Innsbruck, Institut f\"ur Astro- und Teilchenphysik, Technikerstr. 25/8, 6020 Innsbruck, Austria\label{aff133}
\and
Department of Physics and Helsinki Institute of Physics, Gustaf H\"allstr\"omin katu 2, University of Helsinki, 00014 Helsinki, Finland\label{aff134}
\and
Universit\'e Paris-Saclay, Universit\'e Paris Cit\'e, CEA, CNRS, AIM, 91191, Gif-sur-Yvette, France\label{aff135}
\and
Infrared Processing and Analysis Center, California Institute of Technology, Pasadena, CA 91125, USA\label{aff136}
\and
Departamento de F\'isica, Faculdade de Ci\^encias, Universidade de Lisboa, Edif\'icio C8, Campo Grande, PT1749-016 Lisboa, Portugal\label{aff137}
\and
Instituto de Astrof\'isica e Ci\^encias do Espa\c{c}o, Faculdade de Ci\^encias, Universidade de Lisboa, Tapada da Ajuda, 1349-018 Lisboa, Portugal\label{aff138}
\and
Cosmic Dawn Center (DAWN)\label{aff139}
\and
Niels Bohr Institute, University of Copenhagen, Jagtvej 128, 2200 Copenhagen, Denmark\label{aff140}
\and
Universidad Polit\'ecnica de Cartagena, Departamento de Electr\'onica y Tecnolog\'ia de Computadoras,  Plaza del Hospital 1, 30202 Cartagena, Spain\label{aff141}
\and
European University of Technology EUt+, European Union\label{aff142}
\and
Centre for Information Technology, University of Groningen, P.O. Box 11044, 9700 CA Groningen, The Netherlands\label{aff143}
\and
Institute of Space Sciences (ICE, CSIC), Campus UAB, Carrer de Can Magrans, s/n, 08193 Barcelona, Spain\label{aff144}
\and
Institut d'Estudis Espacials de Catalunya (IEEC),  Edifici RDIT, Campus UPC, 08860 Castelldefels, Barcelona, Spain\label{aff145}
\and
Istituto Nazionale di Fisica Nucleare, Sezione di Ferrara, Via Giuseppe Saragat 1, 44122 Ferrara, Italy\label{aff146}
\and
ICSC - Centro Nazionale di Ricerca in High Performance Computing, Big Data e Quantum Computing, Via Magnanelli 2, Bologna, Italy\label{aff147}
\and
INFN-Bologna, Via Irnerio 46, 40126 Bologna, Italy\label{aff148}
\and
Universit\'e C\^{o}te d'Azur, Observatoire de la C\^{o}te d'Azur, CNRS, Laboratoire Lagrange, Bd de l'Observatoire, CS 34229, 06304 Nice cedex 4, France\label{aff149}
\and
Department of Mathematical Physics, Institute of Physics, University of S\~ao Paulo, R. do Mat\~ao 1371, 05508-090, S\~ao Paulo, SP, Brazil\label{aff150}
\and
CEA Saclay, DFR/IRFU, Service d'Astrophysique, Bat. 709, 91191 Gif-sur-Yvette, France\label{aff151}
\and
Univ. Grenoble Alpes, CNRS, Grenoble INP, LPSC-IN2P3, 53, Avenue des Martyrs, 38000 Grenoble, France\label{aff152}
\and
Dipartimento di Fisica, Sapienza Universit\`a di Roma, Piazzale Aldo Moro 2, 00185 Roma, Italy\label{aff153}
\and
Institut d'Astrophysique de Paris, 98bis Boulevard Arago, 75014, Paris, France\label{aff154}
\and
ICL, Junia, Universit\'e Catholique de Lille, LITL, 59000 Lille, France\label{aff155}
\and
CERCA/ISO, Department of Physics, Case Western Reserve University, 10900 Euclid Avenue, Cleveland, OH 44106, USA\label{aff156}
\and
Laboratoire Univers et Th\'eorie, Observatoire de Paris, Universit\'e PSL, Universit\'e Paris Cit\'e, CNRS, 92190 Meudon, France\label{aff157}
\and
Departamento de F{\'\i}sica Fundamental. Universidad de Salamanca. Plaza de la Merced s/n. 37008 Salamanca, Spain\label{aff158}
\and
Center for Data-Driven Discovery, Kavli IPMU (WPI), UTIAS, The University of Tokyo, Kashiwa, Chiba 277-8583, Japan\label{aff159}
\and
California Institute of Technology, 1200 E California Blvd, Pasadena, CA 91125, USA\label{aff160}
\and
Department of Mathematics and Physics E. De Giorgi, University of Salento, Via per Arnesano, CP-I93, 73100, Lecce, Italy\label{aff161}
\and
INFN, Sezione di Lecce, Via per Arnesano, CP-193, 73100, Lecce, Italy\label{aff162}
\and
INAF-Sezione di Lecce, c/o Dipartimento Matematica e Fisica, Via per Arnesano, 73100, Lecce, Italy\label{aff163}
\and
Instituto de F\'isica de Cantabria, Edificio Juan Jord\'a, Avenida de los Castros, 39005 Santander, Spain\label{aff164}
\and
Universit\'e Paris Cit\'e, CNRS, Astroparticule et Cosmologie, 75013 Paris, France\label{aff165}
\and
Department of Computer Science, Aalto University, PO Box 15400, Espoo, FI-00 076, Finland\label{aff166}
\and
Ruhr University Bochum, Faculty of Physics and Astronomy, Astronomical Institute (AIRUB), German Centre for Cosmological Lensing (GCCL), 44780 Bochum, Germany\label{aff167}
\and
DARK, Niels Bohr Institute, University of Copenhagen, Jagtvej 155, 2200 Copenhagen, Denmark\label{aff168}
\and
Department of Physics, Centre for Extragalactic Astronomy, Durham University, South Road, Durham, DH1 3LE, UK\label{aff169}
\and
Astrophysics Research Centre, University of KwaZulu-Natal, Westville Campus, Durban 4041, South Africa\label{aff170}
\and
School of Mathematics, Statistics \& Computer Science, University of KwaZulu-Natal, Westville Campus, Durban 4041, South Africa\label{aff171}
\and
Department of Physics and Astronomy, Vesilinnantie 5, University of Turku, 20014 Turku, Finland\label{aff172}
\and
Serco for European Space Agency (ESA), Camino bajo del Castillo, s/n, Urbanizacion Villafranca del Castillo, Villanueva de la Ca\~nada, 28692 Madrid, Spain\label{aff173}
\and
ARC Centre of Excellence for Dark Matter Particle Physics, Melbourne, Australia\label{aff174}
\and
Centre for Astrophysics \& Supercomputing, Swinburne University of Technology,  Hawthorn, Victoria 3122, Australia\label{aff175}
\and
Department of Physics and Astronomy, University of the Western Cape, Bellville, Cape Town, 7535, South Africa\label{aff176}
\and
MTA-CSFK Lend\"ulet Large-Scale Structure Research Group, Konkoly-Thege Mikl\'os \'ut 15-17, H-1121 Budapest, Hungary\label{aff177}
\and
Konkoly Observatory, HUN-REN CSFK, MTA Centre of Excellence, Budapest, Konkoly Thege Mikl\'os {\'u}t 15-17. H-1121, Hungary\label{aff178}
\and
Brazilian Center for Research in Physics (CBPF), Dr. Xavier Sigaud st. 150, zip 22290-180, Rio de Janeiro, RJ, Brazil\label{aff179}
\and
Institute for Theoretical Particle Physics and Cosmology (TTK), RWTH Aachen University, 52056 Aachen, Germany\label{aff180}
\and
INAF-Osservatorio Astrofisico di Arcetri, Largo E. Fermi 5, 50125, Firenze, Italy\label{aff181}
\and
Dipartimento di Fisica, Universit\`a di Roma Tor Vergata, Via della Ricerca Scientifica 1, Roma, Italy\label{aff182}
\and
INFN, Sezione di Roma 2, Via della Ricerca Scientifica 1, Roma, Italy\label{aff183}
\and
HE Space for European Space Agency (ESA), Camino bajo del Castillo, s/n, Urbanizacion Villafranca del Castillo, Villanueva de la Ca\~nada, 28692 Madrid, Spain\label{aff184}
\and
Center for Frontier Science, Chiba University, 1-33 Yayoi-cho, Inage-ku, Chiba 263-8522, Japan\label{aff185}
\and
Department of Physics, Graduate School of Science, Chiba University, 1-33 Yayoi-Cho, Inage-Ku, Chiba 263-8522, Japan\label{aff186}
\and
Instituto de Astrof\'isica e Ci\^encias do Espa\c{c}o, Faculdade de Ci\^encias, Universidade de Lisboa, Campo Grande, 1749-016 Lisboa, Portugal\label{aff187}
\and
Department of Astrophysics, University of Zurich, Winterthurerstrasse 190, 8057 Zurich, Switzerland\label{aff188}
\and
INAF - Osservatorio Astronomico d'Abruzzo, Via Maggini, 64100, Teramo, Italy\label{aff189}
\and
Theoretical astrophysics, Department of Physics and Astronomy, Uppsala University, Box 524, 751 20 Uppsala, Sweden\label{aff190}
\and
Mathematical Institute, University of Leiden, Einsteinweg 55, 2333 CA Leiden, The Netherlands\label{aff191}
\and
School of Physics \& Astronomy, University of Southampton, Highfield Campus, Southampton SO17 1BJ, UK\label{aff192}
\and
Center for Astrophysics and Cosmology, University of Nova Gorica, Nova Gorica, Slovenia\label{aff193}
\and
Department of Astrophysical Sciences, Peyton Hall, Princeton University, Princeton, NJ 08544, USA\label{aff194}
\and
Fakult\"at f\"ur Physik, Universit\"at Bielefeld, Postfach 100131, 33501 Bielefeld, Germany\label{aff195}
\and
Space physics and astronomy research unit, University of Oulu, Pentti Kaiteran katu 1, FI-90014 Oulu, Finland\label{aff196}
\and
SRON Netherlands Institute for Space Research, Landleven 12, 9747 AD, Groningen, The Netherlands\label{aff197}}    
% portal + stefan email + A Faisst

%
% For a Key Project paper, please use instead:
%
% \title{\Euclid\/ preparation} % or another appropriate pre-phrase
% \subtitle{Your title}
%
% \author{Euclid Collaboration: F.~Author, .....}
%

% \institute{$^{1}$ Dark Energy Institute\\
% $^{2}$ Center for Large-Scale Structure\\
% $^{3}$ Institute for Fundamental Research \\
% $^{4}$ Institute for Astrophysics and Cosmology\\}

% 
% Put your abstract here:
%
\abstract{

We present \fixed{\gradeAB} high-confidence strong lens candidates in Euclid Data Release 1 (DR1), spanning \fixed{\dronearea}\,deg$^2$. These comprise \fixed{\gradeA} grade A (`sure lens') and \fixed{\gradeB} grade B (`probable lens') systems, which we publish alongside a further \fixed{\gradeC} grade C (`possible lens') candidates.
With more high-confidence candidates than all previous catalogues combined, this sample establishes a large, homogeneous data set of galaxy-scale strong lenses with high-resolution imaging, transforming strong lensing from a field of rare discoveries to a robust statistical tool for astrophysics and cosmology. 
Of the roughly 900 million sources in DR1, we searched the 68 million passing our catalogue selection using a bespoke discovery pipeline that synthesises scores from \fixed{five} machine-learning models with visual classifications from around 7000 citizen scientists and \fixed{\experts} strong lensing experts.
Beyond classical Einstein rings, our catalogue includes \fixed{\dspl} candidate double-source-plane lenses and over \fixed{\discs} candidate disc lenses, two exotic classes with few previously known examples, as well as several hundred candidate lensed AGN and quasars.
We deliver automated lens models for the combined grade A+B systems, and photometric redshifts and stellar population properties for the deflector galaxies, providing immediate science-ready data products.
We find \comments{that} the majority of the deflector galaxies are massive and quiescent, with \comments{stellar masses of} \red{\stellarmass} \comments{(assuming a Salpeter initial mass function; the range corresponding to the
$16$th--$84$th percentiles)}.
The \comments{deflector galaxies span a range of photometric redshifts ($z=$\,\red{\photoz})}, extending to higher redshifts than previous ground-based lens samples, and \comments{the lenses have} a \comments{typical} Einstein radius of \red{\thetaE} arcsec.
Extrapolating by area, we expect over \num{100000} high-confidence strong lens \comments{candidates} in the Euclid Wide Survey.
}
%
% Provide up to five key words:
%
    \keywords{Gravitational lensing: strong -- Catalogues -- Galaxies -- Cosmology: observations}
%    from the list in
%     https://www.aanda.org/for-authors/latex-issues/information-files#pop}
%
% Add short versions of title and author list for page headings
%
   \titlerunning{ECLIPSE A -- DR1 strong lens catalogue overview}
   \authorrunning{Euclid Collaboration: Lines et al.}
   
   \maketitle
%
%-------------------------------------------------------------------
%
%
%   Start the main text of your paper here
%

\section{\label{sc:Intro}Introduction}

Strong gravitational lensing is the general relativistic phenomenon in which massive objects warp spacetime sufficiently to bend the light from a background source into multiple images, arcs, Einstein rings, and Einstein crosses \citep{Einstein1936}. These cosmic mirages are unique and powerful probes of astrophysics and cosmology.
Strong lensing is one of the most robust methods of accurately measuring the total mass of galaxies, providing a unique probe of their underlying mass distributions without a reliance on luminous matter.
In addition to producing multiple images, the strong lensing effect acts as a cosmic telescope, magnifying the light from the background source.
This opens new windows into galaxies in the early Universe, enabling the observation of objects too distant and faint to otherwise be detectable. Strong lensing is also sensitive to the composition and expansion history of the Universe, and hence can be used to constrain cosmological parameters, including the Hubble constant when there is a time-variable background source. We refer interested readers to \citet{Shajib2024} for a review of applications of galaxy-scale strong lensing to astrophysics and cosmology.

\begin{figure}
    \centering
    \includegraphics[width=1.\linewidth]{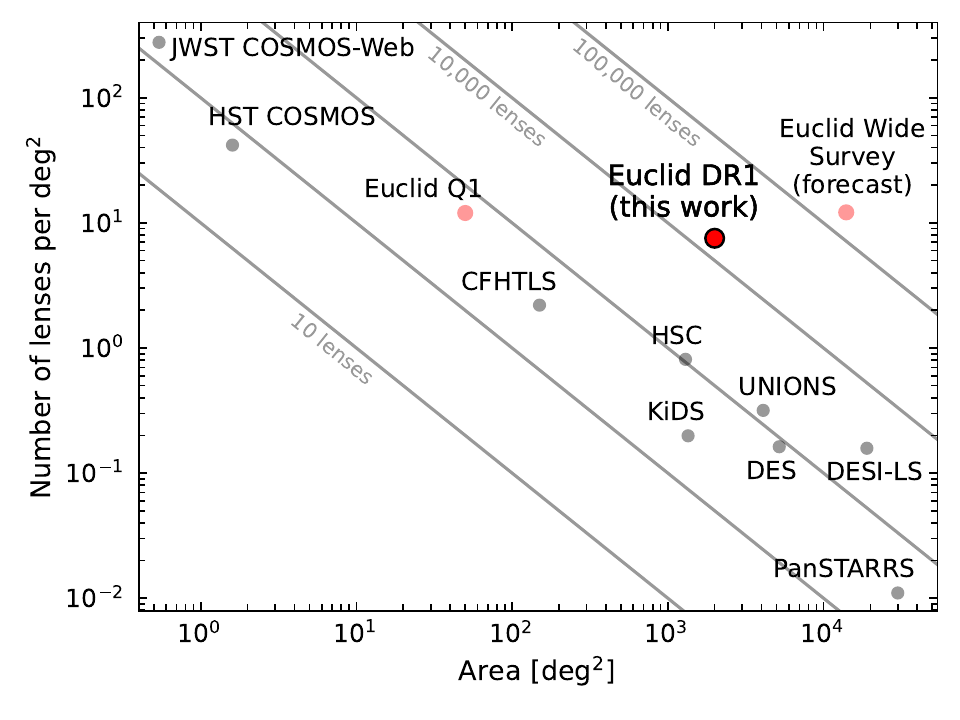}
    \caption{Number of \comments{strong} lens \comments{candidates} found from different \comments{imaging} surveys as a function of survey area and lens \comments{number} density. The Euclid DR1 data point represents this work. All numbers are approximate due to the subjective nature of grading and the varying details of each search approach. \comments{The figure is adapted from \citet{Q1-SP048}.}}
    \label{fig:lens_surveys}
\end{figure}

Historically, strong lensing has been a data-starved field due to its inherent rarity.
Strong lensing requires a near-perfect alignment of the observer, a foreground deflector galaxy, and a background source -- a rare configuration that occurs in fewer than 0.1\% of observed galaxies. 
Additionally, galaxy-scale strong lenses are hard to detect: the average radius of an Einstein ring is \comments{theoretically expected to be} just \ang{;;0.4} \citep{Collett2015}, meaning that a large fraction of strong lenses are unresolvable from the ground. 
The compounded effect of intrinsic rarity and small angular size has limited the growth of strong lens samples \comments{in the optical and near-infrared (NIR)}. Generally, previous \comments{imaging} surveys have either been ground-based (e.g., \citealt{gavazzi2014, Jacobs2019a, Jacobs2019, Petrillo2019, Canameras2020, Storfer2024, Storfer2026}), and therefore limited by resolution, \comments{or were conducted by high-resolution space telescopes, such as \textit{Hubble} (HST) and \textit{James Webb} (JWST; e.g., \citealt{Garvin2022, petit2025, Nightingale2025, Mahler2025, Hogg2025}), which have a limited field of view}. 
\comments{Figure~\ref{fig:lens_surveys} illustrates this trade-off, showing the number of lenses found by previous optical and NIR imaging surveys as a function of survey area and lens number density.}
\comments{Wide-area ground-based searches have so far produced the largest samples, with} the largest homogeneous catalogue previously identified from imaging containing around 3000 lens candidates \comments{across \num{19000}\,deg$^{2}$} \citep{Storfer2024}.

However, \Euclid \citep{EuclidSkyOverview} is perfectly designed to overcome these challenges. 
The Euclid Wide Survey (EWS) will cover an area of \num{14000}\,deg$^2$ over the nominal six-year mission, mitigating the intrinsic rarity, while its high-resolution imaging (with a \ang{;;0.1} pixel scale in \comments{the} VIS \comments{instrument}) is sufficient to resolve a substantial fraction of the lenses that would otherwise go undetected. As a result, \Euclid will revolutionise the field of strong lens discoveries, being forecast to image \comments{around \num{170000}} galaxy-scale strong lenses -- over an order of magnitude more than what was previously known \citep{Collett2015}. 
Euclid DR1 \citep{DR1cite} covers approximately \fixed{\dronearea}\,deg$^2$ and represents an important milestone towards the survey's final data set. Making up 15\% of the full survey area, forecasts suggest that around \num{23000} strong lenses should be discoverable within DR1. 
The prospect of such a massive sample offers unprecedented scientific opportunities, but extracting these rare systems from the tens of millions of non-lenses in the data is a non-trivial task.
With true lenses outnumbered by ordinary galaxies by thousands to one, discovering even a majority of the strong lenses \comments{is} not guaranteed.
Maximising the potential from this wealth of data therefore requires a highly automated, robust, and concerted search effort.

Our DR1 strong lens search builds upon the `Strong Lens Discovery Engine' (SLDE; \citealt{Q1-SP048, Q1-SP052, Q1-SP053, Q1-SP054, Q1-SP059, Q1-Ecker}), a lens-finding pipeline that led to the discovery of approximately 600 galaxy-scale strong lens \comments{candidates} in the Euclid Quick Release 1 (Q1; \citealt{Q1cite}). Q1 consists of just seven days worth of \Euclid observations spanning 63\,deg$^2$, serving as a test-bed for developing and refining the pipeline in preparation for DR1.
While the overall approach of combining multiple complementary machine-learning (ML) models with the power of crowd-sourced visual inspection remains the same, the DR1 search benefits from improved ML models \citep{Q1-Vincken, Q1-SP085, DR1-SL06}, optimisation of visual inspection time \citep{DR1-SL02}, and is informed by lessons learnt from objects missed in the initial Q1 search \citep{Q1-Ecker, Q1-Xu}.

Discovering these systems is also only half of the challenge, since extracting the scientific information from them requires lens modelling.
The most robust approach, fitting physical mass and light profiles with posteriors inferred through sampling, is traditionally a system-by-system manual process that would be fundamentally infeasible at the scale of DR1.
While automated modelling requiring minimal human input has been successfully carried out on roughly 300 systems from Euclid Q1 \citep{Q1-SP048} and 500 systems from JWST \citep{Nightingale2025}, scaling this up to over $\num{10000}$ lenses expected in DR1 demands a further order-of-magnitude shift. 

This paper presents the \comments{Euclid Lens Identification Pipeline and Scientific Exploration (`ECLIPSE')} catalogue for DR1, comprising \fixed{\gradeABC} strong lens \comments{candidates} identified across \fixed{\dronearea}\,deg$^2$ of space-based imaging. The catalogue is the product of the discovery pipeline developed and described by 
% \citet{DR1-SL02} (`ECLIPSE-B'), 
\citeauthor{DR1-SL02} (\citeyear{DR1-SL02}; `ECLIPSE-B'), which leverages the combined power of ML, citizen science, and expert visual inspection. We overview the resulting catalogue here, alongside key results from lens modelling with \texttt{PyAutoLens} \comments{from} \citeauthor{DR1-SL03} (\citeyear{DR1-SL03}; `ECLIPSE-C') and photometric redshift and spectral energy distribution (SED) fitting of the deflector galaxies \comments{from} \citeauthor{DR1-SL04} (\citeyear{DR1-SL04}; `ECLIPSE-D'). Together, these efforts constitute the first end-to-end strong lens discovery and characterisation at this scale.

The paper is structured as follows. Section~\ref{sec:data} introduces the data, and Sect.~\ref{sec:method} overviews the discovery pipeline.
\comments{Section~\ref{sec:products} summarises the lens modelling, photometric redshift estimation, and SED fitting.} The catalogue itself is presented in Sect.~\ref{sec:catalogue}, with notable systems highlighted in Sect.~\ref{sec:lenses-of-interest}. Sample properties and completeness are discussed in Sects.~\ref{sec:properties} and~\ref{sec:completeness}\comments{,} respectively. We end with future prospects in Sect.~\ref{sec:future} and our conclusions in Sect.~\ref{sec:conclusions}.
\comments{Throughout the work we use the \textit{Planck} 2016 flat $\mathrm{\Lambda}$CDM cosmology with $H_0 = 67.74$\,km\,s$^{-1}$\,Mpc$^{-1}$, $\Omega_\mathrm{m} = 0.3089$, and $\Omega_\mathrm{\Lambda} = 0.6911$ \citep{planck}}.

\section{Data}\label{sec:data}
\Euclid produces images from four photometric bands. The VIS instrument \citep{EuclidSkyVIS} produces images in the optical \IE band, while the NISP instrument \citep{EuclidSkyNISP} produces images in the NIR \YE, \JE, and \HE \comments{bands}. 
\comments{NISP reaches point-source depths of $\YE = 24.6$, $\JE = 24.6$, and $\HE = 24.5$ at 5\,$\sigma$, whereas VIS reaches $\IE = 26.0$ for extended sources at 5\,$\sigma$.}
The DR1 data we searched through included single-visit coverage spanning \fixed{\dronearea}\,deg$^2$. 
\comments{Although DR1 includes the deep fields at both wide depth (equivalent to the Q1 release) and full deep-field depth, we excluded these fields from the DR1 search to avoid duplication, to account for the fact that the ML models were trained on Q1, and to keep the search homogeneous.\footnote{The original Q1 catalogue of strong lenses is available at \url{https://doi.org/10.5281/zenodo.15025832}.}}

One of the strengths of \Euclid in the context of galaxy-scale lens discoveries is that it is able to  resolve the large population of strong lenses down to radii of around $\ang{;;0.4}$ \citep{Q1-SP048}. 
\comments{This is primarily enabled by the VIS imaging, which has a pixel scale of $\ang{;;0.1}$ and a point spread function (PSF) full width at half maximum (FWHM) of $\ang{;;0.16}$, compared with the NISP imaging that has a pixel scale of $\ang{;;0.3}$ and PSF FWHM $\ang{;;0.35}$.}
The VIS resolution helps in identifying faint counter-images and arc structure, allowing us \comments{to} better distinguish lenses from non-lenses.
This motivated the design of our search to use images produced using the luminance of VIS for maximal resolution, with relative colours of the pixels informed by the NIR fluxes (following the image processing of \citealt{Q1-SP048}, where example images are shown). 
\comments{For each system, we produced \ang{;;10}$\times$\ang{;;10} cutouts, sufficient for the expected extent of galaxy-scale strong lenses, with a range of processing options: arcsinh and mid-tone transfer function scalings, applied to grey-scale \IE images and to \IE+\YE, \IE+\JE, and \IE+\YE+\JE colour composites. These cutouts formed the input to the search.}

\section{Search method overview}\label{sec:method}

As in Q1, our DR1 lens search takes the \comments{broad strategy}: 
\begin{itemize}
    \item[i)] produce cutouts of objects that could plausibly be lensing galaxies at the catalogue level;
    \item[ii)] apply ML networks to the resulting catalogue;
    \item[iii)] citizen scientists visually inspect the top ML scored objects;
    \item[iv)] experts assign grades to objects the citizens voted as likely lenses.
\end{itemize}
The discovery pipeline is detailed fully in ECLIPSE-B, and here we provide a brief summary.

\subsection{Catalogue selection}
There are approximately 900 million detected sources in DR1 according to the official \Euclid object detection catalogues (MER; \citealt{DR1-TP009}). Of these, a large number are objects that can be ruled out as uninteresting for our search at the catalogue level. Primarily these \comments{are} stars and galaxies too faint to exhibit plausibly detectable lensing features. This motivated our catalogue selection cuts
$$
\left\{
\begin{array}{ll}
\texttt{VIS\_DET} = 1  \\
(\IE < 23)~\textrm{or}~(\HE < 22.5)  \\
\texttt{MU\_MAX} > 15 \\
\texttt{MUMAX\_MINUS\_MAG} > -2.6~, \\

\end{array}
\right.
$$
intended to hold for the putative lens galaxy.
The condition \texttt{VIS\_DET} = 1 ensures that we only select objects that are detected in VIS,
\comments{and we keep those satisfying at least one of $\IE < 23$ and $\HE < 22.5$, the latter to retain massive high-redshift galaxies.}
\texttt{MU\_MAX} > 15 removes any object with a maximum peak surface brightness brighter than $15~\textrm{mag arcsec}^{-2}$, at which point the pixels are saturated, and \texttt{MUMAX\_MINUS\_MAG} $>-2.6$ acts as a compactness filter, removing objects where the total magnitude is dominated by a few pixels.
The VIS detection, saturation, and compactness criteria are the same as the Q1 search, while the magnitude limit
now includes fainter and redder systems than the Q1 limit of $\IE < 22.5$. 
This was motivated by the number of lenses in Q1 that were fainter than expected, and the considerable improvement in the ML models meant that we could be more lenient with the catalogue filters.
After applying this catalogue selection and removing objects in the Q1 and deep field areas, there were 68 million objects to search through.

\subsection{Machine learning and visual inspection}

With an unprecedented number of strong lenses in DR1, it was essential to develop a search approach that maximised lens recovery while retaining as many individually valuable lensing systems as possible. 
Expert visual inspection remains the most reliable method for identifying strong lenses from imaging alone.
\comments{Definitive confirmation generally requires spectroscopy showing distinct deflector and source redshifts, with a consistent lens model providing strong corroborating evidence, but in many cases the quality of \Euclid VIS imaging is sufficient to identify a system confidently as a strong lens from its morphology alone.}
However, inspecting 68 million images is far from viable, even for a large and enthusiastic team of experts. While ML offers a scalable alternative, current networks are not yet as reliable as expert human classifiers for distinguishing true lenses from contaminants \citep{Q1-SP059}. 
Fortunately, there exists a global community of citizen scientists who volunteer their time to help with such large-scale people-powered projects. 
However, the largest visual inspection search for strong lenses has been limited to $10^5$ objects \citep{Marshall2016}, meaning that searching through the full 68 million requires a synergy of ML and visual inspection by both citizens and experts.

For DR1, we developed a bespoke discovery pipeline that optimally synthesises information from these complementary search components. In ECLIPSE-B we describe the development and calibration of this pipeline in full, demonstrating how the optimal combination maximises both the completeness and purity of the final catalogue.
In brief summary, \black{five} ML networks were developed to find strong lenses in DR1. These networks were created by different teams in the Euclid Consortium strong lensing group, each \comments{being} sensitive to different lens morphologies 
(Euclid Collaboration: Vincken et al., in prep.).
% \citep{DR1-SL05}.
These networks scored all 68 million images, with their scores combined based on their performance on a calibrated test set. 
The ML models performed well: for \comments{two} of the top networks, none of the 1000 highest-scoring objects were classified as non-lenses by the experts (in Fig.~\ref{fig:collage} we show the top \comments{117} images from these two ML networks).
These ML scores were then used to prioritise the 
%Any ML flagged system inc 50deg.
\phildone{\num{220000}} most promising candidates, which were then passed to the citizen\comments{-}science project `Space Warps' \citep{Marshall2016, More2016} on the Zooniverse platform. 
In addition to the objects flagged by the ML models, \phildone{\randoms} images were passed to Space Warps that had been randomly selected from the original 68 million objects, allowing us to calibrate an intrinsic rate of detectable strong lenses in the parent sample. 

Both the ML-prioritised and the randomly selected images were passed to Space Warps, with each \comments{initially} assessed by \comments{between \phildone{6} and \phildone{25}} citizens \comments{(more contentious systems required more classifications). The median number of classifications per volunteer was \phildone{25}, with 6975 citizen scientists contributing to the project.} These responses were calibrated and combined using a Bayesian framework to minimise the variability arising from the inherent subjectivity of the classification task \citep{Marshall2016}.
Space Warps was carried out in two stages: the initial `Space Warps Classify' stage, in which volunteers classified each object as either a lens or a non-lens, and a subsequent `Space Warps Refine' stage, in which volunteers assigned lens grades.
%Include anything passed to SW:
Of the images sent to Space Warps, around \fixed{\num{30000}} were flagged as likely to be strong lenses. \comments{These candidates were then sent to a final expert vetting stage (`Galaxy Judges'), where each candidate was assigned a grade by \comments{an average of} \phildone{ten} members of the Euclid Consortium's strong lensing group.}
These expert grades were calibrated before aggregating into a final lens score (see ECLIPSE-B for more details). The experts assigned each candidate one of four grades, \comments{defined as follows: grade A, a
sure lens, showing clear lensing features with no additional information needed; grade B, a probable lens, showing lensing features but requiring additional information to verify; grade C, a possible lens, showing lens-like features that could also be explained without gravitational lensing; and grade X, definitively not a lens. These reflect the confidence of the graders based on the information available, and are not confirmations of the lensing nature of a system.}

\comments{This DR1 lens catalogue is the product of an enormous collective effort, involving \phildone{\num{6975}} citizen scientist volunteers and \phildone{\experts} experts from the Euclid Consortium's strong lensing group.} 
This catalogue would not have been possible without the immense contributions of the citizen scientists who participated in the project. The \comments{names} of all \comments{the} participants who contributed to the Space Warps project \comments{are} available on the Space Warps \href{https://www.zooniverse.org/projects/aprajita/space-warps-esa-euclid/about/team}{`The Team'} webpage and are listed in Appendix \ref{app:sw}.

\section{\comments{Lens characterisation}}\label{sec:products}

\subsection{\comments{Lens modelling}}\label{ssec:modelling}

\comments{Lens modelling provides independent corroborating evidence for the lensing nature of candidates beyond the morphological assessment of visual inspection, and} also enables a wealth of science through reconstruction of the light
of the source galaxies and mass distributions of the deflectors. \comments{The full details of the modelling are covered in ECLIPSE-C, but we provide a brief overview here.}
We attempted to model all grade A+B \comments{systems} using an automated pipeline implemented with \texttt{PyAutoLens} \citep{Nightingale2021}.\footnote{\url{https://github.com/PyAutoLabs/PyAutoLens}} These lens models were initiated by an automated ML-based decomposition of the data into deflector light, source light, and other line-of-sight objects, allowing automated masking and adjustment of model priors 
\citep{DR1-SL07}.
% \citep{DR1-SL07}
The lens models employed a multi-Gaussian expansion (MGE) for the deflector light and a singular isothermal ellipsoid (SIE) combined with external shear for the deflector mass.
Throughout this paper we report the effective Einstein radius  $\theta_{\rm E}$ defined in ECLIPSE-C, which is derived from the area enclosed by the tangential critical curve of the full lens model.
The modelling proceeds in three stages: an initial fit with MGEs for both the deflector and source light, which establishes the mass model and deflector-light subtraction; a second fit that replaces the source with a pixelised reconstruction on an adaptive Delaunay mesh, which refines the mass model and recovers the source morphology; and a final fit with S\'ersic profiles for the deflector and source light, with the mass model held fixed, which provides the deblended photometry and structural parameters used in Sect.~\ref{ssec:phz}. \comments{The set-up assumes a single deflector galaxy and a single source plane, and is therefore not designed to fit multi-galaxy deflector systems (e.g., groups and clusters) or multi-source-plane lenses.}

Each resulting lens model was vetted by at least \red{four} members of the Euclid Consortium's strong lensing group to assess the robustness and accuracy of the individual models, quantify an overall success rate, and understand failure modes. \comments{Each system was assigned one of the following categories.
\begin{itemize}
    \item[\textbullet] Success: the model is sufficiently robust for the inferred
    Einstein radius to be trusted.
    \item[\textbullet] Failure: the system appears to be a lens, but the fit failed.
    \item[\textbullet] Outside model scope: the system is a multi-galaxy deflector or
    a double-source-plane lens.
    \item[\textbullet] Unsure if lens: the modelling casts doubt on whether the system is a
    strong lens at all.
\end{itemize}
Each system received at least four votes, and a classification was adopted when at least 75\% of inspectors agreed (systems without 75\% agreement were recorded as having no consensus).
% While making lens models available during the initial grading would have been informative, this was not viable at this scale: modelling every candidate is computationally expensive, and assessing model quality is slower and requires more specialised expertise than assigning a lens grade alone.
}

\subsection{\comments{Deflector photometric redshifts and SED fitting}}\label{ssec:phz}

Although \Euclid possesses slitless spectroscopic capabilities, it was not optimised for obtaining redshifts in galaxy-scale strong lenses. In particular, the typically small angular separation between the deflector and source leads to significant blending of their light in the slitless spectra. For the lensed sources, \citet{DR1-Bergamini} show that reprocessing the slitless spectra to remove the overlapping continuum can recover emission lines from the brightest sources. This has so far been applied to only a small number of systems, but shows promise for measuring source redshifts. For the deflectors, which are mostly passive early-type galaxies, the \Euclid depth is generally insufficient to obtain spectroscopic redshifts. With external spectroscopy also available for only a small fraction of the sample (Sect.~\ref{sec:catalogue}), we rely on photometry to provide deflector redshifts at scale.

\comments{For systems with successful lens models, the resulting deblended photometry enables robust photometric redshift and SED modelling of the deflector galaxies, avoiding the biases introduced by catalogue-level measurements where the deflector and source light are blended.
ECLIPSE-D presents a detailed analysis of these deblended measurements, which is overviewed here.}
The deblended photometry comes from the final, S\'ersic stage of the lens modelling (Sect.~\ref{ssec:modelling}). The deflector and lensed source are each fitted with a single S\'ersic profile \citep{sersic1963} in the high-resolution \IE band, giving their S\'ersic indices, effective radii, and \IE fluxes. The same fixed mass model and \IE-based mask are then applied to every other available band, with the S\'ersic index and effective radius of each component held constant and only the flux in each band left free.
This produces unconvolved deflector and source images at each observed wavelength, extending beyond the \Euclid bands to ground-based imaging from the Ultraviolet Near Infrared Optical Northern Survey (UNIONS; \citealt{gwyn}) in the north and the Dark Energy Survey \citep[DES;][]{abbott2021} in the south. 
Aperture photometry is then conducted on the PSF-matched model images.
% \comments{The deflector light is modelled in the high-resolution \IE band, and the resulting S\'ersic \citep{sersic1963} index and effective radius are held fixed while fitting the remaining bands, producing deflector model images at each observed wavelength.
% In addition to the \Euclid bands, this is extended to ground-based imaging from the Ultraviolet Near Infrared Optical Northern Survey (UNIONS; \citealt{gwyn}) in the north and the Dark Energy Survey \citep[DES;][]{abbott2021} in the south, giving deblended photometry across a wider range of filters. Because the modelling returns unconvolved model images, these can be convolved directly to the lowest-resolution band, and aperture photometry measured from this. 
Photometric redshift and SED fitting are attempted only for the deflector population, since characterising the lensed sources requires a separate treatment of the lensing configuration and their larger photometric uncertainties, which is left to future work.

%An empirically calibrated term, derived from S\'ersic modelling of a matched sample of non-lens galaxies, is added in quadrature to the catalogue uncertainties to account for the fact that formal uncertainties from S\'ersic fits are unrealistically small.

\comments{Deflector photometric redshifts are derived with the template-fitting code \texttt{Phosphoros} (Paltani et al., in prep.), using the same SED library adopted for the main \Euclid photometric-redshift analysis \citep{Q1-TP005, DR1-TP011}, and validated against the subset of lenses with spectroscopic redshifts. Stellar population properties are then inferred by SED fitting with \texttt{Prospector} \citep{prospector}, which generates model SEDs from the Flexible Stellar Population Synthesis (FSPS) code \citep{conroy2009} with a non-parametric star-formation history and adopting a Salpeter initial mass function (IMF; \citealt{Salpeter}). This allows us to infer stellar masses, star-formation rates, stellar ages, and metallicities for the deflector population.}

% The combination of lensing observables and detailed stellar population measurements establishes this catalogue as a homogeneous data set for investigating both the demographics of galaxy-scale strong lenses and the physical properties of deflector galaxies. 

\input{fig/misc1}

\newpage

\section{Catalogue overview}\label{sec:catalogue}

Our final catalogue contains every system that reached the expert visual inspection stage of the search (Sect.~\ref{sec:method}), together with its grade and, where available, the lens model parameters and deflector properties described in Sect.~\ref{sec:products}. In total, the catalogue contains \fixed{\gradeABCX} unique objects, comprising
\begin{itemize}
    \item[\textbullet] \fixed{\gradeA} grade A lens \comments{candidates};
    \item[\textbullet] \fixed{\gradeB} grade B lens \comments{candidates};
    \item[\textbullet] \fixed{\gradeC} grade C lens \comments{candidates};
    \item[\textbullet] \fixed{\gradeX} grade X (non-lenses).
\end{itemize}
Of these, \fixed{\knownA} grade A, \fixed{\knownB} grade B, and \fixed{\knownC} grade C had been reported as candidates in prior publications (Sect. \ref{ssec:missed}). 
Examples of the lenses found are shown in Fig.~\ref{fig:misc1}, while the top 117 candidates according to the ML networks (all grade A) are shown in Fig.~\ref{fig:collage} and randomly selected examples from each grade are shown in Fig.~\ref{fig:random_grades}.
These grades are inherently subjective, with blurred boundaries between categories, although we expect the vast majority of grade A and a large fraction of grade B candidates to be genuine strong lenses. The catalogue therefore also includes a continuous lens score, aggregated from the individual inspector classifications, allowing readers to set their own boundaries for their use case. This catalogue, together with all data products used and the subsequent lens modelling and SED fitting results, is available on Zenodo.\footnote{\url{https://doi.org/10.5281/zenodo.19654904}}

Following the \comments{automated lens modelling and subsequent} vetting stage (covered in detail in ECLIPSE-C), the experts reached a consensus (at least 75\% agreement) that the model was successful for \red{67\% (43\%)} of grade A (B) systems. 
A further \red{5\% (4\%)} exhibited lensing geometries that fell outside the scope of the adopted galaxy--galaxy model, mostly multi-deflector lenses plus a small number of double-source-plane lenses. These systems are being modelled separately by \black{Wielders et al. (in prep.)} and \citet{dr1dspl}, respectively.
Of the remainder, \red{10\% (13\%)} were flagged as lens model failures, \red{0.1\% (1.4\%)} were flagged as uncertain lenses, and \red{17\% (38\%)} reached no consensus.
Since subtracting the deflector light can reveal a missing counter-image or an unphysical lens configuration, the models also offer an independent check on the visual grades. A successful model corroborates rather than proves a lensing interpretation, and a failed model often reflects the limitations of the automated pipeline, so the clearest test is how often the modelling cast doubt on a system being a lens at all. Even at 50\% agreement, this was the case for only \red{0.4\% (4.8\%)} of grade A (B) systems. By this test the grade A sample appears very pure, and the higher rate among grade B systems reflects the lower confidence of that grade.

\comments{Spectroscopy provides independent evidence on the lensing nature of a candidate. Of the spectroscopic data currently available, the Dark Energy Spectroscopic Instrument (DESI) provides the most extensive usable overlap. Storfer et al. (in prep.) crossmatch this ECLIPSE catalogue with DESI DR1 \citep{desidr1} and vet the spectra for deflector and source redshifts.
Of our grade A, B, and C candidates, \fixed{351} have a deflector redshift only (\fixed{95} grade A, \fixed{146} grade B, and \fixed{110} grade C), \fixed{eight} have a source redshift only (\fixed{four} grade A and \fixed{four} grade B), and a further \fixed{20} have both (\fixed{12} grade A, \fixed{five} grade B, and \fixed{three} grade C).
These deflector redshifts span $\fixed{0.05<z<1.1}$, and the source redshifts span $\fixed{0.3<z<2.7}$.
A detection of two distinct redshifts is highly informative where arc-like features could instead be intrinsic to the deflector (e.g., a ring galaxy or tidal feature), but does little for two clearly aligned objects where the question is whether the configuration is strongly or only weakly lensed. 
Lens modelling is complementary, since subtracting the deflector light can reveal a counter-image blended with the deflector, and so is most informative in the latter case.
Two of the three grade C systems with two redshifts (Appendix~\ref{app:spec}) fall into the former category, and as such the detection of two distinct redshifts provides almost conclusive evidence that these systems are genuine strong lenses.
}
\comments{In addition to vetting the DESI spectra, Storfer et al. (in prep.) present additional optical spectroscopic follow-up of \red{12} systems from this catalogue with the Next Generation Palomar Spectrograph (NGPS). One of these, a grade A candidate, shows evidence of not being a strong lens, based on a slit placed along the arc showing emission lines at the same redshift as the putative deflector, whose redshift was already known from DESI. This is not entirely conclusive since this could have been contamination from the seeing. Nevertheless, it illustrates that the grades reflect the confidence attainable from \Euclid imaging alone, and are imperfect in both directions.}

\section{Individual lenses of interest}\label{sec:lenses-of-interest}

This catalogue represents not only the largest homogeneous catalogue of strong lenses with which population-level statistics are meaningful, but it is also a treasure trove of rarer and more unique lensing configurations.
The majority of galaxy-scale strong lens systems involve a massive red elliptical deflector lensing a blue star-forming source. However, a sample of this size also contains rarer classes of lenses in meaningful numbers, including double-source-plane lenses, disc-galaxy deflectors, and lensed quasars.

\input{fig/dspl}

\subsection{Double-source-plane lenses}
\label{subsec:compound_lenses}

\input{fig/discs}

Double-source-plane lenses (DSPLs) are strong-lensing systems in which multiple background source galaxies at different redshifts are lensed by the same foreground deflector, producing multiple rings at different radii. Strong lenses with multiple source planes provide especially powerful probes of the density profiles and dark matter substructure of galaxies. The second source can help break lensing degeneracies under appropriate assumptions, and provides an additional aperture within which the enclosed mass is tightly constrained \citep[e.g.,][]{Enzi2025,Li2026SalpeterNFWDSPL}. The ratio of the Einstein radii is also related to the cosmological scaling factor and can therefore provide geometric leverage on cosmological parameters \citep{Gavazzi2008,Collett2012,CollettAuger2014,Q1-SP054, ballard26}, although its ability is limited by the mass-sheet degeneracy \citep{Schneider2014}.

Galaxy-scale DSPLs are rare, since the probability of lensing two independent background sources is much smaller than that of producing a single-source-plane lens. Prior to DR1, there were approximately ten discovered galaxy-scale DSPLs. Four of these were discovered in Euclid Q1, demonstrating that such systems can be identified in \Euclid imaging with our search approach. This number discovered in the Q1 area was broadly consistent with the expected rate from forecasts \citep{Q1-SP054}. 
In our DR1 search, we identify \fixed{\dspl} candidate DSPLs. These are systems passed to the final expert inspection stage that showed evidence of two sets of lensed features at distinct Einstein radii. We then checked their configuration with initial lens models built specifically for DSPLs, in which the two sources are fitted on separate source planes \citep{dr1dspl}.
Examples of DSPLs identified in this catalogue are shown in Fig.~\ref{fig:dspls}. Figure~\ref{fig:dspls}\blue{a} shows the clearest example, with two concentric, almost complete Einstein rings of distinct colours. Figure~\ref{fig:dspls}\blue{m}, despite having a low signal-to-noise ratio (S/N), is notable \comments{for} showing signs of two Einstein rings around an edge-on disc galaxy, the first such system of its kind. If confirmed, this system would enable a unique study of the mass profiles of disc galaxies in a way never previously possible.

\subsection{Disc lenses}

\comments{Much of what is known about the mass structure of spiral galaxies comes from rotation curves in the local Universe, which probe deep into the halo. However, this kind of kinematic information becomes more difficult to obtain beyond the local Universe. This is compounded by the intrinsic rarity of massive spirals, whose abundance declines sharply above $M_\ast \simeq 10^{11}\,M_\odot$.} Strong lensing offers a powerful complementary probe of the decomposed mass structure of disc galaxies \comments{in this otherwise inaccessible regime}.
Disc galaxies comprise several superimposed components (discs, bulges or bars, and dark matter haloes), and samples of disc lenses spanning a range of Einstein radii and inclinations can help disentangle their mass contributions. Face-on systems make it easier to separate the bulge or bar from the disc, while in edge-on systems the flattened disc and rounder halo project differently, making them sensitive to the dark matter distribution and the disc mass-to-light ratio, and allowing tests of alternative models of gravity \citep{ale_rev}. Together, these measurements enable studies of the stellar mass-to-light ratio and the IMF \comments{(e.g., \citealt{brewer12})}.
% Disc galaxies typically comprise several superimposed structural components, namely discs (often with spiral arms), bulges or bars, and dark matter haloes. Samples of disc lenses spanning a range of Einstein radii can help disentangle the mass contributions of these components. Individual systems provide measurements of the total projected mass within the Einstein radius, while extended arc morphologies can constrain the slope of the mass profile at larger radii.
% A range of inclination angles is also beneficial. Face-on systems make it easier to separate the mass of the bulge or bar from that of the disc, while in edge-on systems the flattened disc and the rounder halo project differently, making them sensitive to the dark matter distribution and the disc mass-to-light ratio \comments{(\citealt{Q1-Ecker})}, and allow tests of alternative models of gravity \citep{ale_rev}. Together, these measurements enable studies of the stellar mass-to-light ratio and the IMF \comments{(e.g., \citealt{brewer12})}.
\comments{Furthermore, comparing disc and elliptical lenses provides constraints on how the dark matter fraction depends on galaxy morphology, complementing studies of the stellar-to-halo mass relation \citep[e.g.,][]{posti2019, Tortora2019, Tortora2025}, thereby weighing the relative roles of internal physical processes (e.g., supernova and AGN feedback) and mergers in their formation.}

Despite this potential, lensing by disc galaxies is considerably rarer than lensing by ellipticals. Disc galaxies are generally less massive, and there is an additional selection effect in that lensed arcs are harder to distinguish in galaxies with more complex structure \comments{and dust extinction}.
Among disc lenses, inclined and edge-on systems are expected to be more common, since inclination increases the projected mass density and thereby the lensing cross-section.
\comments{Before \Euclid, only a few tens of edge-on disc lenses were known, most of them identified spectroscopically (e.g., \citealt{swells}), and morphological searches in UNIONS \citep{AcevedoBarroso2026} and Euclid Q1 \citep{Q1-SP053} have since roughly doubled this number.}
Face-on disc lenses, in contrast, are less efficient lenses and harder to identify. No complete Einstein rings had previously been identified within a face-on disc galaxy, and the few known face-on spiral deflectors had been spectroscopically selected (e.g., \citealt{spiral_lens}).
In our DR1 search, we identify \fixed{1060} \comments{candidate} disc lenses (including S0 galaxies) with a projected deflector axis ratio $q \leq 0.5$, corresponding to an inclination of $i \gtrsim 60^\circ$. We identify them by morphology alone, using the median $q$ of the deflector light measured in the \IE imaging, the segmentation maps, and the lens-light models, with contaminants removed manually. Because elliptical galaxies rarely appear flatter than $q \approx 0.4$ in projection, this selection reliably identifies inclined discs, but face-on and moderately inclined discs cannot as easily be distinguished from ellipticals, so the sample is a lower limit on the number of disc lenses.
Figure~\ref{fig:discs} shows the most striking disc lenses in the catalogue, with clear lensing arcs alongside resolved internal structure in the deflector, such as spiral arms and dust lanes. These include a small number of face-on disc lens candidates with clearly identifiable lensed arcs, of which Figs.~\ref{fig:discs}\blue{a} and~\ref{fig:discs}\blue{c} are the two clearest examples.

The complex, asymmetric light structure of disc deflector galaxies \comments{and the effects of dust attenuation} complicate not only detection but also lens modelling. 
% Historically, deflector light has been modelled using parametric approaches, typically assuming smooth analytic profiles such as S\'ersic components or combinations of Gaussian light distributions. However, such methods are insufficient for these systems.
{In ECLIPSE-C, many automated lens-model failures are attributed to poor deflector light subtraction, driven by dust or asymmetric structure such as in spiral galaxies.}
Therefore, these particularly valuable systems require a more rigorous treatment to fully exploit the lensing information.
\black{Ecker et al. (in prep.)} present a dedicated modelling project for a subset of edge-on disc lenses, while \black{Enzi et al. (in prep.)} present a detailed analysis of the face-on system in Fig.~\ref{fig:discs}\blue{a}. In this latter case, the availability of both the deflector and source redshifts (provided by the DESI collaboration) enables a joint lensing and photometric analysis, in which the lens model simultaneously decomposes the deflector and source light while SED fitting of the spiral constrains the stellar mass.

\subsection{Lensed \comments{AGN} and quasars}

Lensed quasars and AGN are among the most scientifically valuable classes of strong gravitational lenses. One of their most powerful applications is time-delay cosmography: quasars are intrinsically variable, allowing measurements of the time delay between the multiple images.
These time delays are sensitive to the Hubble constant ($H_0$; \citealt{Refsdal1964}), and have been used to constrain $H_0$ to high precision (e.g., \citealt{tdcosmo2026}).
Additionally, the multiple lines of sight through the deflector galaxy make lensed quasars sensitive probes of dark matter substructure via flux ratio anomalies \citep{Dalal2002}. 
Stellar-mass objects within the deflector galaxy produce microlensing, causing magnification fluctuations that can be used to spatially resolve the quasar structure down to $10^{-6}$ arcsec scales \citep{Vernardos2024}.
As such, a larger sample of lensed quasars from \Euclid would provide a valuable resource for a wide range of science cases.

\input{fig/qso}
\input{fig/misc2}

\begin{figure*}
    \centering
    \includegraphics[width=1\textwidth]{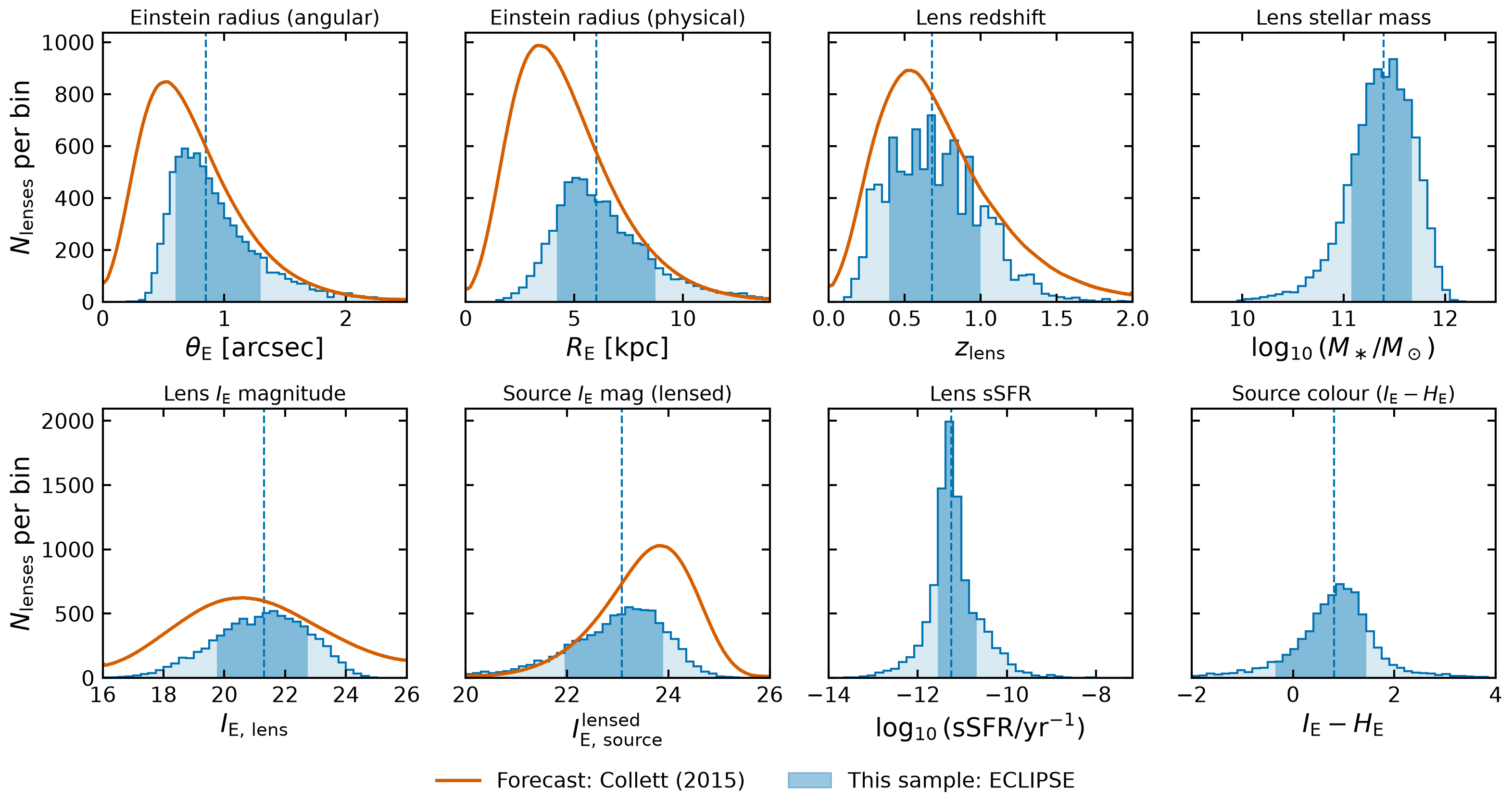}
\caption{Distributions of lens and source properties for the grade A+B systems (blue). Lens redshifts, stellar masses, and sSFR are measured photometrically in ECLIPSE-D for the \angelosmodels systems passing its quality cuts. Einstein radii, magnitudes, and source colours come from the \num{7980} lens models judged successful in ECLIPSE-C, and physical Einstein radii use the \num{6199} systems in both samples. Where predictions are available, comparisons to the forecasts of \citet{Collett2015} are overlaid (orange).
\comments{Vertical dashed lines represent the inferred median, while the darker shaded region represents the inferred $16$th--$84$th percentile range.}}
    \label{fig:lens-properties}
\end{figure*}

Our search specifically targets galaxy--galaxy lenses and not lensed quasars: the catalogue-level selection cuts intentionally exclude compact sources (aimed at excluding stars).
\comments{
Although these cuts act on the deflector rather than the source, in small-separation systems where the lensed quasar images dominate the light, the blended detection can be sufficiently compact to fail the criteria.
A further selection arises from the search itself. The ML networks are primarily trained on galaxy--galaxy lenses and are best
at recognising extended arcs. Lensed quasars are also more difficult for humans to identify, since point-like configurations are much more often false positives.
For these reasons, a dedicated lensed quasar search, using a targeted catalogue selection and tailored ML models, will be presented in future work.}
Nevertheless, lensed quasars with a visibly lensed host galaxy will survive the selection cuts and are likely to be recognised by both the ML models and human inspectors, making them a probable product of this search.

\comments{{Acevedo Barroso et al. (in prep.)} crossmatch this ECLIPSE catalogue with external AGN selections, such as through their optical variability, their mid-infrared colours, or X-ray emission. Around 300 lens systems from this ECLIPSE catalogue are found to have counterparts. However, an AGN counterpart does not necessarily reveal the origin of the activity, which could be the lensed source
(e.g., Figs.~\ref{fig:lqso}\blue{a}, \ref{fig:lqso}\blue{b}, \ref{fig:lqso}\blue{i}, \ref{fig:lqso}\blue{j}), the deflector galaxy (e.g., Figs.~\ref{fig:lqso}\blue{k} and \ref{fig:lqso}\blue{m}), or in a neighbouring galaxy or group. For around \fixed{65} of these $\sim$300 systems, the activity can be confidently tied to the lens system. 
Moustakas et al. (in prep.) take the complementary approach, assembling a sample of four-image lensed quasars and AGN in DR1 from multiple origins, and requiring for each system both an astrometric lens model and independent evidence of nuclear activity from spectroscopy, X-rays, or mid-infrared colours. Around 30 of their systems are in this catalogue, almost all at grade A (examples in Figs.~\ref{fig:lqso}\blue{a}--\blue{h}). These approaches also rediscover several previously studied lensed quasars amongst our sample, including SDSS\,J1524+4409 
\citep{Oguri2008},
DES\,J0408$-$5354 \citep{Lin2017}, DES\,J0420$-$4037 \citep{Shajib2019}, ATLAS\,J2344$-$3056 \citep{Schechter2017},
DES\,J0115$-$5244 \citep{agnello2015}, and the three-image system J0316$-$4106 (Fig.~\ref{fig:lqso}\blue{j}; \citealt{Sheu2026}).}

\comments{Cross-matching \Euclid candidates with external catalogues is constrained by the depth of the catalogues, with many fainter lensed AGN and quasars likely going undetected. Morphology offers a complementary and less conservative route for identification. During visual inspection, around \fixed{500} objects were flagged as potential lensed quasars on the basis of a bright, compact multiple-image morphology.
However, compact morphology alone is not a reliable discriminant of lensed quasars, since such systems could equally be extremely compact galaxies (e.g., blue nuggets; \citealt{Napolitano2020, jaelani2020, ruili2024}).}

\comments{In total, this catalogue contains \fixed{around 65} lens candidates in which the activity can be tied to the lens system, and \fixed{a few hundred} more with either an AGN counterpart of uncertain origin or a quasar-like morphology. These provide a starting sample for a dedicated lensed quasar search, which can build on \Euclid's demonstrated capability for quasar science, including the discovery of some of the highest-redshift quasars known \citep{yang26}.}

% A robust search for lensed quasars candidates should therefore rely strongly on extended-wavelength photometry, spectroscopy, or time-domain observations, none of which are used in the current search.

\subsection{Other rare strong lenses}

In Fig.~\ref{fig:misc2}, we show a collection of some of the other more interesting systems found in the search. In addition to galaxy--galaxy lenses, we find \comments{multi-deflector lenses} (e.g., Fig.~\ref{fig:misc2}, panels a--e), which are explored in more detail in \black{Wielders et al. (in prep.)}. Systems such as Fig.~\ref{fig:misc2}\blue{f} show bright extended arcs with resolved source structure, making them particularly sensitive to mass perturbations in the deflector galaxy such as from dark matter substructure \citep{Despali2022, ORiordan2023, Amvrosiadis2026}.
Such systems also enable studies of individual star-forming regions at spatial scales unattainable at these redshifts in the absence of gravitational lensing.
\comments{Lenses like those in Figs.~\ref{fig:misc2}\blue{g} and \ref{fig:misc2}\blue{h} with satellite galaxies perturbing the lensed arc allow us to probe the mass of satellite galaxies beyond the local group.}

Figures~\ref{fig:misc2}\blue{i},~\ref{fig:misc2}\blue{j}, and~\ref{fig:misc2}\blue{k} show systems with redder sources, either from high-redshift galaxies or from intrinsically redder populations, such as quiescent red nuggets or dust-obscured submillimetre galaxies. We note that our reliance on VIS for high-resolution images (as discussed in Sect.~\ref{sec:data}) means that any sources that are VIS-dropouts (\comments{e.g., systems beyond} $z = 7$ \comments{or very dust-obscured galaxies}) will go undetected. Therefore, searches for such objects will require using different data products (e.g., \citealt{DR1-SL06}), or a targeted search for specific VIS-dropout objects (\black{e.g., Harvey et al., in prep.}).
Another consequence of relying on VIS is that lenses with very red deflector galaxies appear to have low surface brightness in the inspection images, despite the deflectors being clearly detected in the NIR bands (e.g., Fig.~\ref{fig:misc2}\blue{m}).
This is contrasted with a class of objects discovered in our search showing ring-like structures with no central detection from any filter bands (including \Euclid NIR data\comments{,} as well \comments{as} data in other filter bands from ground-based surveys), discussed in detail in Appendix~\ref{app:non-lenses}.

\section{Properties}\label{sec:properties}

The lens models and deblended deflector photometry allow us to characterise the lens population as a whole. Einstein radii, magnitudes, and source colours come from the \num{7980} lens models judged successful in ECLIPSE-C, and deflector redshifts, stellar masses, and specific star-formation rates come from the \angelosmodels systems in the ECLIPSE-D final sample. Because these rely on the deblended photometry from the lens models, ECLIPSE-D excludes systems whose models were flagged as failures, multi-deflector lenses, double-source-plane lenses, or unlikely lenses, and further requires detections in at least six bands, reliable photometric redshifts and SED fits, and a stellar mass above $10^{10}\,M_\odot$. Neither subset is a random sample of the catalogue. Modelling fails preferentially for the systems that are hardest to model, such as those with small Einstein radii or complex deflector morphologies, so the distributions presented here are biased against such systems.

Figure~\ref{fig:lens-properties} presents the distributions of several key measured properties of these lenses, compared with the expected distributions from the forecasts of \citet{Collett2015}.
As expected, we find that our deflector galaxies are primarily quiescent, early-type galaxies, with stellar masses of \red{\stellarmass}, and specific star-formation rates (sSFR) of \red{\ssfr}
\comments{(quoted ranges correspond to the $16$th and $84$th percentile values around the median), while the colours of the lensed sources are consistent with a predominantly star-forming population at higher redshift}.
Although deflector galaxies are expected to be more massive than the underlying population on average, these deflector galaxies are found to be largely representative of the broader population of massive quiescent galaxies. In particular, \red{they follow the expected stellar mass--size relation, with further scaling relations presented in ECLIPSE-D.}

\begin{figure}
    \centering
    \includegraphics[width=0.95\linewidth]{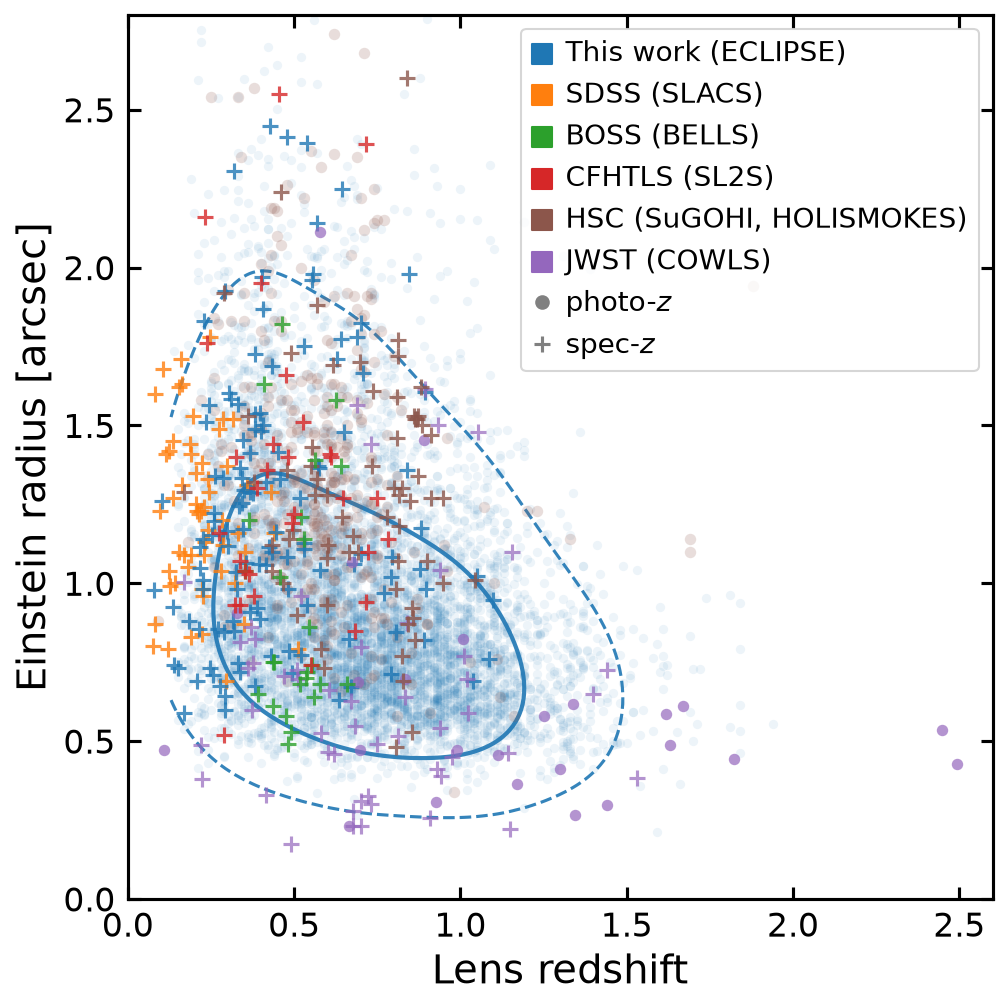}
    \caption{Distribution of lens redshift and Einstein radius of lenses from previous samples (SLACS, BELLS, SL2S, SuGOHI, HOLISMOKES, and COWLS) compared to this work (ECLIPSE). For ECLIPSE we show the \num{6199} grade A+B systems with a successful lens model and an ECLIPSE-D photometric redshift, together with the \num{122} successfully modelled systems with a vetted DESI DR1 spectroscopic lens redshift, and the solid and dashed contours enclose 68\% and 95\% of the photometric-redshift sample. For the HSC lenses we limit the sample to grade A+B lens \comments{candidates}, for COWLS we limit the sample to the lenses of score `S07' and above, and we plot photometric redshifts where spectroscopic redshifts are not available.
    In most cases, Einstein radii are inferred from lens modelling, with the exception of HOLISMOKES where they are inferred by ML.}
    \label{fig:lens-properties-compared-to-other-samples}
\end{figure}

Figure~\ref{fig:lens-properties-compared-to-other-samples} compares the Einstein radii and redshifts of our lenses with those of previous samples, namely the spectroscopically selected Sloan Lens ACS Survey (SLACS; \citealt{bolton2008, auger2009}) and BOSS emission-line lens survey (BELLS; \citealt{brownstein2012}), the ground-based imaging-selected Strong Lensing Legacy Survey (SL2S; \citealt{gavazzi2014, sonnenfeld2013b}), Survey of Gravitationally lensed Objects in Hyper Suprime-Cam (HSC) Imaging (SuGOHI; \citealt{Sonnenfeld2018, Gawade2025}), Highly Optimized Lensing Investigations of Supernovae, Microlensing Objects, and Kinematics of Ellipticals and Spirals (HOLISMOKES; \citealt{holismokes16}), and the JWST COSMOS-Web Lens Survey (COWLS; \citealt{Nightingale2025, Mahler2025, Hogg2025}).
Compared with the ground-based samples, our catalogue extends to smaller Einstein radii (1st--99th percentile range \comments{\red{$0.42<\theta_{\rm E}<2.32$ arcsec}} for systems with a successful lens model) and contains higher-redshift lenses on average (\red{median $z=0.68$}, compared with $z = 0.22, 0.52, 0.46$, and 0.59 for SLACS, BELLS, SL2S, and HSC, respectively). COWLS reaches smaller Einstein radii and higher redshifts still (median $z=0.81$) owing to the depth and resolution of JWST, but over a limited area.
Our deflector stellar masses are typical of the previous spectroscopic samples, lying predominantly in the range $11 \lesssim \logten (M_\ast /M_\odot) \lesssim 12$ (Fig.~\ref{fig:lens-properties}) as found for SLACS and BELLS (\citealt{bolton2012}, also assuming a Salpeter IMF). 
Unlike these spectroscopic samples, our deflectors are relatively faint
(\red{median $\IE \approx 21.3$}), and so our catalogue probes this population of massive early-type galaxies out to higher redshifts, making it well suited to studying their scaling relations and evolution.

% \comments{The photometric redshifts can be compared to the subset with DESI spectroscopy (Sect.~\ref{sec:catalogue}).}
% These spectroscopic redshifts from DESI for the deflector galaxies are in the range $\fixed{0.05<z_{\rm lens}<1.1}$, while the source redshifts are in the range $\fixed{0.3<z_{\rm source}<2.7}$. 
% For the deflector galaxies, the spectroscopic redshifts are \black{broadly consistent} with the photometric redshifts reported here. However, the subset with secure DESI lens redshifts is biased towards brighter lens galaxies and, consequently, towards lower redshifts than the full sample.

Compared to the forecasts of \citet{Collett2015}, our Einstein radii are systematically larger and our lensed sources are systematically brighter by almost a magnitude. This is likely because \citet{Collett2015} assumed the \Euclid lens search would be \IE-band only, using lens-light subtracted images with subtractions limited only by Poisson noise. 
In reality, we used the \Euclid NISP \comments{photometric} data to construct high-resolution colour images: while this bypassed the need for lens-light subtraction, it limited our ability to distinguish faint arcs buried in the cores of bright deflectors. 
\comments{This could also explain why we miss lenses with the brightest deflectors relative to the forecast, since the deflector light makes the lensed images in these systems harder to identify.}
The implications are three-fold. 
First, the close agreement between \comments{the number density of lenses found} \fixed{(around $8\,{\rm deg}^{-2}$) for grade A+B} and the \citet{Collett2015} forecast \comments{(11\,deg$^{-2}$)} is largely coincidental: the two approaches select substantially different, only partially overlapping systems, and the agreement in number belies a real difference in  \comments{the types of} lenses that are found. Second, it follows that a large population of strong lenses -- preferentially those with small image separations and low arc-to-lens contrast -- remains to be discovered in lens-subtracted imaging. 
Third, because the lensed sources in our sample are typically a magnitude brighter than forecast, their spectroscopic follow-up will be easier than planned in the design of the 4MOST Strong Lensing Spectroscopic Legacy Survey (4SLSLS; \citealt{4SLSLS}), a programme to measure source redshifts and deflector velocity dispersions for large samples of strong lenses, including those found by \Euclid.
On the other hand, missing these bright-deflector systems will make it harder for 4SLSLS to measure large numbers of reliable lens velocity dispersions.

% \tom{Groupiness of our sample. (Numerical richness, SW tagged companions). Edge on - face on- LRG ratio.?}

\section{Completeness}\label{sec:completeness}

A statistically large sample of strong lenses is beneficial for population-level science, but requires an understanding of the completeness rate to mitigate selection effects.
Although this catalogue is the largest of its kind to date, it is not exhaustive.
We explore the completeness of our sample in three ways: by estimating the fraction of lenses lost at each stage of the search (Sect.~\ref{ssec:completeness-estimate}), by measuring the selection function of visual inspection with simulated lenses (Sect.~\ref{ssec:visinsp}), and by testing the recovery of previously known lenses (Sect.~\ref{ssec:missed}).

\subsection{\comments{End-to-end completeness}} \label{ssec:completeness-estimate}

\newcommand*{\MLfoundlens}{\ensuremath{\textit{L}_{\rm{{ML \,found}}}}}
\newcommand*{\SWCfoundlens}{\ensuremath{\textit{L}_{\rm{{SWC \,found}}}}}
\newcommand*{\SWRfoundlens}{\ensuremath{\textit{L}_{\rm{{SWR \,found}}}}}
\newcommand*{\MLlostlens}{\ensuremath{\textit{L}_{\rm{{ML \,lost}}}}}
\newcommand*{\SWClostlens}{\ensuremath{\textit{L}_{\rm{{SWC \,lost}}}}}
\newcommand*{\SWRlostlens}{\ensuremath{\textit{L}_{\rm{{SWR \,lost}}}}}

\comments{Based on the randomly selected objects from the parent sample of objects passing the catalogue selection, it is possible to estimate the completeness of the DR1 lens search against expert grading (i.e., estimating the number of lenses we would have found had the experts inspected all 68 million objects passing the catalogue selection, taking their grades as the ground truth).
In doing so, we attribute any incompleteness to systems that failed to exceed the score thresholds imposed on the ML models, or systems that were identified by the ML models but not scored highly enough by the citizen scientists of the Space Warps stage. For these calculations, we also do not consider potential lens candidates missed due to not passing our catalogue selection criteria. 
}

\comments{
Since the expert grades are taken as the ground truth, every lens in this analysis is by construction a system seen by the experts. However, a system could reach the final expert grading stage by one of three routes. Objects with the highest ML scores were sent directly to expert grading, while those with intermediate scores were first passed to Space Warps Classify (SWC). Systems with high SWC scores were then sent directly to expert grading, while those with intermediate SWC scores were passed to Space Warps Refine (SWR), and those passing SWR were in turn sent to expert grading. Each stage can therefore both find lenses, by passing them to the experts, and lose them, by rejecting systems that had passed the previous stage. The number of lenses found and lost can be written as
\begin{equation}
L_{\rm{found}}=\MLfoundlens+\SWCfoundlens+\SWRfoundlens
\end{equation}
and 
\begin{equation}
L_{\rm{lost}}=\MLlostlens+\SWClostlens+\SWRlostlens~,
\end{equation}
respectively. Therefore, the completeness $f$ can be written as
\begin{equation}
    f = \frac{L_{\rm{found}}}{L_{\rm{found}}+L_{\rm{lost}}}~.
\end{equation}
}

\comments{We estimate these numbers of found and lost lenses at each stage using three sets of random objects injected into the pipeline. The first is \phildone{\randoms} objects drawn randomly from those passing the catalogue selection and shown to SWC regardless of ML score, which quantifies the systems missed by the ML classifiers. The second and third are samples of around \phildone{\num{1000}} and \phildone{\num{2000}} objects drawn randomly from those seen by SWC and SWR, respectively. Both were shown in their entirety to the experts, which allows us to quantify the systems missed by the citizens.
While showing the \phildone{\randoms} randoms directly to experts would have been simpler, this would have far exceeded the project's inspection capacity, and
with a base rate of only one lens per several thousand objects, a smaller random sample would instead have been dominated by small-number statistics. We therefore showed the full random sample to the citizens and corrected for the Space Warps completeness separately.
Amongst the initial \phildone{\randoms} randoms, we recovered \phildone{five} grade A, \phildone{16} grade B, and \phildone{34} grade C systems.}

\begin{figure}
    \centering
    \includegraphics[width=0.9\linewidth]{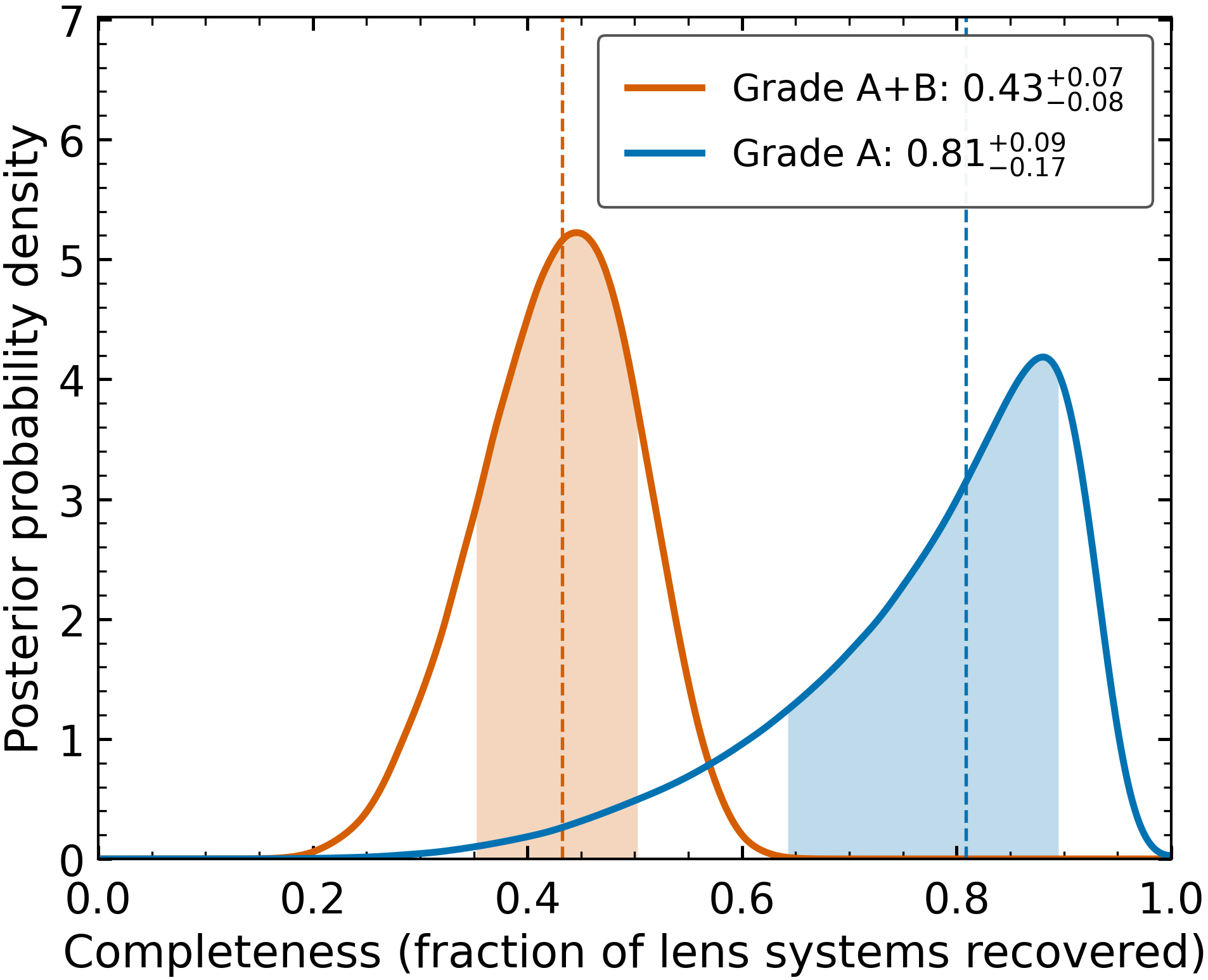}
    \caption{Probability distribution of the completeness of this lens sample, for both the grade A and grade A+B lens \comments{candidates}. These numbers are inferred from the number of lenses found from the randomly selected samples, and use expert visual inspection scores as the ground truth. \comments{The shaded regions represent the inferred $16$th--$84$th percentile range, with the dashed lines representing the median.}}
    \label{fig:completeness}
\end{figure}

\comments{Each lost term is drawn from a Beta posterior (Jeffreys prior) conditioned on the number of lenses recovered in the relevant random sample, and scaled to the number of objects rejected at that stage. The ML term is additionally divided by the completeness of the citizen stages, since the \phildone{\randoms} randoms were inspected by citizen scientists rather than experts. 
% This assumes that volunteers classify objects failing the ML threshold in the same way as those passing it. 
The full derivation is given in ECLIPSE-B. 
For grade~A systems, we infer that \phildone{$490^{+5300}_{-490}$} lenses were lost at the ML stage, \phildone{$270_{-240}^{+1500}$} at the SWC stage, and \phildone{$860_{-230}^{+330}$} at the SWR stage. While we have definitive examples of grade~A lenses that were lost at the SWC and SWR stages, there were notably zero grade A lenses in the random sample of \num{50000} that did not separately pass the ML threshold. The inferred rate of grade~A lenses lost by the ML is entirely prior-dominated and the most likely value is that zero are lost by the ML.}
\comments{For grade~A+B systems, the numbers lost at the ML, SWC, and SWR stages are respectively \phildone{$5300^{+16400}_{-4900}$}, \phildone{$7100_{-3200}^{+4600}$}, and \phildone{$7000_{-650}^{+780}$}. These numbers exclude the lenses found and lost during the initial stages of the DR1 search, as part of the ML training and calibration stages.
Extrapolating from these numbers to the full DR1 area, we infer a completeness of \phildone{$0.81_{-0.17}^{+0.09}$} for grade A and 
\phildone{$0.43_{-0.08}^{+0.07}$} for grade A+B systems. The posterior probability distributions are shown in Fig.~\ref{fig:completeness}.}

\comments{Unlike in smaller searches such as Q1, the completeness here is limited not only by the ML models but also by the capacity for visual inspection. The experts could inspect only 0.04\% of the parent sample, so the threshold for reaching the expert grading stage was necessarily very high. As a result, many candidates that the ML models and Space Warps ranked highly could not be shown to the experts.
The completeness is relatively high for grade A systems but much lower for grade A+B, and notably the number of grade A+B lenses we infer to have been missed exceeds the number of non-lenses graded by the experts. Consequently, even with perfect ML and Space Warps selection, the expert capacity of this project would not have allowed every grade A+B lens to be recovered. 
At this scale, the question is therefore not only how many lenses are found, but which ones, and our search preferentially recovers grade A systems.
With \phildone{\experts} strong lensing experts across the Euclid Consortium already contributing substantial time to this project, expanding expert inspection much further is unlikely to be feasible. 
Scaling the search to later data releases is likely to require automating the final grading stage, for example with ML.}

\subsection{\comments{Visual inspection selection function}}\label{ssec:visinsp}

\comments{Our completeness estimate above uses expert grades as the ground truth, and so cannot reveal systems that the experts themselves would miss. The selection function of the human inspection stages (both Space Warps and expert grading) can instead be measured independently by inserting simulated lenses with known properties into the lens identification pipeline. We used two sets of simulations, described fully in ECLIPSE-B: lenses constructed from real \Euclid galaxies for both the deflector and source \citep{Q1-SP052}, and lenses generated with \texttt{GLAMER} \citep{Metcalf2014} using deflectors from the Flagship simulation \citep{EuclidSkyFlagship} and sources based on Hubble Ultra Deep Field galaxies. Together these span a wide range of Einstein radii and source brightness. The simulations were injected into the lens identification pipeline at the Space Warps stage, and so bypassed the ML selection. Recovery was measured cumulatively, with a simulated lens counting as recovered at a given stage only if it passed that stage's selection threshold and every preceding stage, with recovery at the expert stage requiring a final grade of A or B. Owing to inspection constraints, only a random subset of the simulations passing each stage was forwarded to the next.}

Figure~\ref{fig:selection-function} shows the resulting visual inspection selection function. We quantify the lensed-source brightness through its S/N, taken as the maximum ratio of the cumulative lensed-source flux to the cumulative root-mean-square background noise, and the recovery probability is smoothed with a Gaussian-process regressor.
S/N is the dominant driver of recovery: simulated lenses with S/N $\lesssim 10$ typically have a cumulative recovery probability below 20\%. At fixed S/N, systems with larger Einstein radii are more likely to be recovered, although a large Einstein radius does not compensate for low S/N. The selection is broadly similar at each individual stage, so the effect compounds through the pipeline. This motivated the direct-to-expert routes, in which systems with the very highest ML or SWC scores bypassed the subsequent stages, ensuring that the clearest lenses were not lost to the compounded selection.

The selection function also depends on parameters beyond Einstein radius and S/N, such as the colour contrast between deflector and source, and the recovery probability measured here is therefore an average over the distribution of these parameters in the simulations. Furthermore, it only applies directly to real lenses that resemble the simulated population. The simulations are not optimised for deflectors with complex morphologies, such as discs, mergers, or multi-deflector lenses, and since such systems are also less familiar to both the ML models and the human inspectors, the selection function measured here likely overestimates their recovery probability.

\begin{figure}
    \centering
    \includegraphics[width=1\linewidth]{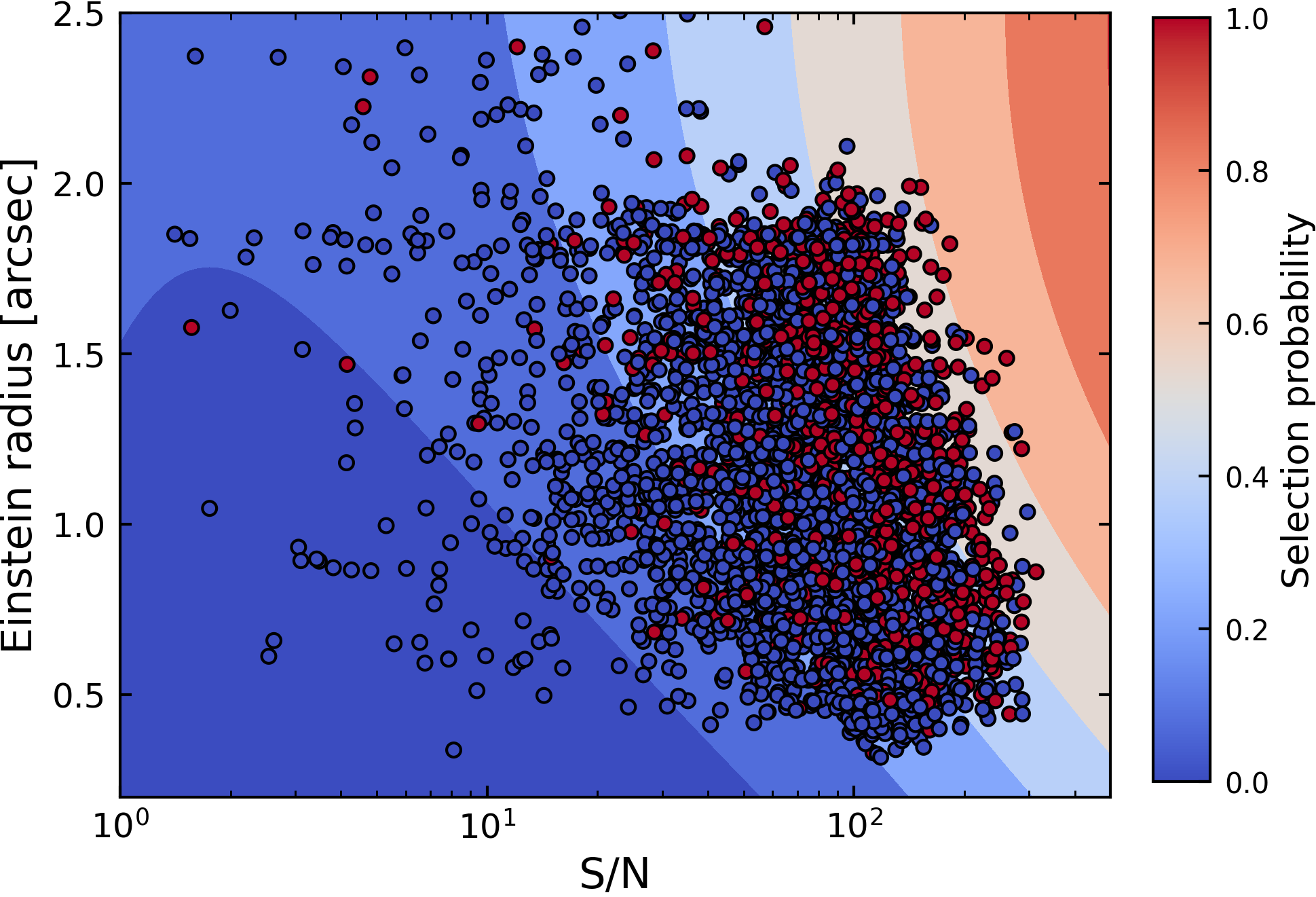}
    \caption{Selection probability for simulated strong lenses passing through the visual-inspection stages, as a function of the lensed-source S/N and the input Einstein radius. Points show individual simulations (red selected, blue not selected), and the background contours show the smoothed selection probability inferred from them. A simulation counts as selected if it passes both Space Warps stages and receives a final expert grade of A or B.}
    \label{fig:selection-function}
\end{figure}

\subsection{\comments{Missed known lenses}}\label{ssec:missed}

Since the random sample contains only a few tens of lenses, all drawn from objects passing our catalogue selection, and the simulations do not necessarily capture the full diversity of real systems, the several hundred previously reported lenses within the DR1 footprint provide a complementary test of completeness. However, some candidates identified as strong lenses in ground-based imaging are revealed in \Euclid data to be clear non-lenses, and we would therefore not expect to recover these. To address this, \black{Rojas et al. (in prep.)} re-inspect previously reported lens candidates that now have \Euclid imaging, reclassifying each as a `sure' or `possible' lens (the original publications are acknowledged in Appendix~\ref{app:previous}). This homogeneously graded sample allows us to test what our systematic search missed and at what stage lenses were lost.

\comments{Of the previously reported candidates that the re-inspection classifies as galaxy--galaxy lenses, 549 lie within our DR1 search footprint (excluding the Q1 area and deep fields, which were not searched), of which 428 (78\%) are recovered in our catalogue. The recovery rate depends strongly on the confidence of the re-inspection: 96\% (290/303) of the `sure' lenses are recovered, compared with 56\% (138/246) of the `possible' lenses, consistent with our search preferentially recovering the clearest systems.}
Out of the 13 `sure' lenses we missed, one never entered our search: an overlapping imaging artefact meant that MER did not detect it, so it was absent from the \Euclid catalogue rather than removed by our selection cuts. Four of the 13 lenses were rejected by the ML models, two of which had image-processing issues and two of which were not fully contained within the search cutout. The remaining eight passed the ML selection but were not passed on by the citizen scientists; half of these had atypically extended, diffuse arcs, while the others lacked obvious counter-images. Of the 108 `possible' lenses that we missed, only seven were lost at the catalogue level: five had no MER entry owing to nearby imaging artefacts, and two were too faint to pass our magnitude cuts (both originally identified in HST imaging). A further 40 were rejected by the ML models and 58 by the citizen scientists, while the remaining three were ambiguous systems with conflicting grades.

Overall, only eight of the 549 lenses (1.5\%) were lost at the catalogue level, and six of these were never detected by MER owing to imaging artefacts rather than being removed by our selection cuts. Our catalogue selection, which the random sample could not test, therefore removes very few genuine lenses. Most missed lenses were instead lost at the Space Warps (66) and ML (44) stages. Since previously reported lenses were mostly found in ground-based imaging, they are biased towards bright systems with large Einstein radii, so the 96\% recovery of `sure' lenses is best regarded as an upper limit on our completeness for the full \Euclid lens population.

\section{Discussion and future prospects}\label{sec:future}

A sample of \gradeAB grade A+B lenses allows strong lensing to be studied at the population level, not only system by system. However, population studies require a thorough understanding of the selection function to account for the incompleteness and biases of the sample.
In Sect.~\ref{ssec:visinsp} we showed that recovery depends strongly on S/N and Einstein radius, and these dependencies must be accounted for to avoid biasing any population-level science results.
That measurement covers only the visual inspection stages, since the simulated lenses were injected at the Space Warps stage and so skipped the ML selection.
Extending the same approach to the ML models is substantially more challenging, since, as shown in \citet{Q1-SP085}, they treat simulations fundamentally differently to real lenses in terms of predicted lens probability. 
Developing robust methods for characterising ML selection functions that take this into account will be the subject of future research. An alternative strategy is to limit the sample to a subset that can reasonably be assumed to be close to complete, such as by \comments{restricting} the sample to large Einstein radii with high S/N, where the lensed arcs subtend a large angle. For example, this approach was used by 
\citet{ale2026} on the Q1 lens sample to study the mass-to-light ratio of elliptical galaxies, taking into account selection effects.

A further complication, affecting both ML and visual inspection approaches, is the definition of ground truth itself. Throughout this work we have used expert visual inspection grades to provide our ground truth, 
but human inspectors are not infallible \citep{Rojas2023}.
These grades are inherently subjective and sensitive to substantial variability between inspectors, making them imperfect as an absolute standard. 
Spectroscopic follow-up provides the most robust means of validating lens candidates (Sect.~\ref{sec:catalogue}), but only a small fraction of this catalogue currently has the necessary spectra.
\comments{The largest planned programme is 4SLSLS, beginning \red{late 2026}, which will measure deflector and source redshift pairs for \num{10000} lenses and velocity dispersions for 5000 deflector galaxies. Beyond anchoring angular measurements to physical quantities and providing independent mass estimates, these data will help us calculate the true positive rate among our grade A, B, and C candidates and serve as a more robust ground truth for data-driven studies of lens populations.
DESI is a comparable wide-field fibre spectroscopic survey that is already ongoing, and although it does not systematically target lenses, it has observed around \fixed{\num{2400}} of our lens candidates, \fixed{950} of which are in DR2.
Once vetted, these are expected to provide deflector redshifts for many systems and a smaller subset of source redshifts (Storfer et al., in prep.).
Beyond these spectroscopic surveys, there is an ongoing effort to obtain deflector and source redshifts for of order 100 lenses, leveraging optical and NIR observations on 8--10\,m class telescopes.
}

Looking further ahead, one of the most exciting prospects is the synergy with \comments{the} Vera C. Rubin Observatory, which is expected to transform the discovery of lensed transients, and in particular lensed supernovae (SNe). The most reliable strategy for finding cosmologically useful lensed SNe is expected to be the systematic monitoring of known lens candidates for transient events \citep{ana2024}, and the catalogue presented here provides a natural watchlist of targets for this purpose.
Lensed SNe offer a promising avenue for time-delay cosmography. Like lensed quasars, lensed SNe allow time-delay measurements, but with several advantages for constraining $H_0$: their lightcurves are well understood, allowing time delays to be measured in weeks rather than years; they disappear, enabling us to resolve the lensed host for lens modelling; and the Type~Ia subset has standardisable luminosities that break modelling degeneracies \citep{Weisenbach2024}. Only a few galaxy-scale lensed SNe have been discovered to date, but by combining Rubin's monitoring with \Euclid's lens catalogues, this sample is set to grow substantially.
\comments{More broadly, \Euclid's large-scale discovery of strong lenses will be complemented by similarly large samples from upcoming surveys, including the Chinese Space Station Telescope \citep{csst_forecast} and the \textit{Nancy Grace Roman} Space Telescope \citep{roman_forecast}, providing an increasingly extensive population of strong lenses for follow-up and cosmological studies.}

\section{Conclusions} \label{sec:conclusions}
We \comments{have} present\comments{ed} the Euclid DR1 strong lens catalogue, the largest homogeneous catalogue of strong gravitational lenses to date, discovered across \fixed{\dronearea}\,deg$^2$ of space-based imaging. The catalogue was constructed using a bespoke discovery pipeline (described in ECLIPSE-B) combining \fixed{five} complementary ML models, \phildone{over $4.5$ million} classifications from \phildone{6975} citizen scientists via the Space Warps project on Zooniverse, and expert grading by \phildone{\experts} members of the Euclid Consortium strong lensing group. Our main conclusions are as follows.

\begin{itemize}
    \item[$\bullet$] Amongst an initial sample of 68 million objects, we discover\comments{ed} \fixed{\gradeA} grade A, \fixed{\gradeB} grade B, and \fixed{\gradeC} grade C strong lens candidates.

    \item[$\bullet$] \comments{Of these DR1 strong lens candidates, 93\% of grade A, 98\% of grade B, and 99\% of grade C systems are new discoveries not previously reported.}
    
    \item[$\bullet$] Automated lens models are available for the combined grade A+B systems (ECLIPSE-C), providing Einstein radii, deblended deflector and source photometry, and pixelised source reconstructions. The model outputs were vetted, with \red{67\%} of grade A and \red{43\%} of grade B systems having a lens model judged to be successful by 75\% of the vetters.
    
    \item[$\bullet$] Photometric redshifts and stellar population properties of the deflector galaxies are provided in ECLIPSE-D. \comments{The deflectors are predominantly massive quiescent galaxies, with \stellarmass, \ssfr, and $z_{\rm phot}=$\,\red{\photoz}.}

    \item[$\bullet$] Vetted DESI spectroscopic redshifts are available for \fixed{379} systems, of which \fixed{20} have both deflector and source redshifts. \comments{For the deflector and source galaxies, these span $0.05<z_{\rm lens}<1.1$ and $0.3<z_{\rm source}<2.7$, respectively.}
    
    \item[$\bullet$] Beyond classical Einstein rings, the catalogue contains rarer and more exotic systems. This includes \fixed{\dspl} double-source-plane lens candidates, over \red{\discs} candidate cases of lensing by disc galaxies (including a small number of face-on spirals), and several hundred lensed AGN and quasar candidates. Together, these constitute the largest sample to date for most of these classes.

    \item[$\bullet$] Using the intrinsic rate of lenses identified from a randomly selected subset of the parent sample, we estimate a completeness of \fixed{$81^{+9}_{-17}\%$} for the grade A lens \comments{candidates}, and \fixed{$43^{+7}_{-8}\%$} for the grade A+B lens \comments{candidates}, relative to what would have been discovered had we inspected everything passing the catalogue selection cut. \comments{The main limitation is the capacity for expert grading rather than the purity of the search stages.}
    
    \item[$\bullet$] \comments{We recover \fixed{96\%} of previously reported galaxy--galaxy lenses in our search area that are independently verified as `sure' lenses in \Euclid imaging.}

\end{itemize}
\comments{This catalogue enables a broad range of science cases, exploiting both the deflector and source populations at a scale not previously possible. These include: the internal structure of deflector galaxies, including dark matter content, density slope, and the stellar IMF; how these properties depend on galaxy morphology, comparing elliptical and disc deflectors; scaling relations between lens properties, stellar mass, and redshift, extending to a wider range of both than previous spectroscopically selected samples; resolved studies of the source population, from extended star-forming galaxies to compact and point-like sources such as quasars; and cosmological constraints from double-source-plane systems and, longer term, from lensed quasars and SNe via time-delay cosmography.}
As the \Euclid survey progresses through its data releases, the scale of the lens sample will grow substantially. Scaling up the number of grade A+B systems to Euclid DR2 (mid 2029) and DR3 (late 2031) by area, we anticipate 
the discovery of \fixed{\drtwo} and \fixed{\ews} cumulative \Euclid strong lens \comments{candidates}, respectively.
\comments{With similarly large samples expected from the \textit{Nancy Grace Roman} Space Telescope and the Chinese Space Station Telescope, Euclid DR1 marks the beginning of the statistical era of strong gravitational lensing. What was once a population of hundreds of individual systems is now becoming a sample of tens of thousands, opening up new possibilities for the systematic study and exploitation of strong lenses for astrophysics and cosmology.
}

\section{Data availability}
The lens catalogue, lens models, photometric redshifts, and stellar population properties are available on Zenodo at \url{https://doi.org/10.5281/zenodo.19654904}. The ECLIPSE catalogue can also be browsed interactively at \url{https://euclid-eclipse-lenses.org}.
The Euclid DR1 data are available at \url{https://www.cosmos.esa.int/web/euclid/data-release-dr1}.

\begin{acknowledgements}

% form + Marusa Bradac + Phil Marshall

NL is supported through a graduate studentship from the UKRI STFC and the University of Portsmouth, and acknowledges travel support provided by Euclid:UK through STFC grant APP91126 and by STFC for UK participation in the Rubin LSST through grant ST/X001334/1.

This work has received funding from the European Research Council (ERC) under the European Union's Horizon 2020 research and innovation program (LensEra: grant agreement No. 945536). TC is funded by the Royal Society through a University Research Fellowship.

PH is funded by the Royal Society (RF/ERE/210033).

JWN is supported by an STFC/UKRI Ernest Rutherford Fellowship, Project Reference: ST/X003086/1.

AN gratefully acknowledges the support of the Belgian Federal Science Policy Office (BELSPO) for the provision of financial support in the framework of the PRODEX Programme of the European Space Agency (ESA) under contract number 4000143347.

CJS acknowledges funding from the NASA Euclid General Investigator Program Grant No. 80NSSC25K7735.

WJRE acknowledges the support by the Italian Ministry of University and Research (grant FIS2023-01611, CUP C53C25000300001).

AS acknowledges support from the National Key R\&D Program of China (No. 2023YFA1607800, 2023YFA1607802), and from the Yangyang Development Fund.

AG acknowledges support by the SNSF through PostDoc Mobility Return grant P5R5-2\_235360.

MG acknowledges support from the Polish National Agency for Academic Exchange grant No. BPN/BEK/2024/1/00325/DEC/1.

RC acknowledges support from the French government under the France 2030 investment plan, as part of the Initiative d'Excellence d'Aix-Marseille Universit\'e -- A*MIDEX (AMX-23-CEI-088).

QH was supported by the Dutch Research Council (NWO) through the grant Open Competitie ENW M21-1.

DJ acknowledges support by the First Rand Foundation, South Africa, and the Centre National de la Recherche Scientifique of France.

MB acknowledges support from the ERC Grant FIRSTLIGHT \#101053208, Slovenian national research agency ARIS through grant P1-0188, and ESA PRODEX Experiment Arrangements No. 4000146646 and 4000149972. 

PB acknowledges financial support from the Italian Space Agency (ASI)
through contract ``Euclid - Phase E'', and INAF Grants ``The Big-Data
era of cluster lensing'' and ``Probing Dark Matter and Galaxy
Formation in Galaxy Clusters through Strong Gravitational Lensing''.

The work of PJM was supported by DOE Contract DE-AC02-76SF00515.

Support was provided by Schmidt Sciences, LLC, for J.H.-H.C.

JAAB's research was supported by an appointment to the NASA Postdoctoral Program at the NASA Jet Propulsion Laboratory administered by Oak Ridge Associated Universities under contract with NASA.

Part of this research was carried out at the Jet Propulsion Laboratory, California Institute of Technology, under a contract with the National Aeronautics and Space Administration (80NM0018D0004).

JP is supported by the ACME, ELSA, and OSCARS projects. ‘ACME: Astrophysics Centre for Multimessenger studies in Europe’, ‘ELSA: Euclid Legacy Science Advanced analysis tools’, and ‘OSCARS: Open Science Clusters’ Action for Research and Society’ are funded by the European Union under grant agreement nos 101131928, 101135203, and 101129751, respectively; ELSA is also funded by Innovate UK grant 10093177. Views and opinions expressed are however those of the authors only and do not necessarily reflect those of the European Union.

Numerical computations were carried out on the SCIAMA High Performance Compute (HPC) cluster, which is supported by the ICG and the University of Portsmouth.

This research makes (extensive) use of ESA Datalabs (datalabs.esa.int), an initiative by ESA’s Data Science and Archives Division in the Science and Operations Department, Directorate of Science.

% Based on observations obtained at the Hale Telescope, Palomar Observatory, as part of a collaborative agreement between the Caltech Optical Observatories and the Jet Propulsion Laboratory [operated by Caltech for NASA].

This research used data obtained with the Dark Energy Spectroscopic Instrument (DESI). DESI construction and operations is managed by the Lawrence Berkeley National Laboratory. This material is based upon work supported by the U.S. Department of Energy, Office of Science, Office of High-Energy Physics, under Contract No. DE–AC02–05CH11231, and by the National Energy Research Scientific Computing Center, a DOE Office of Science User Facility under the same contract. Additional support for DESI was provided by the U.S. National Science Foundation (NSF), Division of Astronomical Sciences under Contract No. AST-0950945 to the NSF's National Optical-Infrared Astronomy Research Laboratory; the Science and Technology Facilities Council of the United Kingdom; the Gordon and Betty Moore Foundation; the Heising-Simons Foundation; the French Alternative Energies and Atomic Energy Commission (CEA); the National Council of Humanities, Science and Technology of Mexico (CONAHCYT); the Ministry of Science and Innovation of Spain (MICINN), and by the DESI Member Institutions: www.desi.lbl.gov/collaborating-institutions. The DESI collaboration is honored to be permitted to conduct scientific research on I'oligam Du'ag (Kitt Peak), a mountain with particular significance to the Tohono O'odham Nation. Any opinions, findings, and conclusions or recommendations expressed in this material are those of the author(s) and do not necessarily reflect the views of the U.S. National Science Foundation, the U.S. Department of Energy, or any of the listed funding agencies.

We express our gratitude to the DESI collaboration for releasing the spectrum of Euclid DR1 J042911.429-204453.48.

\AckDRone
\AckEC

\end{acknowledgements}

\bibliography{Euclid, DR1, bib, extra_bib}

\appendix

\clearpage
\thispagestyle{empty}
\begin{figure*}[h]
    \centering
    \vspace{-1cm}
    \includegraphics[
        width=0.99\textwidth,
    ]{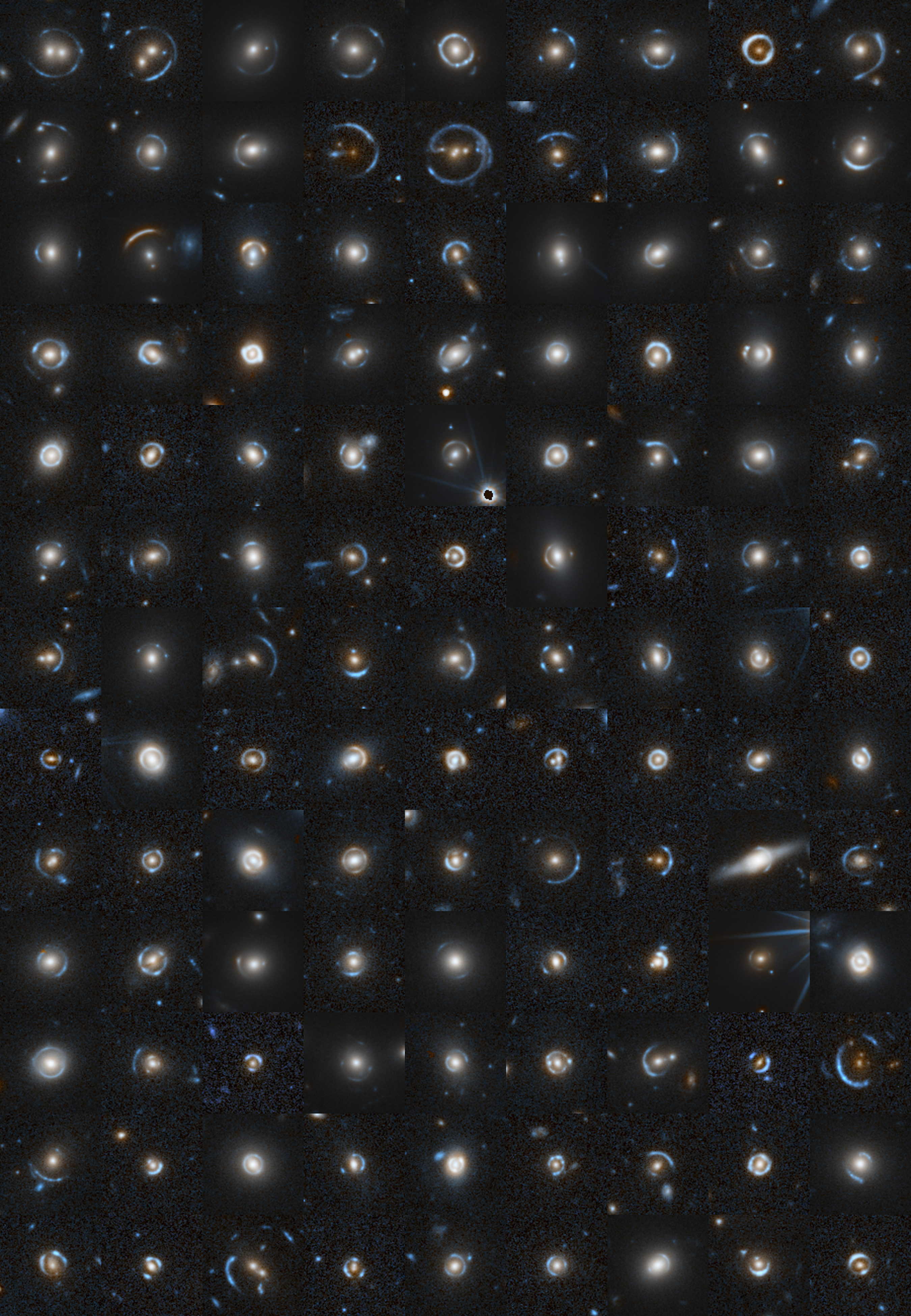}
    \setcounter{figure}{0}
    \renewcommand{\thefigure}{A.\arabic{figure}}
    \caption{Top 117 lens \comments{candidates} in DR1 according to ML scores. All of these were assigned grade A during the expert inspection stage.}
    \label{fig:collage}
\end{figure*}
\clearpage

\captionsetup[subfigure]{font=small}

\clearpage
\thispagestyle{empty}
\begin{figure*}[h]
    \centering
    \begin{subfigure}{\textwidth}
        \centering
        \includegraphics[width=\textwidth]{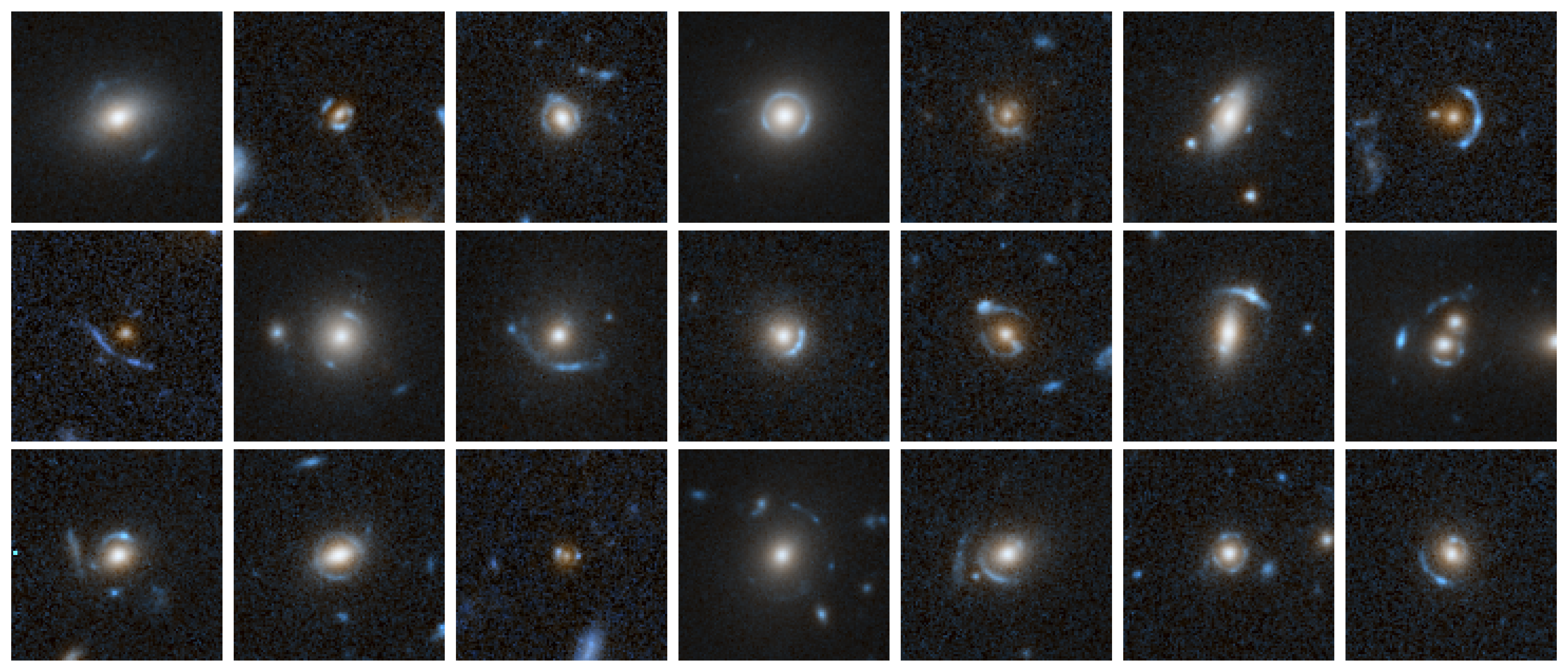}
                \caption{Randomly selected subset of the grade A systems.}

        \label{fig:random_A}
    \end{subfigure}

    \vspace{0cm}

    \begin{subfigure}{\textwidth}
        \centering
        \includegraphics[width=\textwidth]{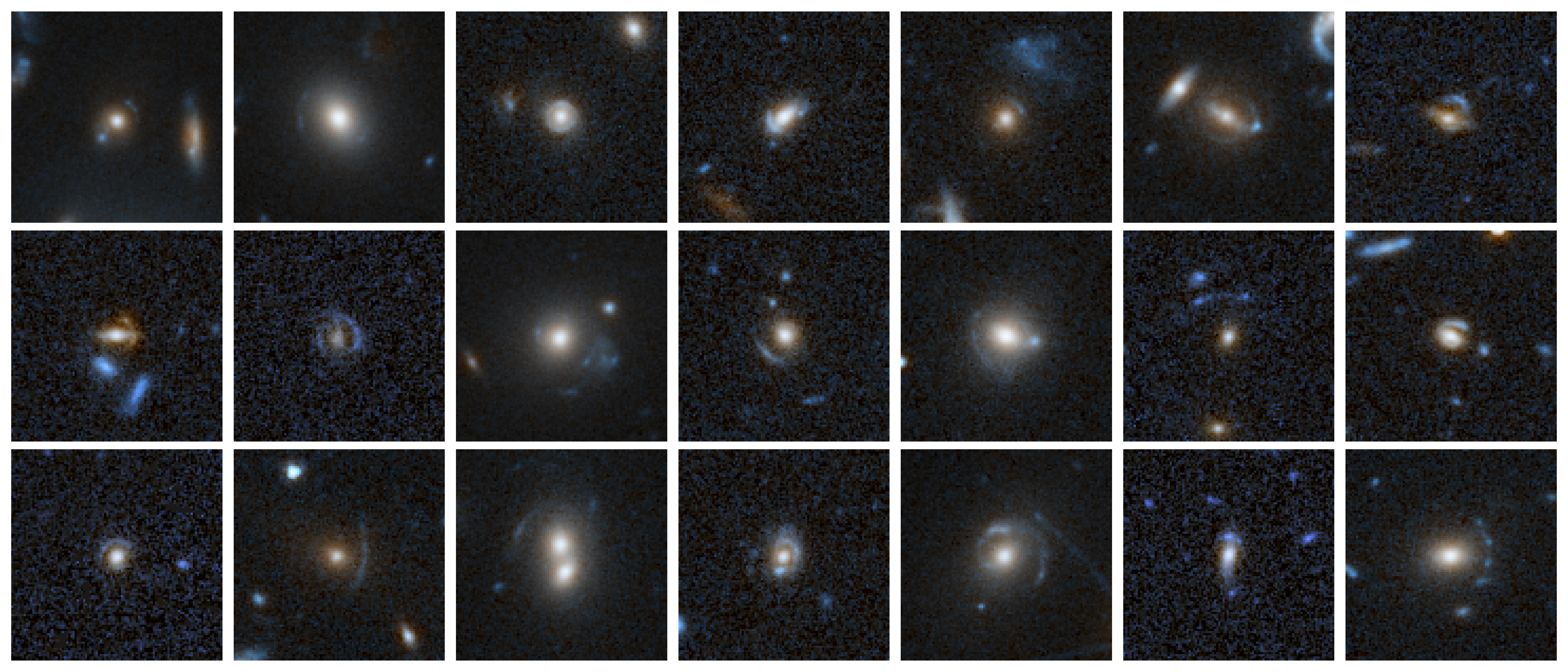}
                \caption{Randomly selected subset of the grade B systems.}

        \label{fig:random_B}
    \end{subfigure}

    \vspace{0cm}

    \begin{subfigure}{\textwidth}
        \centering
        \includegraphics[width=\textwidth]{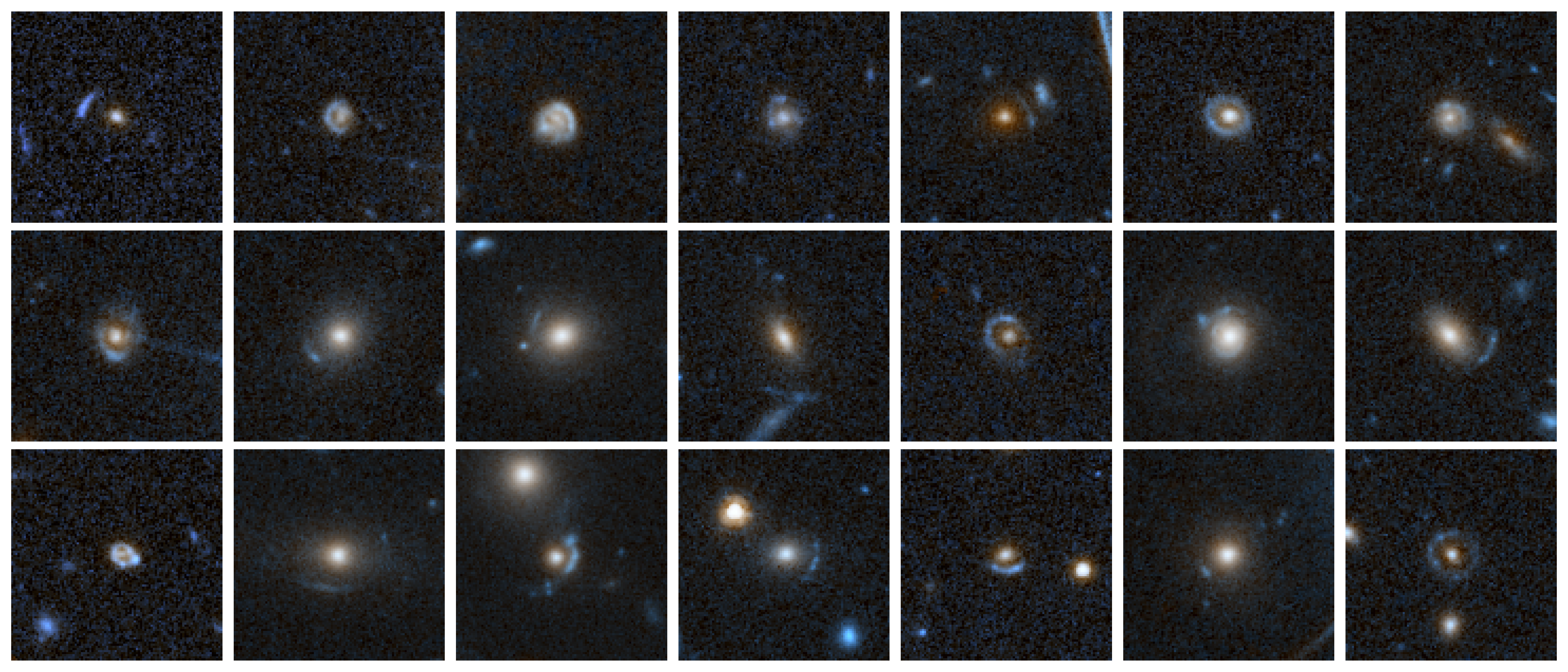}
                \caption{Randomly selected subset of the grade C systems.}

        \label{fig:random_C}
    \end{subfigure}

    \setcounter{figure}{2}
    \renewcommand{\thefigure}{A.\arabic{figure}}
    \caption{\comments{Examples of randomly selected systems assigned grades A, B, and C during the expert inspection stage.}}
    \label{fig:random_grades}
\end{figure*}
\clearpage

\begin{strip}
    \setcounter{figure}{0}
    \renewcommand{\thefigure}{B.\arabic{figure}}
    \centering
    \includegraphics[width=0.9\linewidth]{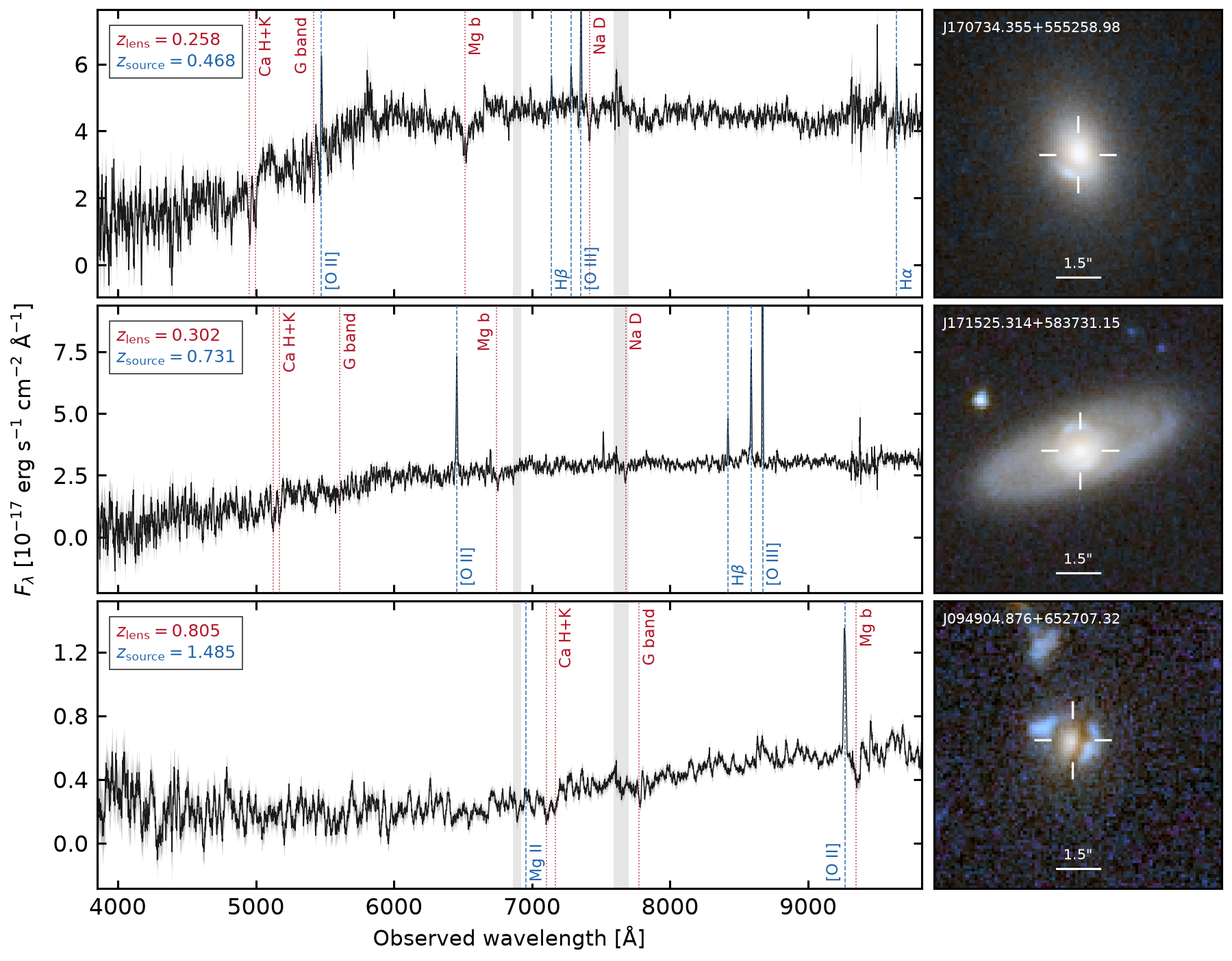}
    \captionof{figure}{\comments{Grade C candidates with both deflector and source redshifts from DESI DR1. The left panels show the smoothed DESI spectra with their $1\sigma$ uncertainties (shaded), with deflector absorption features (red, dotted) and source emission lines (blue, dashed) marked at the indicated redshifts. Vertical grey bands mark the telluric O$_2$ absorption bands. The right images show $\ang{;;10}\times\ang{;;10}$ \Euclid cutouts, with the white ticks marking the edge of the $\ang{;;1.5}$ DESI fibre.}}
    \label{fig:gradeCdesi}
\end{strip}

\section{Representative images}

\subsection{Top images identified by ML}

In Fig.~\ref{fig:collage} we show the top 117 (unique) lenses in DR1 according to the mean score of two ML classifiers (neural networks 4 and 5 of ECLIPSE-B). All of these lenses were classified as grade A, indicating a 100\% purity for these ML classifiers in the top ranks \comments{based on expert grades}.

\subsection{\comments{Random images of each grade class}}

\comments{In Fig.~\ref{fig:random_grades} we show a random selection of systems from each grade assigned during the expert classification. Grade A systems typically show clear arcs or multiple images, while grade B and C systems increasingly include fainter, smaller-separation, or less symmetric features, for which a lensing interpretation is plausible but not certain from the imaging alone. In particular, grade C systems mostly consist of either arcs with no obvious counter-image, or ring-like structures that are not clear Einstein rings.
The boundaries between grades are not sharp, and some systems shown here could reasonably have been assigned an adjacent grade. Readers who require a different trade-off between purity and completeness can instead use the continuous lens score provided in the catalogue.}

\section{\comments{DESI spectra of grade C candidates}}\label{app:spec}

\comments{In Fig.~\ref{fig:gradeCdesi} we show the three grade C candidates for which DESI DR1 spectra include both a deflector and a source redshift (Storfer et al., in prep.). In each case, a single fibre centred on the deflector captures both the deflector continuum and absorption features and the emission lines of a background source.}

\input{fig/empty-rings}

\comments{J170734.355+555258.98 (top) is a bright early-type galaxy at $z=0.258$ with a faint arc close to the deflector light. The spectrum shows [\ion{O}{ii}], \hb, [\ion{O}{iii}], and \ha emission at $z=0.468$, placing a star-forming galaxy behind the deflector and within the fibre. However, without an obvious counter-image, we cannot exclude that the background galaxy is only weakly lensed. This would likely be resolved through lens modelling.
J171525.314+583731.15 (middle) is an inclined disc galaxy at $z=0.302$ with a blue ring close to its centre, for which a spiral arm or star-forming ring could provide a plausible non-lensing explanation. However, strong [\ion{O}{ii}], \hb, and [\ion{O}{iii}] emission at $z=0.731$ shows that the emission does not originate in the disc itself.}
\comments{J094904.876+652707.32 (bottom) has a compact red deflector at $z=0.805$, identified from Ca H+K and G-band absorption, with blue features on either side. The spectrum shows strong [\ion{O}{ii}] emission at 9263\,\AA, placing a star-forming source at $z=1.485$ behind the deflector and within the fibre. Together with the configuration of the blue features, this makes the system a very likely strong lens. These examples show that the grade C sample, while of lower purity than grades A and B, still contains genuine strong lenses.}

\section{Unusual non-lenses discovered in the search}\label{app:non-lenses}

The ML algorithms trained to detect strong lenses \comments{have been shown to focus on arcs} \citep{Q1-Vincken}.
Therefore, we expect that unusual arc-like structures are likely to come up in our search. While the models are trained against more standard contaminants, typically spiral and ring galaxies, there \comments{exist} other atypical arc structures in the Universe.

During the Space Warps inspection phase of the project, \comments{approximately \black{20} objects} were flagged that exhibit ring-like morphologies with seemingly empty centres in all \Euclid \comments{bands} and in available ground-based data\comments{, some of which are displayed in Fig.~\ref{fig:empty-rings}}. Follow-up checks confirmed that the ring structures are not \Euclid artefacts; an example is shown in Fig.~\ref{fig:empty-ring-bands}. \comments{This distinguishes them from red deflectors such as that in Fig.~\ref{fig:misc2}\blue{m}, which are faint in VIS but clearly detected in the NIR.}
Due to their unusual morphologies, these systems were mostly graded as non-lenses.
If such objects were confirmed as strong-lensing systems, they would be of considerable interest.
Rare lensing configurations arising from unstable caustics can produce offset, four-image ring-like structures, known as hyperbolic umbilics \citep{Meena2020}; however, these configurations require highly complex mass distributions and are primarily expected in cluster-scale lenses (e.g., \citealt{Lagattuta2023}), which \comments{does not appear} to be the case for these systems.
More generally, more exotic lensing scenarios involving predominantly dark deflectors (e.g., dark matter haloes or non-luminous compact objects) are not excluded in principle, but there is currently no observational evidence for such configurations at the scales relevant here.

When considering the full set of examples collectively, the interpretation of these systems as potential gravitational lenses is less compelling. The rings are generally more elliptical than is typical for galaxy-scale lenses, and the features are often clumpy, diffuse, and asymmetric in a manner less characteristic of strongly lensed arcs.
We attempted to model one such system, shown in Fig.~\ref{fig:empty-ring-model}. The lens model used an elliptical power law plus shear to model the mass (with the neighbouring galaxy masked out), and attempted to reconstruct the source light on a pixelised grid, as in \citet{Li2026SalpeterNFWDSPL}. While the model was able to reproduce the broad ring-like morphology, it failed to recover the detailed structure of the data while focusing the source in the source plane. Although a more sophisticated lens model with additional degrees of freedom could potentially improve the fit, this exercise serves as a useful diagnostic test and indicates that the system is not straightforwardly explained as a strong lens.

Additionally, the fact that all cases have a nearby companion galaxy suggests that these objects are more likely to be collisional ring systems \citep{Appleton96} -- e.g., see Arp 147 -- than instances of strong gravitational lensing.
\comments{In collisional ring galaxies, a companion passes through the disc of a larger galaxy, driving an expanding density wave that pulls stars outwards and compresses gas into a star-forming ring. The absence of a central detection is not unusual for such systems: the stellar density interior to the wave can be left far below that of the ring, leaving any remaining central population too faint to detect, and an off-centre impact can displace the nucleus into the ring itself.}
\comments{\citet{madore2009} present a catalogue of over 100 collisional ring candidates, many of which resemble the systems highlighted here, and for all of which a plausible companion collider is identified.}

\begin{figure*}
\setcounter{figure}{1}
    \renewcommand{\thefigure}{C.\arabic{figure}}
    \centering
    \begin{subfigure}[c]{0.60\textwidth}
        \centering
        \includegraphics[width=\linewidth]{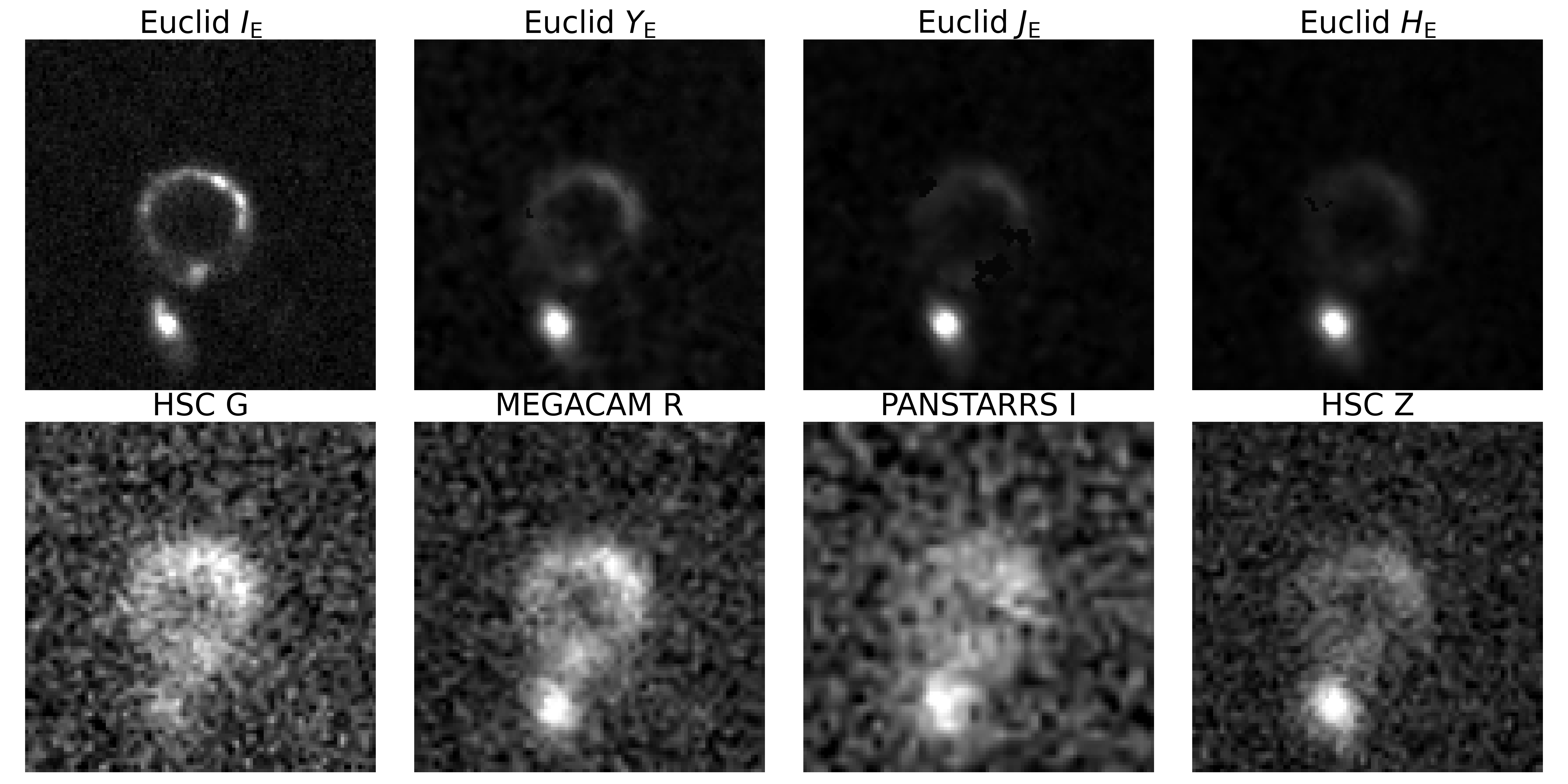}
        \caption{}
        \label{fig:empty-ring-bands}
    \end{subfigure}
    \hfill
    \begin{subfigure}[c]{0.35\textwidth}
        \centering
        \includegraphics[width=\linewidth]{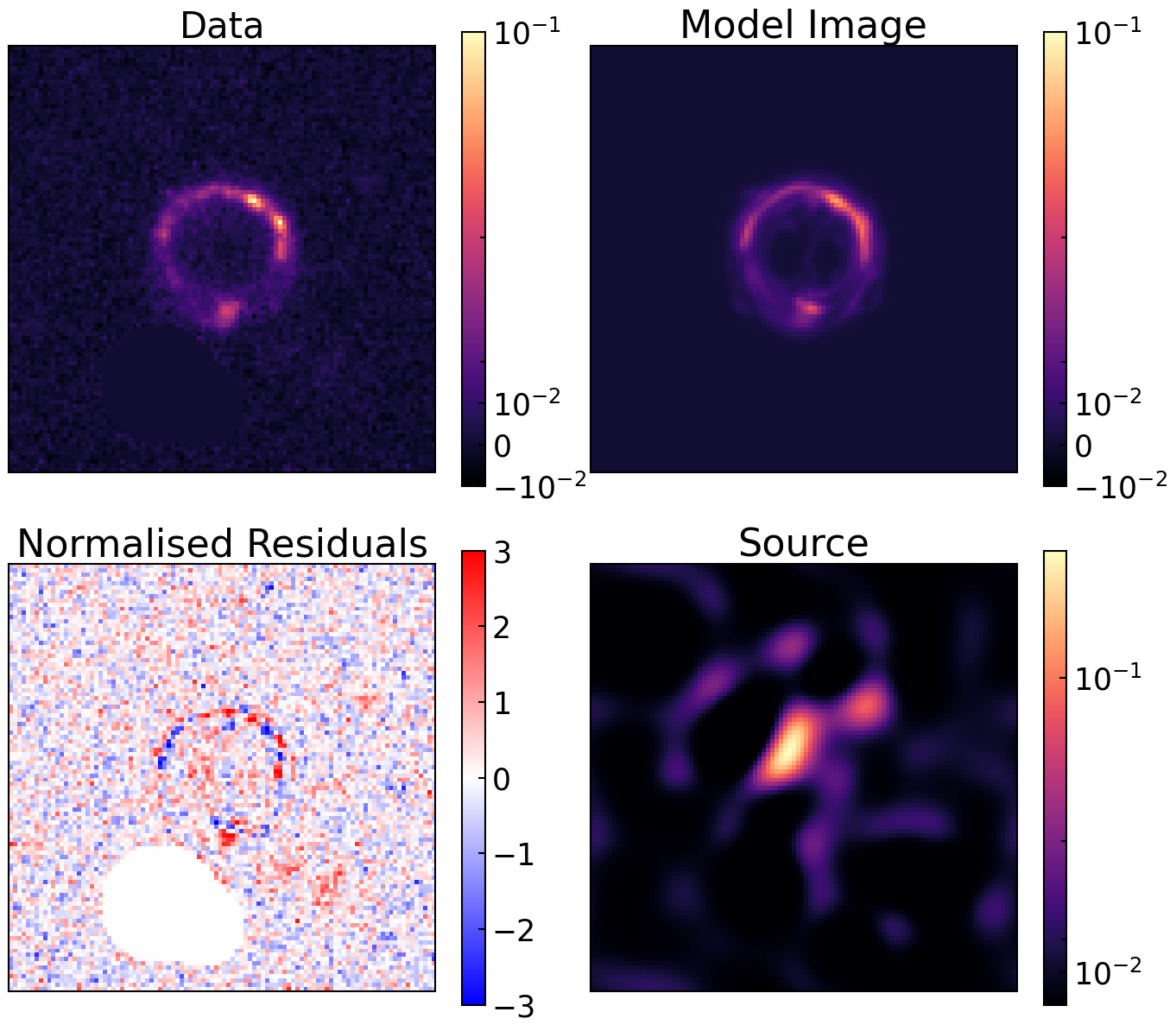}
        \caption{}
        \label{fig:empty-ring-model}
    \end{subfigure}
    \caption{\comments{Empty ring J100937.880+724953.06. (a) $\ang{;;10}\times\ang{;;10}$ cutouts in all eight available filters; although the ground-based resolution is limited, the structure appears similar in every filter, none containing significant central flux. (b) Attempted lens model, showing the data, model image, normalised residuals, and reconstructed source.}}
    \label{fig:empty-ring-case}
\end{figure*}

\comments{Spectroscopy supports this interpretation in the one case where it is available for both a ring and its companion. Two DESI DR1 fibres were placed on J175700.256+631344.17: one on the ring, and one on the compact red galaxy to its north-east.\footnote{\url{https://www.legacysurvey.org/viewer/desi-spectrum/dr1/targetid39633425258253254} (ring) and \url{https://www.legacysurvey.org/viewer/desi-spectrum/dr1/targetid39633425258253267} (companion).} The ring has a blue, star-forming spectrum with [\ion{O}{ii}], \hb, and [\ion{O}{iii}] emission at $z=0.7718$, while the companion has a red continuum with Ca H+K and G-band absorption at $z=0.7732$. The two are separated by only about 240\,km\,s$^{-1}$ in velocity and $\sim$21\,kpc in projection, and are therefore a physically associated pair, strongly favouring a collisional ring over a lensed background source. }

Nevertheless, collisional ring systems remain rare and astrophysically interesting in their own right. It is also encouraging that the ML models do not appear to be strongly biased towards recognising only systems resembling those in the training set, and that unusual arc-like morphologies are nonetheless recovered by the search pipeline.

\section{\comments{Previously reported lenses}}\label{app:previous}

\comments{Of the strong lens candidates presented in this catalogue, \knownA grade A, \knownB grade B, and \knownC grade C systems had been reported in prior publications. For each of these systems, our catalogue includes the earliest publication we could identify in which it was reported. These works, along with those reporting the lenses used to test our completeness in Sect.~\ref{ssec:missed}, span ground-based and space-based imaging as well as spectroscopic searches:
\citet{cabanac2005, Oguri2008, calanog2014, Pawase2014, agnello2015, Nord2016, Carrasco2017, Diehl2017, Schechter2017, Lin2017, desprez2018, repp2018, Wong2018, Jacobs2019a, Jacobs2019, Petrillo2019, Shajib2019, Canameras2020, cao2020, Jaelani2020SuGOHIV, lemon2020, Li2020, Teimoorinia2020, Canameras2021, Huang2021, Li2021, Talbot2021, Garvin2022, ODonnell2022, Rojas2022, Shu2022, Stein2022, Wong2022, Zhong2022, KeerthiVasan2023, lemon2023, Jaelani2024, Storfer2024, Liu2025, Nagam2025, oryan2025, holismokes16, Gonzalez2026, Inchausti2026, Karp2026, Sheu2026, Storfer2026}.}

\section{\comments{Space Warps volunteers}} \label{app:sw}
This project would not be possible without the immense contributions of citizen-scientists. Volunteers who classified more than 100 Euclid images on Space Warps whose names were available for publication are listed below. Zooniverse IDs for all those who classified more than one image are listed on \href{https://www.zooniverse.org/projects/aprajita/space-warps-esa-euclid/about/team}{`The Team'} webpage. We greatly appreciate their commitment, enthusiasm, and time.\\
\begin{spacing}{0.7}
\begin{CJK}{UTF8}{gbsn}
\tinier{\noindent A. Kirkpatrick, Aaira Sharjeel, Aaron Michael Cano, Abbey Jennings, Abby Clevenger, Abegail Chanyalew, Abiel Wong, Abigail Ginsberg, Abigail Zheng, Ada Rosenfield, Adam Setchfield, Adam Vitti, Addison Butz, Adeerah Nair, Adrian Astraeus Powers, Adrian Kai En Tee, Adriana Campos Dominik, Adrit Dongol, Agastya Srivastava, Agustina Villa Vidal, Aidan Conrad Gibbs, Aidan Duncan, Aiden Hummelgaard, Aidyn Cuenca, Akeela Valdes, Al Lamperti, Alaa Mahjoubi, Alan M. Gross, Alan Needham, Alan Pellé, Alan W Upchurch, Alberto Hernadez, Alejandro Morales, Aleks Stoyanov, Aleksandr Timofeev, Aleksandra Bijata, Aleksri Bykov, Alessandro Bejinaru, Alex Alves, Alex Anchain, Alex Arktos, Alex Ceballo, Alex Faber, Alex LaCoss, Alex Znyk, Alexander Benra, Alexander Folkers, Alexander Gagliano, Alexander Ivitsky, Alexander Ramer, Alexander Wehl Nielsen, Alexandra Attikova, Alexandre Hainaut, Alexandre Portefaix, Alexis Goupil, Alexis Louis C. Lebrun, Alexis Reed, Alexsey Gorbkov, Ali Alhawas, Ali Hasan Naser Hasan, Alicia Teo, Alicia Weissman, Alina Schulz, Alina Stefanescu, Alisa Karaneuskaya, Alison Watson, Alissa Strauf, Alkistis Kallinikou, Allie Bender, Allison Espinosa, Alma Quijano, Alondra Lara Cruz, Alyona Kobzeva, Alyssa Adam, Alyssa Ferrazzara, Amanda Demster, Amanda Peters, Amanda Peters, Amelia Booth, Amelia Evans, Amelina Anna, Amera Mostafa, Amit Kumar Tamrakar, Amoli Kakkar, Amy Nerkeliunas, Ana Fernandez Alvarez, Anaaya Chowdhary, Anakin Belarmino, Ananya Yogesh, Andelana O'Hara, Andrea Ceriana, Andrea Schimpke, Andreas Gruber, Andreas Lauth, Andreas Richnow, Andreas Thöny, Andres Eloy Martinez Rojas, Andrew Croager, Andrew Lackoff, Andrew M. Hodges, Andrew St.Clair, Andrew Stanley, Andrey CRB Gorokhov, András Kovács, André Holland, Andrés Abel Cosio Tello, Andy Middleton, Andy Spafford, Angela Jursevics, Angela Karunakaran, Angelin Vincent, Angelo Ricciardo, Angus Geidesz, Aniruddh Herle, Anja Verhöfen, Anmol Gandhi, Anna Fei, Anna Holub, Anna Ivanova, Anna Jungell, Anna Kozak, Anna Nikodym-Bilska, Anna Todorova, Annie Su, Anthony Charest, Anthony Michel, Anthony Yoder, Antonio David Garcia Gil, Antonio Pasqua, Antonín Unucka, Anupreeta More, Anvi Digra, Arabella Brosnan, Arief Al Hadid, Arkhipova Daria, Armando Campos, Armin L. Steyn, Arnaud Geollot, Arnaud Trybou, Arnav Singh, Arnav Srivastava, Arndt Stelter, Arne Peterson, Artem Barto, Artemiy Seleznev, Arthur Rodrigues Oliveira, Arto Inkala, Arto Koukkula, Arturo M. Gtz., Aruzhan Sungatkyzy, Aryan Kate, Aryan Pandey, Aryash Singh, Ashish Rai, Ashley McCann, Ashley Pullen, Ashton Taylor, Ashwini Axl Malayan, Asiah Ath-Thohiroh, Astro Yona VII, Athanasia Vlachou, Athene Bunch, Aubrey Tyson, Audrey Schollier, Augusto Gabanzo, Aurélien Genin, Autumn Faulk, Autumn Pearce, Ava Biasotti, Ava Johnston, Ava K Wozniak, Ava Vanhemert, Avanteesh Motheram, Avisha Gupta, Aviva Christie, Aya Mellaoui, Ayana Sinton, Ayesha Kashif Khan, Aymeric Galan, Baba Karthik Kalapatapu, Bailey Kinner, Balaji Prasath S, Barbara Pavlov, Barbara Téglás, Baris Mouland, Baron Michael Mcdonald, BeBe Wren Kiel, Belbahi Oualid, Ben Adamowski, Ben Apgar, Benjamin Cassan-de Gorostarzu, Benjamin Gill, Benjamin Graetz, Benjamin Keplinger, Benjamin Langley, Benjamin See, Bernd Nikolaus, Bill Richardson, Birte Rosiñol Prieto, Blaise Pyź, Bob Fletcher, Brad Brown, Brad M. Collins, Braden Sharette, Bradford Riley, Bradley Kavanagh, Brett Sorby, Brian Altmeyer, Brian Eckfeldt, Brian Myers, Brian Olivero, Brice Young, Brigitte Wollenweber-Ratzer, Brodie Dines, Brody Hoffman, Brody Ramirez, Bruce Binnie, Bruno Chiaranti, Bryan Ranharter, Bryan Shook, Bryan Tom, Bryce Greco, Byunghee Kim, C Unsworth, C. Graham, C. S. Tolliver, Caden Freimark, Cadence Edkins, Cadence Mestel, Caeley N Carmack, Callum Baird, Callum Blake, Camden Micci, Camelia Mallek, Canteloup Samy, Caoimhe Murphy, Carl Larsson, Carl Noack, Carla Brown, Carlos Gainzarain Armentia, Carlos Gonzalez, Carlos Romero, Carlos Santiago, Carlotta Mevert, Carly Gillespie, Carly Siedman, Carol Sabo, Casey Lambert, Casper Lehmann Aalund Vestergaard, Cassandra J Kurtz, Catherine Sui, Cathleen Robinson, Cathryn Collins, Chaithanya K Iruvaram, Charity Mystique Gregory, Charles Kime, Charlie Granger, Charlie Parks, Charlie Rayfield, Charlotte Smith, Charlotte Williams, Charlotte Williams, Cheng Yılın, Cherie Spielman, Cheyenne Best, Chhavy Loch, Chiara Cisternino, Chimalume Eze, Chioma Anyasodo, Chiriac Eduard Cristian, Chloe Hyndes, Chloe Reyes, Chonglin Ran, Chris Calleja, Chris Clarkson, Chris Ensign, Chris Harshaw, Chris McConnell, Chris Molloy, Chrissy Holwerda, Christian Malbroux, Christian Stork, Christine Rödel, Christine Steller, Christine V. Macmillan, Christophe Sims, Christopher B. Edwards, Christopher Doran, Christopher Pemberton, Christopher Tanner, Christopher Vittore, Ciro Sirio Perrella, Claire Bonnett, Claire Duguin, Claire Watts, Clara Stanton, Clare J Flood, Claude Cornen, Claude Revilliod, Claudio Pedrazzi, Clément Imbert--Youssef, Colby Jones, Colette Boever, Colin Bilsland, Colin Chandler, Collien Stone, Colter Kashian, Connor Bruce Campbell, Connor Choi, Connor J Gibson, Connor Wallace, Cooper Bell, Corina Trifu, Cory Nathan Fessler, Cory O'Brien, Coy Morris, Craig Janda, Crescenzo Tortora, Damien Cherouvrier, Dan Prins, Dani Muller, Daniel Cunningham, Daniel Gascón Jiménez, Daniel Hearnden, Daniel Lopes, Daniel Molina Fernandez, Daniel Motherway, Daniel Pantoja, Daniel Piaget, Daniel Pishyar, Daniel Robinson, Daniel Skarednov, Daniela Mantula, Danielle Hoss, DanielR. Gigot, Daniil Filatov, Danika Belzile, Danilo Blagojević, Danny Muskens, Danny Westphal, Daria Dawson, Darla Dangcil, Darrell Bettany, Dave Hardy, Dave Santos, David Aldred, David Alexander, David Browne, David Carozza, David Hockaday, David Klemmer, David Lafuente, David M Williams, David Meierhenry, David Mumma, David N. Saunders, David Osten, David Plesl, David Serrano, David Stegemann, David Svinning, David Truckenmiller, David W. Rodgers, David Warrilow, David “Tony” Fantone, Davin Brown, Dawn Sunrise, Dayne Svendsen, Declan Ora, Deepti Sharma, Denis Korostelevs, Denis Terskov, Deniz Ege Emre, Dennis Kuenzel, Dennis R. Patten, Sr., Dennis Stanescu, Dennis Ward, Dennis Werkes, Derek Blauth, Derek Jones, Derek Sperry, Dereon Robertson, Devanandana Mohan, Dhananjay Singla, Diana Osipov, Diego A. Díaz, Diego Ivan Reynolds Fernandez, Diego Lorenzo de los Ríos, Diego Pirrami, Dieter Haile, Dillan Hayes, Dina Guttmann, Dinesh Patel, Djoulane Bentarcha, Dmitriy Emelyanov, Dmitry Babin, Dmitry Ivanov, Doha Bedda, Domenico Avino, Dominic Savard, Dominikus Koelbl, Don J Vavasseur, Don T. Beakunt, Donald K. Lorenzo, Donna Charles, Donna Frostick, Dr Richard Nitzsche, Dr. Gundolf Lange, Duncan Jones, Durando Nathan, Dylan Jennings, E Chillington, E.V. Valentine, Edith Freundl, Eduardo da Costa Valadão, Edward Silvia, Edward Zhou, Eka Thea Sharma, Elad Ben-Ezra, Eleanor Aalders, Eleanor Cox, Eleni Hope, Elina Litvinova, Elisa Casale, Elisa Cipriani, Elisa Goni, Elisabeth Baeten, Elizabeth Beckwith, Ella Cowan, Elmar Delvos, Elmar Pohl, Elspeth Edwards, Elwood Downey, Emese Kocsis, Emily A Caffarel, Emily Bollich, Emily Jayne Bean, Emily Lutterloh, Emily Sharo, Emma Bazemore, Emma Cucca, Emma Eloise Thompson, Emma Gerntke, Emma Short, Emma Taylor, Eoin Neylon, Eric Balestra, Eric Cooper, Eric Fabrigat, Eric Henrique Marques da Silva, Eric Le, Eric Petersen, Erick F. Del Cid Borja, Erik Bowman, Erik Gasteiger, Erik Martinez, Erjon Buzo, Ernest Courtois, Ernest Jude P. Tiu, Esraa samy Saad, Essi Kultanen, Ethan Hernandez, Ethan Jeffris, Ethelinda McCall, Etienne Sevrin, Eva Gimenez Jelic, Evan Thompson, Eve England, Evelyn Kelley, Evelyne Stanton, Eyser Choquet, Fabio Rollo, Fabrice Bongiardini, Fabrizio Trara Genoino, Faiz Qamar, Fan Baoyi, Farha Ahmed Morsy, Federico Desii, Felicia Vinciguerra, Felix S. Firmbach, Fermor Baird, Fernanda Freitas, Fernando A. Alfaro, Filip Vdović, Fintan Sargent, Fiona Barbolak, Floortje van Santen, Florent Perrin, Florian Arina, Florian Busch, Florian Mansion, Flávia Nathalia da Silva Santos, Fox James, Frances and Stuart Holloway, Francesca Cottone, Francesco Simeoni, Franck Pochet, Frank Falor, Franziska Bauer, Fred Linley, Freddie Jr. O. Cabrejas, Frederick Wilner, Freya Isaksson, Fruzsina Jelen, G. Greene, Gabriel Taylor, Gabriela Madera Pena, Gabriele Pavan, Gael Lubin, Gail Gibson-Smith, Gaivoronskiy Stanislav Vadimovich, Galini Michaela Chronaki, Gary Buss, Gary Rowland, Gaurav Kumar, Gavin Rider, Gayathri Manoj, Geoff Kirby, George Deng, Gerald W. Nash Jr., Gerhard Petz, Gina Grogg, Giro Neil C. Generao, Giulia Despali, Gordon Eldon, Grace Scura, Grady Fitzpatrick, Graham Darlington, Graham Dungworth, Greg Balnis, Greg Jenkins, Greg Roth, Gregory Avramenko, Gregory Drzymkowski, Gregory Kaptcov, Griffin Cook, Guido Bucchi, Guido Lauer, Guilherme Pereira Brandão, Guillaume Penaud, Guillaume Quenouille, Gunnar Russell, Guoyou Sun（孙国佑）, Gyongyi Plank, Génesis Pimentel, Gérault Nikita, Günter Steinbach, H. A. Güler, H. W. Riley, Haley Willits, Hams Gadalla, Hanna Lebedzevich, Hannah Kempken, Hannes Lundvall, Haotian Yang, Haoxuan Zhang, Harold Price, Harry Moreland, Harshitha Vijayakumar, Haruka Yanagisawa, Haseeba Zia, Hassan Namwar, Hau Rava, Hayoon Kang, Heather Bassett, Heiko Gericke, Helen Wetherbee, Helena Yeung, Helga Haase, Hendrik Teixeira, Henry Clark, Henry Williams, Henryk Krawczyk, Heraldo Henrique Felix de Moraes, Hiago Jorge Silva, Hibah Arshad, Hiroki Fukuyama, Hiya Singh, Hrvoje Krizic, Huang Rui Guo, Hudson Fu, Hudson Varkony Cook, Hugo Mazer, Hugo Mendes, Hugo Podsieldik, Huub Mooren, Ian Carley, Ian Russo, Iaroslav Mokroguz, Icaro Melo Ferreira, Igor Alvise Justo, Ilhan Abdul Rahim, Ilja Dotšar, Ina Thamke, Ingo Holetzek, Ingrid Nurminen, Inika Sridhar, Ioana-Alexandra Nedela, Irina Aerith, Irina Stocker, Iris Ham, Isabella Ana Cuan, Isabella Eva Bigatti, Isabella Roberts, Isabella Skaggs, Isabella Wang, Isabelle Brander, Isabelle Graves, Isaiah Thomas Varghese, Ishana Parmar, Ishita Mukherjea, Isla Gray, Isobel Frew, Ivan A. Terentev, Ivana Varsányiová, İrem İlhan, J. Rocco, J.J. Carrol, J.T. Magee, Jaana Kemppainen, Jace Spence, Jackie Branc, Jackson Daniel Griffith, Jackson Keener, Jackson VanHall, Jacob Amato, Jacob Andrusyszyn, Jacob Brown, Jacob Fielding, Jacob Reed, Jacqueline Elliston, Jacqueline Hirschfeld, Jacquelyn Higinbotham, Jafar Hasanjee, Jaiden Lopez, Jakub Jarý, Jakub Michalak, James Auld, James Betts, James C Mullikin, James Edmunds Mace, James Lambourne, James Pearson, James Perdue, Jan Sjoerd de Vries, Janae Anderson, Jane Sloan, Janya Singh, Jared Thompson, Jari Taavitsainen, Jarne Loos, Jason Linke, Jason MacDonald, Jason Spencer, Jasper Buan, Jaxon Sowder, Jay Alexander, Jay Maleev, Jay Rexroad, Jay Valdes, Jean Nimes, Jean Rheault, Jean-Francois Peyron, Jean-Noel Terry, Jeff Taggart, Jeffrey Sgrignoli, Jehu Campos, Jelly Penalosa, Jen Cotrone, Jennifer D Mesa, Jennifer Rosebrock, Jennifer Saar, Jens Mario Schmidt, Jeremiah Obinna Odaba, Jeremias Peltola, Jeremie Muller, Jeremy Maniago, Jerried Ramon, Jerry Nguyen, Jesse Prenatt, Jessica Hallberg, Jessica Hevia, Jessica Keane, Jessica L Esteves, Jessie Huang, Jessie R. Hayes, Jesús Pasten Arvizu, Jetro Ketola, Jett Amsden, Jhoanna Jimena Chanta Flores, Jiashuo Zhang, Jieun Bok, Jim Locke, Jimmy Simpson, Jo Hofmann, Joanna Maria Dulowska, Joanne Villani, Joanne-Denise Nickel, Joceline Castrejon, Jocelyn Huang, Joe Mertes, Joe Reape, Joe Schmidt, Joey Serber, Johan Bertrand, Johan Le Menn, John Barcelona, John Bragg, John Davis, John Fellows, John Kotas, John Michael, John Parks, John R DuPont III, John V Morris, John Webb, Johnny B. Cowart Jr, Johnny Reid, Jon-Paul  Merry, Jonah Turner, Jonathan Grossauer, Jonathan Tielve, Jonathan Verrier, Jordan A. Simpson, Jordi Solaz, Jordyn Hutchins, Jorge Sobradelo Silva, Joris Beelen, Jose Diego, Jose Luis Navarro, Jose Palafox, Jose Pinto, Joseph Downs, Joseph Henderson, Josh Gebelein, Josh Hill, Josh Wilde, Joshua Fitzpatrick, Joshua McArthur, Joshua Ridgill, José Manuel Leal Quesada, José Victor de Matos Lima, Juan Alberto Cano Díez, Jude Garbe, Jude Wechsler, Judelin Joasilus, Judith Waschelitz, Judy Garstone, Juli Fowler, Julia Brazolim dos Santos, Julia Cavanaugh, Julia de Montfort, Julia H. Stickney, Julia Katona, Julian Asialo Kioa, Julian Collis, Juliana Newman, Julie Paynter, Julie Potasiewicz, Julien Blary, Julien Frutozo, Julien Lühning, Julien Schaer, Jur Vogelzang, Justin Klingele, Justine Lefebvre, Justus Maximilian Gebauer, Jérémie Cuny, Jérôme F. Delaunay, Jérôme Rafani, K Siva Srinath Reddy, Kai (Connor) Young, Kai Kaufmann, Kai Schindler, Kai Thomas, Kaitlyn Randall, Kaja Sinur, Kal Marshall, Kalvin Dewayne Rush Sr, Kamil Dębowski, Kamila Vesela, Kareem Hany, Karen L Jones, Karolína Minaříková, Karthika Kannan, Karthikeyan Duraisamy, Katharina Stowe Martin, Katherine Collins, Katherine Erin Calton, Kathleen Del Prince, Kathrin Loch, Katie Stigter, Katrina Wong, Katya Shlyakhova, Kaustubhi Joshi, Kaylie Scorza, Kazi Rejwan Islam, Kelsey Ryan, Kelson Sisk, Kelvin Salter, Kemo Sidibeh, Ken Davis, Ken Lyman, Ken Nakada, Ken O'Neill, Kendall Hall, Kevin J. Daly B.E., Ph.D., Kevin Moreau, Kevin Peter, Kevin Quesada, Khushi Lalit, Kieran Smith, Kieron Nate S. Presbitero, Kiersten Hoffman, Kievon Ferguson, Kiran Kumar, Kiran Nandkumar Suryarao, Kirill Sakharov, KK Maclean, Klajdi Qinami, Klaudia Solska, Klaus Rohrich, Kohsheen Kachru Naik, Kommal Tariq, Korbin Spencer, Kostula Papandrianu, Kristen Ess, Kristie Yorkston, Kristiina Rimpeläinen, Ksenia Nikonova, Ksenia Smirnova, Kumar Utukuri, Kyara Haffane, Kyle Hartman, Kyle Nelson, L. Van Warren, La Wun Eain, Laetitia Martel, Laila Gregory, Lamberto Veraldi, Lara Ertaul, Lara Maia, Lara Nelson, Laura Dionysius, Laura J. Martin, Laura Phillips, Laura Turiano, Lauren Li, Lauren Murphy, Laurie B. Gaines, Laëtitia Renard, Le Garrec Jean-François, Leah Sun, Leah Yang, Leen Rawad Hilal, Leighton Yanity, Leila Billingham, Leishangthem Debajit Singh, Lena Blomeyer, Leon Roman Ecker, Leona Kleinemeier, Leonardo Rodrigues, Leonie van Vliet, Leslie Ruiz, Leslie Tolentino, Lev Gribakin, Leyna Bajaj, Li Gu, Liam Buki, Liana Greyber, Liana Pavlovica, Lianne Louise Manalili, Liew En Quan, Lily Mastin, Lincoln Hamilton, Linda Saffer, Linh Nguyen, Lino Thomas, Lioba Elisabeth Happel, Lisa Daxer, Logann Ibarra-Ortiz, Lohit Basavaraju, loic Divay, Lois Holmes, Lorenzo Garza, Lorin Monck, Lottiee Haslam-Simmons, Louis Howarth, Louisa Kupke, Loïc Baert, Luan Bermudez, Luca Andreula, Lucas Babuk, Lucas Eilola, Lucas Lévêque, Lucia Hecht, Lucia Maria Ferreira Lima, Luciana Navarrete Gualteros, Lucie Golding, Ludovic Toucheau, Luigi Guzzo, Luka Layton, Luke Carreras, Luke Hanson, Luke Scanlan, Luminita Georgescu, Lynnette Kiluwe, Lyra Glass, Lysandros Wasilopoulos, M. Anderson, M. Zizlsperger, M.S. Batukov, Maciej Woloszka, Madeleine Tuck, Madeline Oxley, Maeve O'Brien, Magdalena Maszewska, Maggie Jardine, Mahmoud Elmagrby, Mai Tanaka, Maia G. Mohsen, Maisie Pattison, Maja Bergmann, Maksym Savenko, Malek Zeidieh, Malia Bouvard Jourde, Mallory Thomsen, Manahil, Mannat Lamba, Manuel Leister, Mara Ioana Șneapotă, Marc Hanset, Marc Raff, Marcela Sepulveda, Marcelino Alvarez, Marcelo Gomes Viana Lopes, Marco Bacchi, Marco Callegari, Marcus Schulz, Margaret Weeks, Maria Eduarda Alves de Araújo, Maria Fragoso, Maria Isabella Mesina Mendoza, Maria Martin, Maria Smirnova, Maria Tubol, Maria Wicher, Maria-Olivia Torcea, Marialena Pouliou, Mariasole Melara, Mariella Gozzi, Marika Jandrey, Marikaitlynn VanZandt, Mario Henrique, Mario Keller, Marion Moffatt, Marius Agafitei, Marius Theissen, Mark A. Fortuna, Mark S Fletcher, Mark Stypczynski, Marlene Schulz, Marsha Speevak, Martha Murray Zinn, Martin Belles, Martin Bergmann, Martin Hall, Martin Miller, Martin Soto, Martin Ungless, Martina Cucchi, Martín Orlando Gil Cardona, Martín Zamora Rodríguez, Mary Vance, Maryceline Espanol, Masi Steele, Mason Gay, Mason Shawe, Massimo Tonioli, Matayah G, Mathias Bergmann, Mathis Noizet, Matilda Wagner, Matt Romriell, Matteo Cianfaglione, Matteo Pedani, Matthew Alexander White, Matthew Allis, Matthew Given, Matthew Helms, Matthew Loomis, Matthew Moon, Matthew Quiroz, Matthew Rhys George, Matthew Seth Marcus, Matthew Thomas Bates, Matthias Kohl-Himmelseher, Matti Sironen, Mauri Simon, Maurizio Mandarino, Maxence Chaminade, Maxence Delmotte, Maxime Fouris, Maxime Guarino, Maximus LaBrunda, Maxine Lamont, Maya Sandberg-Smith, McKenzie Hall, Meera Desai, Meera Zaki, Megan Cooper, Mehmed Emin, Melanie A. Downes, Melanie Whitehouse, Melisa Díaz Rivero, Melissa Arsaut, Melissa Reed, Melissa Wyrostek, Meredith Navin, Meriel McManemy, Merle Mia Drews, Meyra Sert, Mia Hanert, Mia Holcombe, Michael A. Capraro, Michael B Williams, Michael Beckham, Michael J. Brodnicki, Michael Krol, Michael Laughton, Michael Mulhern, Michael Redfield, Michael Tepen, Michal Sivak, Michaël Crevoisier, Michał Biesiada, Michał Smółka, Michel Cavayé, Michela Pallaver, Michele T. Mazzucato, Michelle Zolina, Migachev Dmitry, Miguel Buzzega Mira, Mikaela Birdsong, Mike Barrett, Mike Beer, Milhann Chodorowski, Milot Sélène, Mina Harker, Mina Nedeva, Minal Bhosale, Mirza Adnan Mohtashim, Mitchell Rogers, Mittapelly Krishna Reddy, Mohamed Ahmed, Mohamed Altawel, Molly Scott, Molly Zhang, Monica Backes, Monika Seniut, Moran Thebault, Morijah Stutz, Mostafa Arafat, Mrunalini Prasad, Muditmayank Mishra, Muhammad Hadi Jawad, Muhammad Muzammil Anwar, Muhammad Ridha Akhsanu Sujiono, Munira Azzout, Murat Calikoglu, Murray Leslie, Mustafa Fayyaz, Myat Thura, Myra Gbesso, Myranda Keightley, Nabhya Gandhi, Nadia Buza, Nadir Chettah, Nadya Neklioudova, Nainpreet Uppal, Nash Gilbert, Nasser Lavaud, Natalie Freund, Natalie Lines, Natalie Savelyeva, Natalie Sebastian, Nathan Corey Bell, Nathan Dales, Nathaniel Hammon, Nathália Pichinin de Oliveira Martins, Nazar Silko, Neeti N Patil, Neil Winterbauer, Nerea Ruiz Martín, Nevena Niagolova, Nicholas Taylor, Nicola Denti, Nicolas Belhoste, Nicolas Denand, Nicolas Prevot, Nicolas Royer, Nicole Humphries, Nicolò Spigariol, Nik Cain, Nikita Kuzminykh, Nikolas Kontaxakis, Nikolaus Hermanspahn, Niladri S Kar, Nimal Kumar, Nina Grkovic, Nina Trubanova, Ninad Muley, Nino Otto, Nithyasri Senthilkumar, Nitika Verma, Noah Hansler, Noah R, Nolan Reket, Nolann Barbeau, Norah Simdon, Notarpietro Sabrina, Océane Rege Nero, Oleksandr Fil, Oliver Rawding, Olivia Lindahl, Olivia Park, Olivia Steger, Olivier Gaytán, Olivier Lecrenier, Olivier Marty, Oussama Guarraz, Ovalle Pérez Paloma, Owen Butler, Oxana Schmidt, P A Gwatkin, P Gopalan, P Tilbury, P. Seal, P.A. Labud, Paige Charman, Pamela Allen, Paola Bungaro, Paola Coves Puelma, Paramita Mohanty, Parizel--Montardi Mickaël, Parker Stasak, Pascal Dicop, Patrick Dorp, Patrick Mahaffey, Patrick Ngo, Patrick Veyssière, Paul Drye, Paul Henry, Paul Niermaréchal, Paula Pascua García, Pavel Atanasov, Pavel Fedyaev, Pedro Augusto Schimidt Leao, Pedro Bergevoet, Peggy Lehmann, Peter Ansorge, Peter Carr, Peter Csitkovics, Peter Dzwig, Peter Everett, Peter Haveland, Peter O'Toole, Phil Andriole, Philip Downey, Philip Holloway, Philippe Laferriere, Pierre Astor, Pietro Guillén Ferreira, Pietro Maiorana, Pilar Fife, Piotr Konecki, Piper Stafford, Prasad Vengurlekar, Preetha Pushkarini B N, Pritam Raikar, Przemysław Kowalski, Pásztor Benedek, Qi Liu, Quentin Guelle, Rachel Alice Ford, Rachel Binnie, Rachel Matt, Raeanne Nichol, Ragushin Aleksandr, Raimo Michaelis, Rainer Gorissen, Rainer Töpler, Rajshekhar Gupta, Rajwa Khansa Dzakirah, Ralf Lengers, Ralf Stephan, Ralitza Radoukanova, Ramita Kemavit, Random Howard, Rapaka Raja Ratnam, Raphael Eckersley, Raphaël Jimenez, Raul Torres, Ravshan Asatov, Raymond Gardocki, Raymond Wilkinson, Razvan Tanase, Rebecca Brown, Rebecca Denae, Rebecca Eid, Rebecca S. Warren, Rebekah F, Renat Saifullin, Renata Pataki, Renato Raduka, Rene-Michel Paquet, Renos Anastasi, René Eichelbaum, Reta Nissan, Retaj Atef Elsobky, Rhianna Cohen, Rhiannon Ford, Rhys Morris, Richard Allen, Richard Goodwin, Richard Huziak, Richard J. Acton, Richard Knijnenburg, Richard Renner, Richard S Blasi, Richard Scott, Richard Trubshaw, Rick Barnwell, Rida Islam, Riley Keppel, Rishaan Akula, Rishi Tangirala, Rob Garrison, Robert Desmond Maher, Robert E. Wheat, Robert J. Adamson, Robert J Heyden, Robert Neil Haeuser, Robert Schwartz, Roberto Aretxaga-Burgos, Roberto Bonamico, Roberto Gorelli, Roberto Haver, Robyn Keats, Robyne Major, Rodolphe DI MAIO, Roger Bowden, Roger Long, Roger Marrotte, Rohan Krishna, Roland Heringhaus, Roman Rakityanskiy, Ron Usry, Ronan Badel, Ronan Hughes, Rosa Bermejo, Rosemary Billington, Roshini Gade, Rowan Tekampe, Ruby Cline, Rudra Protap Nandi, Ruslan Migranov, Russell Parman, Russell Reed, Ruth Kevess-Cohen, Ruth Tucker, Ryan Chen, Ryan Craft, Ryan Fick, Ryan Li, Ryan Mavis, Ryan Ross, Rylee Funseth, S THILAK, S. Ominami, Saadet Polat, Saber Williamson, Sadewo Abdi Panuntun, Sai Likhith Poreddy, Sajeev T Prabhakaran, Sakari Vedenpää, Sallyann Chesson, Sam A. Fazelinia, Sam Deen, Sam Dulys, Sam Lee, Samanta Holdaway, Samantha Henry, Samantha Schenk, Samuele Meschini, Sana Farhat Abdul Qayyum, Sandra Lee Harris, Sandro José de Souza, Sandro Montagner, Sandupa Sanudaya, Sanjit Kolluru, Santi Ranjan Atta, Santiago Clements-Rodriguez, Santino Giacometti, Sara Capacchione, Sara I. Estrada, Sara Kurdy, Sara Vardanyan, Sarah Grissett, Sarah Hünermund, Sarah Musumeci, Sarah Sayer, Sarah Szita, Sarah White, Sarp Akar, Satanay Makhmud, Sauro Gaudenzi, Scott Grant, Scott Joseph Sokalski, Scott Richards, Sean Miller, Sebastian Lefebvre, Sebastian Oppermann, Sehr Almas Noordin, Serge Robert, Sergio Borrás Pardo, Seth Smithey, Shannen Chovanec, Sharlotte Phifer, Sharon Dinh, Shaunie Beswick, Shay Shalit, Sherie Kuek, Shirley Xie, Shivani Hancock, Shneor Gubits, Shogo Noma, Shrestha Sarkar, Shrihaan Sindhushri, Shrujal H. Nandurkar, Shuming Wang, Shuning Bian, Siddharth Mehrotra, Silke Jonda, Silke Wachholz, Simon Basler, Simon Nicholson, Simon Tarin, Simon Vincent Götz, Simone Glau, Siqi Yao, Siya Singh, Siyona Das, Skye Grimm, Sofia Yaroshenko, Sofiia Gudkova, Sophia Berinski, Sophia Ottner, Sophie Markman, Sovan Acharya, Soveros Arctos, Spriha Pudasaini, Sreeja Pal, Srujanika Rudra, Stacey Cornell, Stacey Cornell, Stacey E. Farr, Stacey Lewis, Stan Smallman, Stefanie Seidel, Stefanos Drakoulis, Stephana Bregovich, Stephanie Ferguson, Stephanie Garcia, Stephanie Potter, Stephen Halston, Stephen Lowcock, Stephen Morris, Steve Mealy, Steve Saunders, Steven Tan, Sudhakar Subramaniam, Sue Sivonen, Sung Kei Li, Susan Feder, Susan Magyarody, Susannah Stubbs, Sushant Pandit, Svetoslav Alexandrov, Swapnasish Ghorui, Sydney Saha, Sydney Saltzman, Syeda Sukaina, Sylvain Martineau, Sylvain Paszkowski, Sylwia Plenzner, Tada Wang, Takshil Doshi, Tammy Checketts, Tamás Gyurka, Tanja Kuusela, Tanusha Subramaniam, Taos Medjoub, Taryn Allie, Tashfin Shifan, Taylor Thornton, Teemu Oinasvirta, Teresa Israels, Terrance P Felegie, Sr, Terry Berryhill, Thanvi Gorla, Thibault Adam, Thijn Kriesels, Thomas Brindle, Thomas E Collett, Thomas Gravekamp, Thomas Rosa, Thomas Wigbels, Thorben Ruthke, Thorsten Eschweiler, Théo Janvier, Théo Neveu, Tiffany Lloyd, Tim Albrecht, Tim Pilz, Timur Podornicov, Tina Mathews, Tirthankar De, Titiksha Pandit, Tobias Elsner, Tom Bickle, Tom Gilde, Tom Udding, Tommy Reynolds, Tone Iversen, Tony Frost, Tora Frankic, Tora Frankic, Trace Jeffries-Ortiz, Tristan Marchand, Tyler Atkinson, Udaiveer Singh, Ushakov Rostislav Konstantinovich, Uswa Fatima, Vaishnavi Shinde, Vajda János, Valasia Anna Valavosiki, Valentino Dalla Verde, Valeria Bondarenko, Valerie Sousa, Vanisha Thiyyagura, Vansh Bahety, Veda Vishal, Veronica Couñago Vidal, Veronica Motta, Veronika Sivayeva, Victor Alteirac, Victor Gabriel Bibé, Victor Juan Garcia Porcel, Victor Teodor Cozea, Victor Yu, Victor-Francois Ros, Victoria Alicja Machnik, Viera Holik, Vika Manjgaladze, Ville Mänttäri, Ville Tissari, Vincent Decker, Vincent Parmentier, Vincenzo Manto, Virgilio Gonano, Vishakha Jingar, Vivian Jin, Vlad Andrei Manolescu, Vladimir Ezhelyy, Vladimir Komarov, Vladimir Tishkin, Vladimír Šintaj, Walter Feichtinger, Walter MacDonald, Wania Latif, Warren Kinninger, Wasif Amin, Wayne Stripe III, Wendy Hawkins, Wesley Webb, Wiesław Feduniewicz, Wihan van Heerden, William Balmer, William Beaudoin, William Hawkesworth, William MacLennan, William Russell, Winfried Schoberth, Wolfgang G. Kurtz, Xi Liao, Xiangyu Huang, Xincheng Zhu, Xingping Lin, Xuanlin Jiang, Yamil A. Montañez Díaz, Yang Feichi, Yann Lelievre, Yannick Facolle, Yaroslava Brilli, Yassen Mahmoud, Yevheniia Furutina, Yi Ying Mok, Yi-Wen Lasure, Yurii Andryshin, Zachary Benfield, Zachary Kronlage, Zachary Swift, Zachary Wood, Zain R.-Schill, Zainab Ahmad, Zainah Imran Waheed, Zakariya El Jilali, Zhao Long Dong, Zharov Dmitriy, Zipora Morgan, Ziyi Zhou, Zoe Posnette, Zoe Spasaro, Zoltán Huszár, Zoran Von Orel Bronte, Đoàn Khánh Hưng, Łukasz Hyrchel-Sakowicz, Šimon Urbaník, \foreignlanguage{greek}{ΒΑΣΙΛΙΚΗ ΠΕΤΡΟΠΟΥΛΟΥ}, \foreignlanguage{kazakh}{Банияз Сағи}, \foreignlanguage{russian}{Дмитрий Сергеевич Фёдоров, Дмитриков Федор, Евгений Юрченко, Исаков Евгений Дмитриевич, Костя Гоман, Мясников Иван Артемович, Панаиотова Наталья Александровна, Сапожникова София, Цыро Снежана, Юлия Емельянова}, 余映逸(Ying Yat,Yu), 片山洋臣 Yoshifumi Katayama}
\end{CJK}
\end{spacing}

\end{document}